\documentclass[sigconf,nonacm]{acmart}

\usepackage{subcaption}

\usepackage{todonotes} 

\usepackage{csquotes} 

\usepackage{setspace} 

\usepackage{libertine}

\usepackage{mathtools} 

\usepackage[most]{tcolorbox}
\usepackage{listings, color}

\usepackage{xcolor}

\usepackage{booktabs}

\usepackage[most]{tcolorbox}
\usepackage{listings, color}
\usepackage{paralist}
\usepackage[inline]{enumitem} 
\usepackage{fancyvrb}
\usepackage{hyperref}

\usepackage{xcolor}

\usepackage[noend,linesnumbered]{algorithm2e}

\usepackage[english]{babel} 

\usepackage{tikz}
\usepackage{ulem}
\newif\iflon
\lontrue                        
\iflon
\newcommand{\iflong}[1]{#1}
\newcommand{\ifshort}[1]{}
\else
\newcommand{\iflong}[1]{}
\newcommand{\ifshort}[1]{#1}
\fi

\newcommand{\shortlong}[2]{\ifshort{#1}\iflong{#2}}

\newtheorem{remark}{Remark}
\newtheorem{property}{Property}

\newtheorem{theorem}{Theorem}
\newtheorem{example}{Example}

\iflon
\usepackage[appendix=inline]{apxproof} 
\else
\usepackage[appendix=strip]{apxproof}
\fi

\newtheoremrep{lemmax}[theorem]{Lemma}
\newtheoremrep{theox}[theorem]{Theorem}
\newtheoremrep{corx}[theorem]{Corollary}

  {\small \em \tt ccc\begin{table} aaa bbb}%
  {\end{table} ddd}
  
\newcommand{\Tr}[1]{\BeginTr{#1}\EndTr}  
\newcommand{\BeginTr}{\ensuremath{\langle}}
\newcommand{\EndTr}{\ensuremath{\rangle}}

\newcommand{\json}{JSON}
\newcommand{\kw}[1]{\textbf{#1}}
\renewcommand{\kw}[1]{\ensuremath{\mathtt{#1}}}
\newcommand{\qkw}[1]{\ensuremath{\mathtt{\QQ{#1}\QQ}}}
\newcommand{\key}[1]{\ensuremath{\mathit{#1}}}
\newcommand{\qkey}[1]{\ensuremath{\mathit{\QQ{#1}\QQ}}}
\newcommand{\akey}[1]{\ensuremath{\mathsf{#1}}}
\newcommand{\qakey}[1]{\ensuremath{\mathsf{\QQ{#1}\QQ}}}

\newcommand{\xnot}{\kw{not}}
\newcommand{\qnot}{\qkw{not}}
\newcommand{\xtrue}{\kw{true}}

\newcommand{\xfalse}{\kw{false}}

\newcommand{\xnull}{\kw{null}}
\newcommand{\qnull}{\qkw{null}}
\newcommand{\xone}{\kw{oneOf}}
\newcommand{\qone}{\qkw{oneOf}}
\newcommand{\xany}{\kw{anyOf}}
\newcommand{\qany}{\qkw{anyOf}}

\newcommand{\qall}{\qkw{allOf}}

\newcommand{\qmin}{\qkw{minimum}}

\newcommand{\qmax}{\qkw{maximum}}
\newcommand{\xreq}{\kw{required}}
\newcommand{\qreq}{\qkw{required}}

\newcommand{\qtype}{\qkw{type}}

\newcommand{\qexmin}{\qkw{exclusiveMinimum}}

\newcommand{\qexmax}{\qkw{exclusiveMaximum}}
\newcommand{\xprops}{\kw{properties}}
\newcommand{\qprops}{\qkw{properties}}
\newcommand{\xpropN}{\kw{propertyNames}}
\newcommand{\qpropN}{\qkw{propertyNames}}
\newcommand{\xpattProps}{\kw{patternProperties}}
\newcommand{\qpattProps}{\qkw{patternProperties}}

\newcommand{\qmaxP}{\qkw{maxProperties}}
\newcommand{\xthen}{\kw{then}}
\newcommand{\qthen}{\qkw{then}}
\newcommand{\xif}{\kw{if}}
\newcommand{\qif}{\qkw{if}}
\newcommand{\xelse}{\kw{else}}
\newcommand{\qelse}{\qkw{else}}

\newcommand{\qite}{\qif-\qthen-\qelse}
\newcommand{\xaddProps}{\kw{additionalProperties}}
\newcommand{\qaddProps}{\qkw{additionalProperties}}

\newcommand{\xaddIts}{\kw{additionalItems}}
\newcommand{\qaddIts}{\qkw{additionalItems}}

\newcommand{\qmof}{\qkw{multipleOf}}

\newcommand{\xmaxL}{\kw{maxLength}}
\newcommand{\qmaxL}{\qkw{maxLength}}
\newcommand{\xminL}{\kw{minLength}}
\newcommand{\qminL}{\qkw{minLength}}
\newcommand{\xpatt}{\kw{pattern}}
\newcommand{\qpatt}{\qkw{pattern}}
\newcommand{\xuniqIt}{\kw{uniqueItems}}
\newcommand{\quniqIt}{\qkw{uniqueItems}}
\newcommand{\xcont}{\kw{contains}}
\newcommand{\qcont}{\qkw{contains}}

\newcommand{\xminC}{\kw{minContains}}
\newcommand{\qminC}{\qkw{minContains}}
\newcommand{\xmaxC}{\kw{maxContains}}
\newcommand{\qmaxC}{\qkw{maxContains}}

\newcommand{\qminIt}{\qkw{minItems}}
\newcommand{\xmaxIt}{\kw{maxItems}}

\newcommand{\xit}{\kw{items}}
\newcommand{\qit}{\qkw{items}}

\newcommand{\xdeps}{\kw{dependencies}}
\newcommand{\qdeps}{\qkw{dependencies}}

\newcommand{\xenum}{\kw{enum}}
\newcommand{\qenum}{\qkw{enum}}
\newcommand{\xconst}{\kw{const}}
\newcommand{\qconst}{\qkw{const}}
\newcommand{\xdref}{\kw{\$ref}}
\newcommand{\qdref}{\qkw{\$ref}}

\newcommand{\xdefs}{\kw{definitions}}
\newcommand{\qdefs}{\qkw{definitions}}

\newcommand{\xformat}{\kw{format}}

\newcommand{\qstr}{\qakey{string}}

\newcommand{\qboolean}{\qakey{boolean}}
\newcommand{\xid}{\kw{id}}

\newcommand{\xdid}{\kw{\$id}}

\newcommand{\xda}{\kw{\$anchor}}

\newcommand{\DefWithKey}[3]{{#1}{#2}:{#3}}
\newcommand{\Def}[2]{\DefWithKey{\DefKey\ }{#1}{#2}}
\renewcommand{\Def}[2]{\DefWithKey{}{#1}{#2}}
\newcommand{\Doc}[2]{{#1}\ \Defs\ ({#2})}
\newcommand{\Defs}{\akey{defs}}

\newcommand{\RRef}[1]{\key{#1}}

\newcommand{\gcomment}[1]{}
\newcommand{\oldversion}[1]{}
\newcommand{\hideforspace}[1]{}
\newcommand{\hide}[1]{}
\newcommand{\code}[1]{}

\newcommand{\Iff}{\Leftrightarrow}
\newcommand{\Implies}{\Rightarrow}
\newcommand{\RevImplies}{\Leftarrow}
\newcommand{\Or}{\vee}
\newcommand{\BigOr}{\bigvee}

\renewcommand{\comment}[1]{}
\newcommand{\ROBDDT}{ROBDDTab}

\renewcommand{\And}{\wedge}
\newcommand{\BigAnd}{\bigwedge}

\newcommand{\Not}{\neg}
\newcommand{\True}{{\bf t}}
\newcommand{\False}{{\bf f}}
\newcommand{\Type}{\akey{type}}
\newcommand{\Num}{\akey{Num}}
\newcommand{\Str}{\akey{Str}}
\newcommand{\Int}{\akey{Int}}
\newcommand{\Arr}{\akey{Arr}}
\newcommand{\Obj}{\akey{Obj}}
\newcommand{\Bool}{\akey{Bool}}
\newcommand{\Null}{\akey{Null}}
\newcommand{\TT}[1]{\Type({#1})}
\newcommand{\CT}[1]{\{\Type({#1})\}}
\newcommand{\TNum}{\TT{\Num}}
\newcommand{\TStr}{\TT{\Str}}

\newcommand{\TArr}{\TT{\Arr}}
\newcommand{\TObj}{\TT{\Obj}}
\newcommand{\TBool}{\TT{\Bool}}
\newcommand{\TNull}{\TT{\Null}}
\newcommand{\CTNum}{\CT{\Num}}
\newcommand{\CTStr}{\CT{\Str}}

\newcommand{\CTArr}{\CT{\Arr}}

\newcommand{\CTBool}{\CT{\Bool}}
\newcommand{\CTNull}{\CT{\Null}}
\newcommand{\Mof}{\akey{mulOf}}
\newcommand{\NotMof}{\akey{notMulOf}}
\newcommand{\Pat}{\akey{pattern}}

\newcommand{\Uni}{\akey{uniqueItems}}
\newcommand{\NotUni}{\akey{repeatedItems}}
\newcommand{\CProp}[2]{\Props(\key{#1}:{#2})}
\newcommand{\CProps}[1]{\Props(#1)}
\newcommand{\Props}{\akey{props}}

\newcommand{\IteK}{\akey{items}}
\newcommand{\Ite}[2]{\IteK(#1;#2)}
\newcommand{\PreIteK}{\akey{item}}
\newcommand{\PreIte}[2]{\PreIteK({#1}:{#2})}
\newcommand{\PostIteK}{\akey{items}}
\newcommand{\PostIte}[2]{\PostIteK({#1}^+:{#2})}

\newcommand{\Bet}{\akey{betw}}
\newcommand{\XBet}{\akey{xBetw}}

\newcommand{\Ex}{\#}

\newcommand{\ContAfterK}{\akey{contAfter}}
\newcommand{\ContAfter}[2]{\ContAfterK({#1}^+:{#2})}

\newcommand{\Pro}{\akey{pro}}
\newcommand{\Nam}{\akey{pNames}}
\newcommand{\keykey}[1]{\key{\underline{#1}}}
\newcommand{\Con}{\akey{const}}
\newcommand{\Enu}{\akey{enum}}

\newcommand{\IBT}{\akey{isBoolValue}}
\renewcommand{\IBT}{\akey{ifBoolThen}}
\newcommand{\Inf}{\infty}

\newcommand{\PReq}{\akey{pattReq}}

\newcommand{\Req}{\akey{req}}

\newcommand{\VerThree}{Draft-03}
\newcommand{\VerFour}{Draft-04}

\newcommand{\VerSix}{Draft-06}

\newcommand{\VerEight}{Draft 2019-09}
\newcommand{\VerTwenty}{Draft 2020-12}
\newcommand{\JS}{JSON Sche\-ma}
\newcommand{\Nat}{\mathbb{N}}
\newcommand{\IntSet}{\mathbb{Z}}

\newcommand{\POfS}{\key{PattOfS}}
\newcommand{\TrueP}{\True^{\ptt}}
\renewcommand{\TrueP}{\ensuremath{.*}}
\newcommand{\FalseP}{\False^{\ptt}}
\renewcommand{\FalseP}{\CoP{\TrueP}}
\newcommand{\AndP}{\And^{\ptt}}
\newcommand{\BigAndP}{\BigAnd^{\ptt}}

\renewcommand{\AndP}{\sqcap}
\renewcommand{\BigAndP}{\bigsqcap}

\newcommand\Overline[2][1pt]{%
    \begin{tikzpicture}[baseline=(a.base)]
      \node[inner xsep=0pt,inner ysep=1.5pt] (a) {$#2$};
      \draw[line width= #1] (a.north west) -- (a.north east);
    \end{tikzpicture}
    }

\newcommand{\CoP}[1]{\overline{({#1})}}
\renewcommand{\CoP}[1]{\Overline[0.75pt]{#1}}
\newcommand{\CoPP}[1]{\CoP{(#1)}}     

\newcommand{\To}{\rightarrow}

\newcommand{\custcom}[2]{\marginpar{\tiny #1: {#2}}}

\newcommand{\GG}[1]{\custcom{giorgio}{#1}}

\newcommand{\M}{\ |\ }

\newlength{\NL}
\newlength{\SaveNL}
\newcommand{\StoreNL}{\setlength{\SaveNL}{\NL}}
\newcommand{\UpdateNL}[1]{\setlength{\SaveNL}{\NL}\setlength{\NL}{#1}}
\newcommand{\RestoreNL}{\setlength{\NL}{\SaveNL}}

\newcommand{\NotVarFun}{\key{co}}
\newcommand{\NotVar}[1]{\NotVarFun(\key{#1})}
\newcommand{\Sat}[2]{{#1}\vDash{#2}}

\newcommand{\NN}{\ensuremath{\ \hat{}\ }}
\newcommand{\QQ}{\textnormal{\textquotedbl}}
\newcommand{\Set}[1]{\{\,{#1}\,\}}
\newcommand{\SetTo}[1]{\{1..{#1}\}}

\newcommand{\SetOpen}{\{\!|}
\newcommand{\SetClose}{|\!\}}
\renewcommand{\Set}[1]{\SetOpen{#1}\SetClose}

\newcommand{\Parts}{\mathcal{P}}
\newcommand{\Vars}{\key{Vars}}

\newcommand{\NR}[1]{\RRef{\NotVar{#1}}}

\newcommand{\nule}{\nullv}
\newcommand{\num}{q}
\newcommand{\nullv}{\ensuremath{\text{null}}}
\newcommand{\str}{s}

\newcommand{\J}{{J}}

\newcommand{\truev}{\ensuremath{\text{true}}}
\newcommand{\falsev}{\ensuremath{\text{false}}}

\newcommand{\semt}{\ensuremath{\mathit{JVal}}}
\newcommand{\semcapar}[3]{[\![ #1 ]\!]_{#2}^{#3}}
\newcommand{\semcap}[2]{\semcapar{#1}{#2}{p}}
\newcommand{\semcai}[2]{\semcapar{#1}{#2}{i}}
\newcommand{\semca}[2]{[\![ #1 ]\!]_{#2}}
\newcommand{\E}{E}

\newcommand{\semas}[2]{\langle\!\langle #1 \rangle\!\rangle_{#2}}

\newcommand{\setst}[2]{\{ #1 \ | \ #2\}}
\renewcommand{\setst}[2]{\SetOpen\, #1 \,\mid\, #2 \, \SetClose}   
\newcommand{\rlan}[1]{L(#1)}

\newcommand{\TGIN}[1]{\akey{#1}}
\newcommand{\TG}[1]{\{\TGIN{#1}\}}

\newcommand{\Bo}{\TGIN{\SBo}}

\newcommand{\St}{\TGIN{\SSt}}
\newcommand{\Ob}{\TGIN{\SOb}}

\newcommand{\SBo}{{Bool}}

\newcommand{\SSt}{{Str}}
\newcommand{\SOb}{{Obj}}

\newcommand{\PPP}[1]{\NN{#1}\$}
\newcommand{\Next}{\kw{next}}
\newcommand{\EOV}{\kw{EOV}}
\newcommand{\DOV}{\kw{DOV}}
\newcommand{\Pop}{\emph{Populated}}
\newcommand{\Open}{\emph{Open}}

\newcommand{\RR}{\ensuremath{\mathcal{R}}}

\newcommand{\jsonsch}{JSON Schema} 

\newcommand{\CMMK}{\akey{cont}}
\newcommand{\CMM}[3]{\ensuremath{\CMMK_{#1}}^{#2}({#3})}

\newcommand{\JSet}{\semt(*)}

\newcommand{\Crcombination}{Cr-combination}
\newcommand{\crcombination}{cr-combination}    
\newcommand{\crcombine}{cr-combine}
\newcommand{\crcombined}{cr-combined}

\renewcommand{\Crcombination}{Preparation}
\renewcommand{\crcombination}{preparation}    
\renewcommand{\crcombine}{prepare}
\renewcommand{\crcombined}{prepared}

\newcommand{\PN}{\mathit{poly(N)}}
\newcommand{\Dep}{\delta}
\newcommand{\XTrue}{x_{\True}}

\newcommand{\TO}{TO}
\newcommand{\ITO}{ITO}
\newcommand{\TE}{TO}

\newif{\ifMarginalComments}
 \MarginalCommentstrue   

 \definecolor{orange}{rgb}{1,0.45,0}
\definecolor{darkblue}{rgb}{0,0.2,0.8}
\definecolor{darkgreen}{rgb}{0.1,0.4,0.1}
\definecolor{darkred}{rgb}{0.8,0,0}
\definecolor{lightblue}{rgb}{0.9,0.9,1}
\definecolor{lightgreen}{rgb}{0.9,1,0.9}

\newcommand{\mrevtwo}[2]{\textcolor{orange}{#1}\marginpar{\textcolor{orange}{#2}}}
\newcommand{\mrevthree}[2]{\textcolor{darkgreen}{#1}\marginpar{\textcolor{darkgreen}{#2}}}
\newcommand{\mrevmeta}[2]{\textcolor{darkblue}{#1}\marginpar{\textcolor{darkblue}{#2}}}

\newcommand{\crevmeta}[1]{\textcolor{darkblue}{#1}}

\newcommand{\bcolorthree}{\color{darkgreen}}
\newcommand{\bcolormeta}{\color{darkblue}}
\newcommand{\ecolormeta}{\color{black}}
\newcommand{\ecolor}{\color{black}}

\iflon
\renewcommand{\bcolorthree}{\color{black}}
\renewcommand{\bcolormeta}{\color{black}}
\renewcommand{\ecolormeta}{\color{black}}
\renewcommand{\ecolor}{\color{black}}

\renewcommand{\mrevtwo}[2]{#1}
\renewcommand{\mrevthree}[2]{#1}
\renewcommand{\mrevmeta}[2]{#1}

\renewcommand{\crevmeta}[1]{{#1}}
\fi

\newcommand{\nop}[1]{{}} 

\makeatletter
\newcommand\querysize{\@setfontsize\querysize\@vipt\@viipt}
\makeatother

\lstdefinestyle{query}{
  stepnumber=1,
  numbersep=3pt, 
  tabsize=4,
  showspaces=false,
  showstringspaces=false,
  basicstyle=\linespread{1}\fontfamily{lmtt}\selectfont\querysize,
  keywordstyle=\color{blue},
  stringstyle=\color{purple},
  upquote=true,
  breaklines=true,
  commentstyle=\color{CadetBlue}
}

\definecolor{mygray}{rgb}{0.643,0.643,0.643}

\newtcolorbox{querybox}[2][]{%
  sidebyside align=top,
  enhanced,
  boxsep=0pt,
  arc=0pt,
  top=-3pt, bottom=-3pt,
  left=2pt, right=0pt,
  colback=white,
  colframe=mygray,
  boxrule=0.5pt,
  leftrule=12pt,
  overlay unbroken and first ={%
    \node[rotate=90,
          minimum width=0.5cm,
          anchor=south,
          font=\small\rmfamily,
          yshift=-13pt,
          white]
    at (frame.west) {#2};
  }
} 
\newtcolorbox{querybox2l}[2][]{%
  sidebyside align=top,
  enhanced,
  boxsep=0pt,
  arc=0pt,
  top=-7pt, bottom=-10.5pt,
  left=2pt, right=0pt,
  colback=white,
  colframe=mygray,
  boxrule=0.5pt,
  leftrule=10pt,
  overlay unbroken and first ={%
    \node[rotate=90,
          align=center,
          minimum width=0.5cm,
          anchor=south,
          font=\small\rmfamily,
          yshift=-12pt,
          white]
    at (frame.west) {#2};
  }
}

\makeatletter
\AtBeginDocument{\def\libertine@figurealign{}\libertineOsF}
\newcommand\libertineTabular{\def\libertine@figurealign{T}\libertineLF}
\makeatother

\hypersetup{draft} 

\begin{document}

\title{Witness Generation for JSON Schema} 

\author{Lyes Attouche}
\affiliation{%
  \institution{Universit\'e Paris-Dauphine -- PSL} 
  \country{}
}
\email{lyes.attouche@dauphine.fr}

\author{Mohamed-Amine Baazizi}
\affiliation{%
  \institution{Sorbonne Universit\'e, LIP6 UMR 7606}
  \country{}
}
\email{baazizi@ia.lip6.fr}
\author{Dario Colazzo}
\affiliation{%
  \institution{Universit\'e Paris-Dauphine -- PSL} 
  \country{}
}
\email{dario.colazzo@dauphine.fr}

\author{Giorgio Ghelli}
\affiliation{%
  \institution{Dip. Informatica, 
                  Universit\`a di Pisa}
                  \country{}
}
\email{ghelli@di.unipi.it}

\author{Carlo Sartiani}
\affiliation{%
  \institution{DIMIE, Universit\`a della Basilicata}
  \country{}
}
\email{carlo.sartiani@unibas.it}
\author{Stefanie Scherzinger}
\affiliation{%
  \institution{Universit{\"a}t Passau}
  \country{}
}
\email{stefanie.scherzinger@uni-passau.de}

\begin{abstract}
\iflong{
{\jsonsch} is \shortlong{a}{an important, evolving} standard schema language for families of JSON documents.
It is based on a complex combination of structural operators,
Boo\-le\-an operators, including full negation, 
and mutually recursive variables.
The static analysis of {\jsonsch} documents comprises practically relevant problems,
including schema satisfiability, inclusion, and equivalence.
These three can be reduced to witness generation: 
given  a sche\-ma, generate an element of the schema --- if it exists --- otherwise report failure.
Schema satisfiability, inclusion, and equivalence have been shown to be decidable, by reduction to reachability in alternating tree automata.
However, no witness generation algorithm has yet been formally described.
We contribute a first, direct algorithm for {\jsonsch} witness generation.
We study its effectiveness and efficiency, in experiments over several schema collections, including thousands of real-world schemas.
\iflong{Our focus is on the completeness of the language (where we only exclude the $\quniqIt$ operator)
and on the ability of the algorithm to run in reasonable time on a large set of real-world examples,
despite the exponential complexity of the problem.}}

\ifshort{{\jsonsch} is a schema language for JSON documents, based 
on a complex combination of structural operators,
Boo\-le\-an operators (negation included), 
and recursive variables.
The static analysis of {\jsonsch} documents comprises practically relevant problems,
including schema satisfiability, inclusion, and equivalence.
These problems can be reduced to witness generation: 
given  a sche\-ma, generate an element of the schema --- if it exists --- and report failure otherwise.
Schema satisfiability, inclusion, and equivalence have been shown to be decidable.
However, no witness generation algorithm has yet been formally described.
We contribute a first, direct algorithm for {\jsonsch} witness generation, and study its effectiveness and efficiency in experiments over several schema collections, including thousands of real-world schemas.
\iflong{Our focus is on the completeness of the language (where we only exclude the $\quniqIt$ operator)
and on the ability of the algorithm to run in a reasonable time on a large set of real-world examples,
despite the exponential complexity of the underlying problem.}}
\end{abstract}

\keywords{JSON Schema, witness generation, inclusion, equivalence}

\maketitle

%

\section{Introduction}

\iflong{This paper is about witness generation for {\jsonsch}~\cite{jsonschema},
the de-facto standard schema language for {\json}~\cite{DBLP:conf/www/PezoaRSUV16,DBLP:conf/edbt/BaaziziCGS19,DBLP:conf/pods/BourhisRSV17,DBLP:conf/sigmod/BaaziziCGS19}.}

{\jsonsch} is a schema language
based on a set of \emph{assertions} that describe features of the {\json}
values described and on logical and structural combinators for these assertions. 

\iflong
{
The semantics of this language can be subtle.
For instance, the two schemas below differ in their syntax, but are in fact equivalent. 
Schema~a) explicitly states that any instance must be an object, 
and that a property named \enquote{foo} is not allowed.
Schema~b) implicitly requires the same: the \xreq\ keyword has implicative semantics,
stating that \emph{if} the instance is an object, it must contain a property named \enquote{foo}.
Via negation, it is enforced that the instance must be an object, where a property named \enquote{foo}
is not allowed. 
\iflong{While this specific example is artificial, it exemplifies the most common usage of \xnot\ in {\jsonsch}~\cite{DBLP:conf/er/BaaziziCGSS21}.}

\begin{center}
\begin{minipage}{0.65\linewidth}
\begin{querybox}{(a)}
\begin{lstlisting}[style=query]
{ "type": "object",
  "properties": { "foo": false } }
\end{lstlisting}
\end{querybox}
\vspace{-0.15cm}
\begin{querybox}{(b)}
\begin{lstlisting}[style=query]
{ "not": { "required": ["foo"] } }

\end{lstlisting}
\end{querybox}

\end{minipage}
\end{center}

}

\iflong{\emph{Validation} of a {\json} value $J$ with respect to a {\jsonsch} schema~$S$, denoted $\Sat{J}{S}$, is 
a well-understood problem that can be solved in time $O(|J|^{2}|S|)$ 
\cite{DBLP:conf/www/PezoaRSUV16}. The JSON Schema Test Suite~\cite{json-schema-test-suite}, a collection of validation tests,
lists over 50 validator tools, at the time of writing.
Yet there are static analysis problems, equally relevant, 
where we still lack well-principled tools.
We next outline these problems, and then point out that they can be ultimately reduced to {\jsonsch} witness generation, the focus of this work.}

\ifshort{While \emph{validation} of a {\json} value $J$ with respect to a {\jsonsch} schema~$S$, denoted $\Sat{J}{S}$, is 
a well-studied problem for which the JSON Schema Test Suite~\cite{json-schema-test-suite}
lists over 50 validator tools at the time of writing, 
for the main static analysis problems, which we describe below, 
we still lack well-principled tools.}

\emph{Inclusion}  $S \subseteq  S^{\prime}$: does, for each value $J$,  $\Sat{J}{S} \Implies \Sat{J}{S^{\prime}}$?
Checking schemas for inclusion (or containment)
is of great practical importance: if the output format of a tool is specified by a schema~$S$,
and the input format of a different tool by a schema~$S'$, the problem of format compatibility is equivalent to schema inclusion $S \subseteq  S^{\prime}$; given the high expressive power of {\jsonsch}, this ``format'' may actually include detailed information about the range of specific parameters.
For example, 
the IBM ML framework LALE~\cite{baudart_et_al_2020-automl_kdd} adopts an incomplete  inclusion checking algorithm for {\jsonsch}, to improve safety of ML pipelines~\cite{DBLP:conf/issta/HabibSHP21}.

\iflong{Schema inclusion also plays a central role in schema evolution, with questions of the kind: 
will a value that respects the new schema still be accepted by tools designed for legacy versions?}
\iflong{If not, what is an example of a problematic value?}

\emph{Equivalence} $S \equiv  S^{\prime}$: does, for each value $J$, $\Sat{J}{S} \Iff \Sat{J}{ S^{\prime}}$? Checking equivalence builds upon inclusion, and is relevant in designing workbenches for schema analysis and simplification~\cite{DBLP:conf/vldb/FruthDS21}.

\emph{Satisfiability} of $S$: does a value $J$ exist such that  $\Sat{J}{S}$?

\iflong{%
Note that the above problems are strictly interrelated. Indeed, as {\jsonsch} includes the Boolean algebra, schema inclusion and satisfiability are equivalent: $S \subseteq  S^{\prime}$ if and only if $S \And \Not S^{\prime}$ is not satisfiable, and $S$ is satisfiable if and only if $S \not\subseteq  {\xfalse}$, where ${\xfalse}$ is the schema that no {\json} document can match.
}

\emph{Witness generation} for $S$, a constructive generalization of satisfiability:
given $S$, generate a value $J$ such that $\Sat{J}{S}$, or return ``unsatisfiable'' if no such value exists. In the first case, we call $J$ a \emph{witness}.
Schema inclusion $S \subseteq  S^{\prime}$ can be immediately 
reduced to 
witness generation for $S \And \Not S^{\prime}$, but with a crucial advantage: if a witness~$J$ for $S \And \Not S^{\prime}$ is generated, we can provide users with an 
explanation: $S$ is not included in $S^{\prime}$ \emph{because} of values such as $J$. 
We can similarly solve a ``witnessed'' version of equivalence: given
 $S$ and $S^{\prime}$, either prove that one is equivalent to the other, or provide an explicit witness $J$ that belongs to one, but not to the other.

\ifshort{The techniques and the notions that we present in this paper can be also useful for the design of 
 \emph{example generation} algorithms, that is, algorithms that do not just generate one arbitrary witness, but
generate many of them, according to some heuristics aimed to fulfill criteria of ``completeness'' and ``realism''.
}

\iflong{A witness generation algorithm, besides its use for the solution of witnessed inclusion, is the first step in the design of 
\emph{complete enumeration} and \emph{example generation} algorithms.
Here, \emph{complete enumeration} is any algorithm, in general non-terminating, that, for a given $S$, 
enumerates every $J$ that satisfies $S$. With \emph{example generation}, we indicate any enumeration algorithm that
is not necessarily complete, but pursues some 
``practical'' criterion in the choice of the generated witnesses, such as the ``realism'' of the 
base values, or some form of coverage of the different cases allowed by the schema.
Example generation is extremely useful in the context of test-case generation, 
and also as a  tool to understand complex schemas through realistic examples.}

\bcolormeta
\paragraph{Open challenges}
\ecolormeta

\mrevmeta{
Witness generation for {\JS} is difficult. Existing tools are incomplete and struggle with this task (as we will show in our experiments).
First of all, {\JS} includes conjunction, disjunction, negation, modal
(or \emph{structural}) operators, recursive second-order variables, and recursion under negation.
Secondly, 
for each JSON type, the different structural operators have complex interactions,
as in the following example, where \qreq\ and the negated \qpattProps\ force the presence of fields whose names match 
$\qkw{\NN{a}}$ and $\qkw{\NN{abz}\$}$ (this is explained in the paper),
\qmaxP\ : 1 \ forces these two fields to be one, and, finally,  \qpattProps\ forces the value of that field
to satisfy \emph{var2}, since \qkw{abz} also matches \qkw{z\$}.
}{M.1}

\bcolormeta

{\small
\begin{verbatim}
{"required":["abz"],
 "not":{"patternProperties":{"^a":{"$ref":"#/$defs/var1"}}},
 "maxProperties":1,
 "patternProperties":{"z$":{"$ref":"#/$defs/var2"}},
 "$defs" : ...
}
\end{verbatim}
}

\hide{code to test:

{"required":["abz"],
  "not":{"patternProperties":{"^a":{"$ref":"#/$defs/var1"}}},
  "maxProperties":1,
  "patternProperties":{"z$":{"$ref":"#/$defs/var2"}},
  "$defs" : {"var1" : {"type" : "number", "multipleOf" : 10},
                  "var2" : {"type" : "number", "multipleOf" : 5}
                }
}

}

\hide{
\begin{center}
\begin{minipage}{1\linewidth}
\begin{querybox}{}
\begin{lstlisting}[style=query]
{ "not": {"patternProperties" : {"^a":  {"$ref": "#/$defs/var1"}}},
  "required": ["abz"],
   "patternProperties": {"z$": {"$ref": "#/$defs/var1"}},
  "maxProperties": 1
}
\end{lstlisting}
\end{querybox}
\vspace{-0.15cm}

\end{minipage}
\end{center}}

Each aspect would make the problem computationally intractable by itself.
Their combination exacerbates the difficulty of the design of a \emph{complete} algorithm that is \emph{practical}, that is, of an algorithm
that is correct and complete by design, but is also 
able to run in a reasonable time over the vast majority of real-world schemas.

\ecolormeta

\hide{Bouhris et al.\  \cite{DBLP:conf/pods/BourhisRSV17}, based on results of Pezoa et al.\ in~\cite{DBLP:conf/www/PezoaRSUV16}, 
 proved that satisfiability (and, hence, inclusion) is  EXPTIME-complete for 
schemas \emph{without} {\Uni}, by translating {\jsonsch} onto an equivalent modal logic, \emph{recursive JSL}, for which a
class of alternating tree automata is defined.
This is extremely important, as it quantifies the intrinsic complexity of the problem. However, the actual design
of a complete and practical algorithm for these static analysis problems remains open.}

\paragraph{Contributions}

 \mrevmeta{The main contribution of this paper is an original sound and complete algorithm for checking the satisfiability of an input schema~$S$, generating a witness~$J$ when the schema is satisfiable. Our algorithm supports the whole language without {\Uni}. 
While the existence of an algorithm for this specific problem follows from the results in 
\cite{DBLP:conf/pods/BourhisRSV17}, where the  problem is proved to be EXPTIME-complete,
we are the first to explicitly describe an algorithm, and specifically one that has the potential
to work in reasonable time over schemas of  realistic size.
Our algorithm is based on a set of formal manipulations of the schema, some of which,
such as \emph{\crcombination}, are unique to {\jsonsch}, and have not been proposed before in this form.
Particularly relevant in this context is the notion of \emph{lazy and-completion}, which we will describe later.
In this paper, we detail each algorithm phase, show that each is in~$O(2^{\PN})$, and focus on  {\crcombination} and generation of \shortlong{objects}{objects and arrays}, the phases completely original to this work.
}{M.4, 2.6}

\mrevmeta{The practical applicability of our algorithm is proved by our experimentation,
which is another contribution of this work.
Our experiments are based on  four real-world datasets, on a synthetic dataset,
and on a handwritten dataset. Real-world datasets comprise 6,427 unique schemas extracted, through an extensive data cleaning process, 
from a large corpus of schemas crawled from  GitHub~\cite{schema_corpus} and curated by us for errors and redundancies;
the other datasets, already used in \cite{DBLP:conf/issta/HabibSHP21}, are related to specific application domains and originated from Snowplow\cite{snow}, The Washington Post~\cite{wp}, and Kubernetes~\cite{kuber}.}{M.4, 2.1}
The synthetic dataset is synthesized from the standard schemas provided by \emph{JSON Schema Org}~\cite{json-schema-test-suite}, 
from which we derive schemas that are known to be satisfiable or unsatisfiable by design~\cite{DBLP:conf/er/AttoucheBCDFGSS21}.
The handwritten dataset is specifically engineered to test the most complex aspects of the {\jsonsch} language.
The experiments show that our algorithm is complete, and that, despite its exponential complexity, it behaves quite well even on schemas with tens  of thousands of nodes.
\iflong{Overall, we can show that  our contributions advance the state-of-the-art.}

\iflong{Our implementation of the witness generation algorithm is available as open source.
The code is part of a fully automated reproduction package~\cite{repro_package},
which contains all input data, as well as the data generated in our experiments.
For convenience, our implementation is also accessible as an interactive web-based tool~\cite{onlinetool}.}

\shortlong{\subsubsection*{Paper outline}}{\subsection*{Paper outline}}

\ifshort{
In Section~\ref{sec:relworks} we review related work. In Section~\ref{sec:prelim} we  describe our algebraic representation of {\jsonsch}.  In Sections~\ref{sec:witness}--\ref{sec:recgen}, we describe the algorithm. In Section \ref{sec:exp} we present our experimental evaluation.  We conclude in
Section \ref{sec:concl}.
}

\iflong{The rest of the paper is organized as follows.
In Section~\ref{sec:relworks} we analyze related work. In Section~\ref{sec:prelim} we briefly describe {\json} and {\jsonsch}. In Sections~\ref{sec:algebra} and~\ref{sec:translation}  we introduce our algebraic framework.  In Sections~\ref{sec:witness}, \ref{sec:firstphases}, and \ref{sec:recgen}, we describe the structure of the algorithm, the initial phases, and the last phases. In Section \ref{sec:exp} we present an extensive experimental evaluation of our approach.  In Section \ref{sec:concl}, we draw our conclusions.
}

\section{Related Work}\label{sec:relworks}

Overviews over schema languages for {\json} can be found 
in~\cite{DBLP:conf/www/PezoaRSUV16,DBLP:conf/edbt/BaaziziCGS19,DBLP:conf/pods/BourhisRSV17,DBLP:conf/sigmod/BaaziziCGS19}.
 Pezoa et al.~\cite{DBLP:conf/www/PezoaRSUV16} introduced the first formalization of {\jsonsch} and showed that it cannot be captured by MSO or tree automata because of the {\Uni} constraints. While they focused on validation and proved that it can be decided in  $O(|J|^{2}|S|)$ time, they also showed that {\jsonsch} can simulate tree automata. Hence, schema satisfiability is EXPTIME-hard.

In~\cite{DBLP:conf/pods/BourhisRSV17} Bourhis et al.\ refined the analysis of Pezoa et al. They mapped {\jsonsch} onto an equivalent modal logic, called recursive JSL, and proved that satisfiability is 
\iflong{PSPACE-complete for schemas without recursion and {\Uni}, it is in EXPSPACE for non recursive schemas with {\Uni}, it is }%
EXPTIME-complete for recursive schemas without {\Uni}, and it is in 2EXPTIME for recursive schemas with {\Uni}.
Their work is extremely important in establishing complexity bounds.
Since they map {\jsonsch} onto recursive JSL logic, and provide a specific kind of alternating tree automata for this
logic, 
they already provide an indirect indication of an algorithm for witness generation.
\mrevmeta{
However, classical reachability algorithms for alternating automata are designed to prove complexity upper 
bounds, not as practical tools. They are typically based on the exploration of all subsets of the state set of the automaton
\cite{comon:hal-03367725}, hence on a sequence of complex operations on a set of sets whose dimension may be in the realm of $2^{10,000}$.
While exponentiality cannot be avoided in the worst case, it is clear that we need a different approach when designing a practical algorithm.
}{M.4, 2.6}
\hide{
\mrevmeta{
However, classical reachability algorithms for alternating automata are designed to prove complexity upper 
bounds, not as practical tool. They are typically based on the exploration of all subsets of the state set $Q$ of the automaton
\cite{tata}, hence on a sequence of complex operations a set of sets whose dimension may be in the realm of $2^{100,000}$ \cite{tata}.
In terms of reachability, our technique may be regarded as a technique where we do not generate every subset of $Q$, but
we first create the sets-of-states that are singletons $\Set{q}$, we bring the rule of $q$ in a DNF $\theta_1\Or\ldots\Or\theta_n$,
and, for any $\theta_i$ that involves a non-singleton set of states $S=\Set{q_1,\ldots,q_m}$, we create a new
set-of-states $S$, we reduce in DNF the conjunction of their rules, and we continue until no more set-of-states.
Here, the difficult issue is how to transform a rule that combines {\JS} object assertions into something that is analogous
to a DNF, since these assertions involve quantification over the children nodes that is mediated by a pattern-based
description of the children label. This issue is explored in Section \ref{}. There, we define a notion of
\emph{choice} and of \emph{viable disjoint solutions}. Choices are used to describe the children of a node, and they are
the bases of state completion, and the
the set of all disjoint solutions of a prepared object group can be regarded as a DNF expression of its rule.
}{M.4}
}

\hide{
Actually, we designed our algorithm from scratch, but there are elements of our algorithm 
that may be
traced back to classical algorithms for reachability in alternating automata, such as DNF normalization, or the technique that
we defined to eliminate negation through not-completion, that has some similarities with complementation of
alternating automata. In this context, one may even relate our and-completion (Section~\ref{sec:objprepgen}) to the idea of
analyzing sets of states at a time, typical of reachability algorithms for alternating automata. 
\iflong{If one wanted to pursue this analogy, it would be interesting to translate
our ``laziness'' in and-completion, that is crucial for our results, to some corresponding idea of laziness in 
the context of reachability for alternating automata, but we leave that for future work.}}

\hide{
\RestyleAlgo{ruled}
\begin{algorithm}\label{alg:reachability}
\footnotesize
\caption{Reachability test}
\SetKwProg{Fn}{}{}{end}
\SetKw{kwWhere}{where}
\SetKwFunction{Reachable}{Reachable}
\SetKwFunction{SetUp}{SetUp}

\Fn{\Reachable{A}}{ 
  setOfSets := \SetUp(E)\;
  \While{something changes}{  
    \For{set in setOfSets}{
       \For{qn in Q}{
           if qn is reachable when set is reachable than add 
         }
    }
   } 
   \Return ($A[x]$)\;
} 
\end{algorithm}
}

To the best of our knowledge, the only tool that is currently available to check the satisfiability of a schema is the containment checker described by Habib et al.~\cite{DBLP:conf/issta/HabibSHP21}. While it has been designed for schema containment checking, e.g., $S_{1} \subseteq S_{2}$, it can also be exploited for schema satisfiability since $S$ is satisfiable if and only if $S \not\subseteq S^{\prime}$, where $S^{\prime}$ is an empty schema. 
The approach of Habib et al.\ 
bears some resemblances to ours, e.g., schema canonicalization has been first
presented there, but its ability to cope with negation is very limited as well as its support for recursion. 

Several tools (see \cite{jsongen} and \cite{faker}) for example generation exist. They generate {\json} data starting from a schema. These tools, however, are based on a trial-and-error approach and cannot detect unsatisfiable schemas.
We compare our tool with \cite{jsongen} in our experiments.
\iflong{There are also grammar-based approaches for generating JSON values.
The tool by Gopinath et al.\ allows for data generation under Boolean constraints~\cite{DBLP:conf/icse/GopinathNZ21}, which have to be specified \emph{manually}.}

\iflong{In \cite{DBLP:conf/erlang/EarleFHM14}, Benac Earle et al. present a systematic approach to testing behavioral aspects of Web Services that communicate using {\json} data. In particular, this approach builds a finite state machine capturing the schema describing the exchanged data, but this machine is only used for generating data and is restricted to atomic values, objects and to some form of boolean expressions.}

\paragraph{Own prior work.}
In our technical report~\cite{maybetcs}, we discuss negation-completeness for {\jsonsch}, that is, we show how
pairs of {\jsonsch} operators such as $\qpattProps$-$\qreq$ and $\qit$-$\qcont$ are \emph{almost} dual under negation,
as $\And$-$\Or$ or $\forall$-$\exists$ are, but not exactly. In the process, we define an algorithm for not-elimination%
\shortlong{.}{, that
we actually developed for its use in the witness generation algorithm that we describe here. 
In Section~\ref{sec:notelim} we will rapidly recap this algorithm.
}

\ifshort{\mrevtwo{A preliminary version of the algorithm described in the current paper has been presented in~\cite{bda}. In that paper we provided an hint on the different phases of the algorithm, while here we go in much more detail.}{2.6}}
An earlier prototype implementation has been presented in tool demos~\cite{DBLP:conf/edbt/AttoucheBCFGLSS21,bda_demo,DBLP:conf/vldb/FruthDS21}.
\iflong{Meanwhile, we have optimized our algorithm, and formalized the proofs, as presented in this paper.}

\ifshort{This paper is accompanied by a full version~\cite{attouche2022witness}, containing detailed proofs and additional experiments.
}
\iflong{A preliminary version of the algorithm described in the current paper has been presented in~\cite{bda} (informal proceedings).}

\ifshort{\section{JSON Schema and the Algebra}\label{sec:prelim}}
\iflong{\section{Preliminaries}\label{sec:prelim}}

\subsection{JSON data model}\label{sec:datamodel}

Each {\json} value belongs to one of the six {\jsonsch} types: nulls, Booleans, decimal numbers 
$\Num$\shortlong{,}{
(hereafter, we just use \emph{numbers} to refer to decimal numbers),}
strings $\Str$, objects, arrays.
\iflong{Objects represent sets of members, each member being a name-value pair,
where no name can be present twice,
and arrays represent ordered sequences of values.}
\[
\begin{array}{lllllllllll}
\J ::= & \!\!\! B \mid O \mid A                       & &\text{\bf {\json} expressions}  \\
B ::=	 & 
     \multicolumn{2}{l}{\!\!\! \nule \mid \truev \mid \falsev \mid \num \mid \str}   \\
	&  & \num\in\Num, \str\in\Str  & \text{\bf Basic values} \\ 
O ::= &  
    \multicolumn{2}{l}{\!\!\! \{l_1:\J_1,\ldots,l_n:\J_n \}}     \\
	 & &  n\geq 0, \ \  i\neq j \Rightarrow l_i\neq l_j & \text{\bf Objects} \\
A ::= &  \!\!\! [\J_1, \ldots, \J_n ] & n\geq 0  &  \text{\bf Arrays} \\
\end{array}
\]

\begin{definition}[\shortlong{{\json} objects}{Value equality and sets of values}] 
We interpret a {\json} object $\{l_1:\J_1,\ldots,l_n:\J_n \}$ as a \emph{set} of pairs
(\emph{members}) $\{(l_1,\J_1),$ $\ldots,(l_n,\J_n)\}$, where $i\neq j \Rightarrow l_i\neq l_j$, 
and an array $[\J_1, \ldots, \J_n ]$ as an ordered list; {\json} value equality is defined
accordingly, that is, by ignoring member order when comparing objects.

\iflong{Sets of {\json} values are defined as collections with no repetition with respect to this notion of equality.
}

\end{definition}

\subsection{JSON Schema}

\shortlong{%
We base our work on  {\jsonsch} {\VerSix}~\cite{Draft06},
as it supports virtually all schemas that we could crawl from GitHub~\cite{schema_corpus}.
The successive {\VerEight} \cite{Version09} made validation dependent on annotations, a questionable semantic shift that we prefer not to embrace for now. However, we include in our algebra the operators $\qminC$ and $\qmaxC$ introduced with {\VerEight},
since they are very interesting in the context of witness generation, and their semantics does not depend on annotations.
}{%
{\jsonsch} is a language for defining the structure of {\json} documents.
Many versions have been defined for this language, notably 
{\VerThree} of November 2010, 
{\VerFour} of February 2013 \cite{Draft04}, 
{\VerSix} of April 2017 \cite{Draft06}, 
{\VerEight} of September 2019 \cite{Version09}, 
and {\VerTwenty} of December 2020 \cite{Draft12}.
{\VerEight} introduced a major semantic shift, since it made assertion validation dependent on annotations,
and has not been amply adopted up to now, hence we decided to base our work on {\VerSix}.
However, we decided also to include the operators $\qminC$ and $\qmaxC$ introduced with {\VerEight}
since they are very interesting in the context of witness generation and they do not present the problematic
dependency on annotations of the other novel operators.
}

{\jsonsch} uses {\json} syntax. A schema is a {\json} object that collects \emph{assertions} that are
members, i.e.,\ name-value pairs, where the name indicates the assertion and the value collects its parameters,
as in $\qminL:3$, where the value is a number, or in $\qit:\{\qtype:[\qboolean]\}$, where the value for 
$\qit$ is an object that is itself a schema\shortlong{.}{, and the value for $\qtype$ is an array of strings.}
\ifshort{%
We next describe {\jsonsch} by giving its translation into an algebra.
} 

\hide{
\begin{figure}
\begin{lstlisting}[style=query]
{"anyOf" : [{"type": "object",
             "properties": {"b": { "type": "array"}},
             "required" : ["b"]
            },
            { "properties": { "a": { "type": "number"}}},
            {
              "patternProperties": {"^a" : 
                 {"anyOf" : [{"type" : "array"},
                             {"not" : {"$ref": "#"}}]}}
            }   
	]        
}
\end{lstlisting}
\caption{A {\jsonsch} document.}
\label{fig:running}
\end{figure}
}

\iflong{%
A {\jsonsch} document (or \emph{schema}) denotes a set of {\json} documents (or \emph{values})
that satisfy it.
The language offers the following abilities.
\begin{compactitem}
\item Base type specification: it is possible to define complex properties of collections of base type values,
   such as all strings that satisfy a given regular expressions ($\qpatt$), all numbers that are multiple of a given  numbers ($\qmof$)
   and included in a given interval ($\qmin$, $\qmax$,\ldots).
\item Array specification: it is possible to specify the types of the elements for both uniform arrays and
  non-uniform arrays ($\qit$), to restrict the minimum and maximum size of the array, 
  to bound the number of 
  elements that satisfy a given property ($\qcont$, $\qminC$, \ldots), and also to enforce uniqueness 
  of the items ($\quniqIt$).
\item Object specification: it is possible to require for certain names to be present or to be absent,  to 
  specify the schemas of both optional or mandatory members, all of this by denoting classes of names using 
  regular expressions (via $\qprops$, $\qpattProps$, and $\qreq$).
  It it possible to specify that some assertions depend on the presence of some members ($\qdeps$), and it is 
  possible to limit the number of members that are present.
\item Boolean combination: one can express union, intersection, and complement of schemas
   ($\qany$, $\qall$, $\qnot$), and also a generalized form of mutual exclusion ($\qone$).
\item Mutual recursion: mutually recursive schema variables can be defined ($\qdefs$, $\qdref$).
\end{compactitem}

In the next section we describe {\jsonsch} by giving its translation into a simpler algebra.
}

\iflong{\section{The algebra}}\label{sec:algebra}

{\subsection{The \emph{core} and the \emph{positive} algebras}\label{sec:syntax}}

In {\jsonsch}, the meaning of some assertions is modified by the surrounding
assertions, making formal manipulation much more difficult.
Moreover, the language is rich in 
redundant operators\iflong{, such as $\qite$ and $\qdeps$, which can both be
easily translated in terms of $\qnot$ and $\qany$}.
\ifshort{In our implementation, we therefore map an input schema onto an algebraic representation based on a \emph{core algebra}, an algebraic version of {\jsonsch} with less redundant operators. We then eliminate negative expressions through not-elimination (Section \ref{sec:notelim}), by using a \emph{positive algebra} without negation but with three new operators:
$\NotMof(n)$, $\PReq(r : S)$, and $\ContAfter{i}{S}$.

Our algebras extend {\jsonsch} regular expressions with external intersection
and complement operators $r \AndP r'$ and  $\CoP{r}$; 
this extension is discussed in Section \ref{sec:regexp}. \iflong{Our core algebra is similar 
to the recursive JSL logic defined in \cite{DBLP:conf/pods/BourhisRSV17}, but has a different aim:
While JSL is a tool for 
theoretical research, hence is elegant and minimal,
our algebra is an implementation tool, a sublanguage of {\jsonsch} itself, obtained by removing 
obvious redundancies, but remaining as faithful as possible to {\jsonsch} peculiarities.} The syntax of the two algebras, \emph{core} and \emph{positive},
\iflong{which are expressive enough to capture  all {\jsonsch} assertions of {\VerSix}, plus the extra operators $\qminC$ and $\qmaxC$ of {\VerEight}, }%
is presented \shortlong{below.}{in Figure \ref{fig:core-syntax}.}}

\iflong{For these reasons, in our implementation, we translate {\jsonsch} onto a
\emph{core algebra}, that is an algebraic version of {\jsonsch} with less redundant operators.
\ifshort{%
This algebra is similar to the recursive JSL logic defined in \cite{DBLP:conf/pods/BourhisRSV17}, but has a different aim.
While JSL is a tool for 
theoretical research, hence is elegant and minimal,
our algebra is an implementation tool, a sublanguage of {\jsonsch} itself, obtained by removing some
obvious redundancies, but remaining as faithful as possible to {\jsonsch} peculiarities.
}
\iflong{%

This algebra is very similar (apart the syntax) to the recursive JSL logic defined in \cite{DBLP:conf/pods/BourhisRSV17}, but has a different aim.
While JSL is an elegant and minimal logic upon which {\jsonsch} is translated, and an excellent tool for 
theoretical research,
our algebra is an implementation tool with two aims:
\begin{compactenum}
\item simplify the implementation by its algebraic nature and its reduced size;
\item simplify the formal discussion of the implementation.  
\end{compactenum}
Both aims are facilitated by the algebraic nature and the reduced size of the algebra, but we also value 
a certain degree of adherence to {\jsonsch}.
}

The first step of our approach is the translation of an input schema into an algebraic representation,
and the second step is not-elimination (Section \ref{sec:notelim}).
For the first step we use a \emph{core algebra} that is defined by a subset of {\jsonsch} operators.
For not-elimination, we use a \emph{positive algebra} where we remove negation 
but we add three new operators:
$\NotMof(n)$, $\PReq(r : S)$, and $\ContAfter{i}{S}$.
Our algebras extend {\jsonsch} regular expressions with external intersection
$\AndP$ and complement $\CoP{r}$ operators; this extension 
is discussed in Section \ref{sec:regexp}.
The syntax of the two algebras, \emph{core} and \emph{positive},
\iflong{which are expressive enough to capture  all {\jsonsch} assertions of {\VerSix}, plus the extra operators $\qminC$ and $\qmaxC$ of {\VerEight}, }%
is presented \shortlong{below.}{in Figure \ref{fig:core-syntax}.}}


\newcommand{\IN}{\!\in\!}

\iflong{\begin{figure}[t]}
{
{
\UpdateNL{0.2ex}
$$
\begin{array}{llll}
\multicolumn{4}{l}{
m\IN\Num^{-\Inf}, M\IN\Num^{\Inf},
 l\IN\Nat_{>0}, i \IN \Nat, j \IN \Nat^{\Inf}, q \IN \Num, k \IN \Str 
 \qquad\qquad\qquad\qquad
 }\\[\NL]
T & ::= & \Arr \M \Obj \M \Null \M \Bool \M \Str  \M \Num \\[\NL]
r & ::= &  \text{Any regular expression} \M \CoP{r} \M r_1 \AndP r_2 \\[\NL]
b & ::= & \xtrue \M \xfalse \\[\NL]
S & ::= & \IBT(b)\M \Pat(r) \M \Bet_{m}^{M} \M \XBet_{m}^{M}    \\[\NL]
&&   \M \Mof(q) \M \Props(\key{r} : S)   \M \Req(\key{k}) \M \Pro_{i}^{j}  \\[\NL]
&&  \M \PreIte{l}{S} \M \PostIte{i}{S} 
  \M \CMM{i}{j}{S}  \\[\NL]
 & &   \M \Type(T) \M \RRef{x}  \M \ S_1 \And S_2 \M S_1 \Or S_2 \\[2\NL]
\multicolumn{2}{r}{\text{core:}}&  \M \Not S  \\[\NL]
\multicolumn{2}{r}{\text{positive:}}&
\M \NotMof(q) \M \PReq(r : S) \M \ContAfter{i}{S} \\[\NL]
\E & ::= & \Def{x_1}{S_1} , \ldots, \Def{x_{n}}{S_{n}} \\[\NL]
D & ::= & \Doc{S}{\E} \\[\NL] 
\end{array}
$$
\RestoreNL}}
\iflong{
\caption{Syntax of the \emph{core} and \emph{positive} algebras.}
\label{fig:core-syntax}
\end{figure}}

\ifshort{$\Num^{-\Inf}$ are the decimal numbers extended with ${-\Inf}$, and similarly
for $\Num^{\Inf}$ and $\Nat^{\Inf}$. $\Nat_{>0}$ is $\Nat$ without $0$, used in  $\PreIte{l}{S}$.}

\iflong{In $\Mof(q)$, $q$ is a number.
In $\Bet_{m}^{M}$ and in $\XBet_{m}^{M}$,
$m$ is either a number or $-\Inf$, $M$ is either a number or $\Inf$. 
In $ \Pro_{i}^{j}$, in $\PostIte{i}{S}$, in $\CMM{i}{j}{S}$, and in $\ContAfter{i}{S}$,
$i$ is an integer with $i \geq 0$, and $j$ is either an integer with $j \geq 0$,
or $\Inf$, 
while  in  $\PreIte{l}{S}$,  $l$  is an integer with $l \geq 1$,
and $k$ in $\Req(k)$ is a string.}


%

We distinguish Boolean operators ($\And$, $\Or$ and $\Not$), variables ($\RRef{x}$),
and \emph{Typed Operators} ({\TO} --- all the others).
All {\TO}s different from $\Type(T)$ have an implicative semantics: ``if the instance belongs to the type 
$T$ then \ldots'', so that they are trivially
satisfied by every instance not belonging to type $T$.
We say that they are \emph{implicative typed operators} ({\ITO}s).

The operators of the core algebra strictly correspond to those of {\jsonsch}, and
in particular to their implicative semantics. \iflong{The exact relationship between core
algebra and {\jsonsch} is discussed in Section \ref{sec:translation}.
}

Informally, an instance $\J$ of the core or positive algebra satisfies an assertion $S$ if:
\begin{compactitem}
\item $\IBT(b)$: \emph{if} the instance $\J$ is a boolean, \emph{then} $\J=b$.
\item $\Pat(r)$: \emph{if} $\J$ is a string, \emph{then} $\J$ matches~$r$.
\item $\Bet_{m}^{M}$: \emph{if} $\J$ is a number, \emph{then} $m\leq \J \leq M$. 
$\XBet_{m}^{M}$ is the same with extreme excluded.
\item $\Mof(q)$: \emph{if} $\J$ is a number, \emph{then} $\J=q\times i$ for some integer~$i$. $q$ is any number,
i.e., any \emph{decimal} number (Section \ref{sec:datamodel}).
\item $\Props(\key{r} : S)$  \emph{if} $\J$ is an object and \emph{if} $(k,\J')$ is a member of $\J$ where
$k$ matches the pattern $r$, \emph{then} $\J'$ satisfies $S$.  Hence, it is satisfied by any instance that is not an object and also by any object where
no member name matches $r$. 
\item $\Req(\key{k})$: \emph{if} $\J$ is an object, \emph{then} it contains at least one member whose name is $k$.
\item $\Pro_{i}^{j}$: \emph{if} $\J$ is an object, \emph{then} it has between  $i$ and $j$ members.
%
\item $\PreIte{l}{S}$: \emph{if} $\J$ is an array $[\J_1,\ldots,\J_n]$ ($n \geq 0$) and \emph{if} $l \leq n$, \emph{then} $J_l$ satisfies $S$.  Hence, it is satisfied by any $\J$ that is not an array and also by any array that is strictly shorter than~$l$\iflong{, such as the empty array}:
it does not force the position $l$ to be actually used.

\item $\PostIte{i}{S}$: \emph{if} $\J$ is an array $[\J_1,\ldots,\J_n]$, \emph{then} $J_l$ satisfies $S$
for every $l > i$.  Hence, it is satisfied by any $\J$ that is not an array and by any array shorter than $i$.
\item $\CMM{i}{j}{S}$: \emph{if} $\J$ is an array, \emph{then} the total number of elements that satisfy $S$ is included between
$i$ and~$j$. 

%
%
\item $\Type(T)$ is satisfied by any instance belonging to the predefined JSON type $T$ ($\Str$, $\Num$, $\Bool$, $\Obj$, $\Arr$, and $\Null$). 
\item $\RRef{x}$ is equivalent to its definition in the environment $\E$ associated with the expression.
\item $S_1 \And S_2$: both $S_1$ and $S_2$ are satisfied.
\item$S_1 \Or S_2$: either $S_1$, or $S_2$, or both, are satisfied.
\item $\Not S$: $S$ is not satisfied.
\item $\NotMof(n)$: \emph{if} $\J$ is a number, \emph{then} is not a multiple of $n$.
\item $\PReq(r : S)$: \emph{if} $\J$ is an object, \emph{then} it contains at least one member $(k,\J)$ where $k$  matches $r$ and $\J$ satisfies $S$ 
\item $\ContAfter{i}{S}$: \emph{if} $\J$ is an array $[\J_1,\ldots,\J_n]$, \emph{then} it contains at least one element $\J_j$ with $j>i$ that satisfies $S$.
%
\iflong{\item An environment $\E = \Def{x_1}{S_1} , \ldots, \Def{x_{n}}{S_{n}}$
defines $n$ mutually recursive variables, so that $\RRef{x}_i$ can
be used as an alias for $S_i$ inside any of $S_1 , \ldots, S_n$. }
\item $D = \Doc{S}{\Def{x_1}{S_1} , \ldots, \Def{x_{n}}{S_{n}}}$: 
 $\J$ satisfies $S$ when every $x_i$ is interpreted as an alias
for the corresponding $S_i$.
\end{compactitem}

Variables in $\E=\Def{x_1}{S_1} , \ldots, \Def{x_{n}}{S_{n}}$ are mutually recursive,
but we require recursion to be \emph{guarded}. 
Let us say that $x_i$ 
\emph{directly depends} on $x_j$ if some occurrence of $x_j$
appears in the definition of $x_i$ without being in the scope
of an ITO. 
\iflong{For example, in \enquote{$\Def{x}{(\Props(\key{r}:\RRef{y}) \And  \RRef{z})}$},
$x$ directly depends on~$z$, but not on~$y$.}
Recursion is not guarded if the transitive closure of the relation ``directly depends on'' contains
a reflexive pair $(x,x)$. Informally, recursion is guarded iff every cyclic chain of dependencies traverses an \ITO. \ifshort{An environment $\E = \Def{x_1}{S_1} , \ldots, \Def{x_{n}}{S_{n}}$ is \emph{guarded} if recursion
is guarded in $\E$.
An environment $\E = \Def{x_1}{S_1} , \ldots, \Def{x_{n}}{S_{n}}$ is \emph{closing} for $S$ if all variables in $S_1,\ldots,S_n$ and in $S$ are included in $x_1,\ldots,x_n$. }

\ifshort{
The three operators added in the positive algebra do not directly correspond to {\jsonsch} operators, 
but can still be expressed in {\jsonsch}, through 
the negation of 
$\Mof$, $\Props$, and $\PostIteK$,
as follows,
where $S_1 \Implies S_2$ is an abbreviation for $\Not S_1 \Or S_2$:
\ifshort{
\UpdateNL{0.1ex}
}
\[\begin{array}{lll}
\NotMof(n) &=&\Type(\Num) \Implies \Not \Mof(n)  \\[\NL]
\PReq(r : S) &=&\Type(\Obj) \Implies \Not \Props(\key{r} : \Not S)  \\[\NL]
\ContAfter{i}{S} &=&\Type(\Arr) \Implies \Not \PostIte{i}{\Not S}  \\[\NL]

\end{array}\]
}

\ifshort{%
In the full paper~\cite{attouche2022witness} we formalize the official {\jsonsch} semantics by defining a function 
$\semca{S}{\E}$ that associates a set of {\json} values to any assertion $S$ whose variables
are defined by the guarded schema $\E$, also in cases of mutual recursion under negation.
}

\ifshort{%
Hereafter we will often use the redundant operators $\True$ and $\False$, where 
$\True$ is satisfied by any {\json} value, and $\False$ is satisfied by none.
}

\iflong{%
Hereafter we will often use the derived operators $\True$ and $\False$. 
$\True$ stands for
``always satisfied'' and can be expressed, for example, as $\Pro_0^{\Inf}$, which is satisfied by any instance.
$\False$ stands for ``never satisfied'' and can be expressed, for example, as $\Not \True$.
}

\iflong{
\subsection{Semantics of the core algebra}\label{sec:semantics}

The semantics of a schema $S$ with respect to an environment $\E$ is the set of {\json} instances  $\semca{S}{\E}$
that \emph{satisfy} that schema, as specified in Figure \ref{tab:corealg-sem}.
Hereafter,  $\E(x)$ indicates the schema that $\E$ associates to $x$. 
$\rlan{r}$ denotes the regular language generated by $r$.
For $T$ in $\Null,\Bool,\Str,\Num,\Obj,\Arr$, $\semt(T)$ is the set of {\json} values of that
type, and
$\JSet$ is the set of all {\json} values. $\IntSet$ is the set of all integers.
Universal quantification on an empty set is true, and the set $\SetTo{0}$ is empty.

\ifshort{
The definition can be read as follows:
the semantics of $\Props(\key{r} : S)$ specifies that 
$\J\in\semca{\Props(\key{r} : S)}{\E}$ $\Iff$ \emph{if} $\J$ is an object, \emph{if} $(k_i:\J_i)$ is a member
where $k_i$ matches $r$, \emph{then} $\J_i\in\semca{S}{\E}$. 
}

\iflong{
The definition can be read as follows (ignoring the index $p$ for a moment):
the semantics of $\Props(\key{r} : S)$ specifies that 
$\J\in\semca{\Props(\key{r} : S)}{\E}$ $\Iff$ \emph{if} $\J$ is an object, \emph{if} $(k_i:\J_i)$ is a member
where $k_i$ matches $r$, \emph{then} $\J_i\in\semca{S}{\E}$, as informally specified in the previous section. 
}

\ifshort{
\renewcommand{\semcap}[2]{\semca{#1}{#2}}
\renewcommand{\semcapar}[3]{\semca{#1}{#2}}

}

\begin{figure}[ht]
\[
\begin{array}{lcl}
   \semcap{\IBT(b)}{\E} & = & \setst{J}{J\in\semt(\Bool) \Rightarrow  J=b }\\
   \semcap{\Pat(r)}{\E} & = & \setst{J}{J\in\semt(\Str) \Rightarrow J \in  \rlan{r}} \\
   \semcap{\Bet_{m}^{M}}{\E} & = & \setst{J}{ J\in\semt(\Num) \Rightarrow  m \leq J \leq M } \\
   \semcap{\XBet_{m}^{M}}{\E} & = & \setst{J}{ J\in\semt(\Num) \Rightarrow  m < J < M } \\
  \semcap{\Mof(q)}{\E} & = & \setst{ J}{ J\in\semt(\Num) \Rightarrow  \\
     &&  \exists i\in\IntSet.\ J = i\cdot q} \\
\semcap{\Props(\key{r} : S)  }{\E} & = & \setst{J}{J = \{(k_1:J_1),\ldots,(k_n:J_n)\} \Rightarrow  \\
     &&
   \forall i\in \SetTo{n}.\ k_i\in\rlan{r} \ \Rightarrow J_i \in \semcap{S}{\E} } \\
\semcap{\Req(k)}{\E} & = & \setst{J}{J = \{(k_1:J_1),\ldots,(k_n:J_n)\} \Rightarrow  \\
     &&
   \exists i\in \SetTo{n}.\ k_i = k } \\
   \semcap{\Pro_{i}^{j} }{\E} & = & \setst{J}{J = \{(k_1:J_1),\ldots,(k_n:J_n)\} \Rightarrow\\
   &&    i \leq n \leq j} \\
 \semcap{  \PreIte{l}{S} }{\E} & = & \setst{J}{ J =  [\J_1, \ldots, \J_n ]   \Rightarrow \\
     &&
      n \geq l  \Rightarrow J_l \in   \semcap{S}{\E} }\\
 \semcap{  \PostIte{i}{S} }{\E} & = & \setst{J}{ J =  [\J_1, \ldots, \J_n ]  \Rightarrow \\
     &&
     \forall j\in\SetTo{n} .\ j > i \Rightarrow J_j \in   \semcap{S}{\E} }\\
\semcap{  \CMM{i}{j}{S}  }{\E} & = & \setst{J}{ J =  [\J_1, \ldots, \J_n ] 
       \Rightarrow \\
     &&
      i \leq \ |\setst{l}{\J_l\in \semcap{S}{\E}}|\ \leq j }\\
 \semcap{\Type(T)}{\E}
   & = & \semt(T)\\
 \semcap{S_1 \And S_2}{\E} & = &  \semcap{S_1 }{\E}  \cap   \semcap{ S_2}{\E} \\
 \semcap{S_1 \Or S_2}{\E} & = &  \semcap{S_1 }{\E}  \cup   \semcap{ S_2}{\E} \\
\semcap{\Not S}{\E}  & = & \JSet \setminus \semcap{S}{\E}  \\
\iflong{
 \semcapar{\RRef{x}}{\E}{0}  &= & \emptyset \\
 }
\semcapar{\RRef{x}}{\E}{p+1}  &= & \semcap{\E(x)}{\E}\\
\iflong{
\semca{S}{\E} &=& \bigcup_{i\in \Nat}\bigcap_{p \geq i}\semcapar{S}{\E}{p} \\
}
\semca{\Doc{S}{\E}}{}  & = & \semca{S}{\E}  \\
 \end{array}
\]

\caption{Semantics of the algebra with explicit negation.}
\label{tab:corealg-sem}
\end{figure}

\ifshort{%
The equation $\semca{\RRef{x}}{\E} =  \semca{\E(x)}{\E}$ is not inductive, since $\E(x)$ is in general
bigger than $x$. In the full paper we prove that this system of equations defines a unique semantics whenever $\E$ is guarded, by exploiting an indexed version of this semantics.}
}

\iflong 
{%
The index $p$ is used since otherwise the definition
$\semca{\RRef{x}}{\E} =  \semca{\E(x)}{\E}$ would not be inductive: $\E(x)$ is in
general bigger than $x$, while the use of the index makes the entire definition inductive on the lexicographic 
pair $(p,|S|)$. However, we need to define an appropriate notion of limit for the sequence $\semcap{S}{\E}$.
We cannot just set $\semca{S}{\E}=\bigcup_{p\in \Nat} \semcap{S}{\E}$, since,
because of negation, this sequence of interpretations is 
not necessarily monotonic in $p$. For example, if we have a definition $\Def{y}{\Not(x)}$, then
$\semcapar{y}{\E}{0}$ contains the entire $\JSet$. However, since the interpretation converges when
$p$ grows, we can  extract an exists-forall limit from it, by stipulating that an instance 
$J$ belongs to the limit $\semca{S}{\E}$ if an $i$ exists such that $J$ belongs to every interpretation 
that comes after $i$:
$$
\semca{S}{\E} = \bigcup_{i\in \mathbb{N}}\bigcap_{j \geq i} \semcapar{S}{\E}{j}
$$

Now, it is easy to prove that this interpretation satisfies {\jsonsch} specifications, since, for guarded schemas,
it enjoys the properties expressed in Theorem~\ref{theo:bigone}, stated below.

\begin{definition}\label{def:closing}
An environment $\E = \Def{x_1}{S_1} , \ldots, \Def{x_{n}}{S_{n}}$ is \emph{guarded} if recursion
is guarded in $\E$.
An environment $\E = \Def{x_1}{S_1} , \ldots, \Def{x_{n}}{S_{n}}$ is \emph{closing} for $S$ if all variables in $S_1,\ldots,S_n$ and in $S$ are included in $x_1,\ldots,x_n$. 
\end{definition}

\begin{lemma}
[Convergence]\label{lem:convergence}
There exists a function $I$ that maps every triple $\J,S,\E$, where $\E$ is guarded and closing for $S$, 
to an integer $i=I(\J,S,\E)$ such that:
$$\begin{array}{llll}
(\forall j \geq i.\ \J \in\semcapar{S}{\E}{j})
\Or
(\forall j \geq i.\ \J \not\in\semcapar{S}{\E}{j})
\end{array}$$
\end{lemma}

\begin{proof}
For any guarded $\E$, we can define a function $d_{\E}$ from assertions to natural numbers such that,
when  $x$ directly depends on $y$, then $d_{\E}(x) > d_{\E}(y)$.
Specifically, we define the degree $d_{\E}(S)$ of a schema $S$ in $\E$ as follows.
If $S$ is a variable $x$, then $d_{\E}(x) = d_{\E}(E(x))+1$.
If $S$ is not a variable, then $d_{\E}(S)$ is the maximum degree of all unguarded variables in $S$ and,
if it contains no unguarded variable, then $d_{\E}(S)=0$.
This definition is well-founded thanks to the guardedness condition.
We now define a function $I(\J,S,\E)$ with the desired property
by induction on $(\J,d_{\E}(S),S)$, in this order of significance.
\\[\NL]

\noindent(i) Let $S=x$. We prove that  $I(\J,x,\E)=I(\J,\E(x),\E)+1$
has the desired property.
We want to prove that
\\[\NL]$
(\forall j \geq I(\J,\E(x),\E)+1.\ \J \in\semcapar{x}{\E}{j})
\Or
(\forall j \geq I(\J,\E(x),\E)+1.\ \J \not\in\semcapar{x}{\E}{j})$
\\[\NL]
We rewrite $\semcapar{x}{\E}{j}$ as $\semcapar{\E(x)}{\E}{j-1}$:
\\[\NL]$
(\forall j \geq I(\J,\E(x),\E)+1.\ \J \in\semcapar{\E(x)}{\E}{j-1})
\Or
(\forall j \geq I(\J,\E(x),\E)+1.\ \J \not\in\semcapar{\E(x)}{\E}{j-1})$
\\[\NL]i.e.,\ $
(\forall j \geq I(\J,\E(x),\E).\ \J \in\semcapar{\E(x)}{\E}{j})
\Or
(\forall j \geq I(\J,\E(x),\E).\ \J \not\in\semcapar{\E(x)}{\E}{j})$
\\[\NL]
 This last statement holds by induction, since $d_{\E}(x)=d_{\E}(\E(x))+1$,
hence the term $\J$ is the same but the degree of $\E(x)$ is strictly smaller than that of $x$.\\[\NL]

\noindent(ii) Let $S=\Not S'$. We prove that  $I(\J,\Not S',\E)$ defined as $I(\J,S',\E)$
has the desired property.
We want to prove that, for any $\J$:\\[\NL]
$(\forall j \geq I(\J,S',\E).\ \J \in\semcapar{\Not S'}{\E}{j})
\Or
(\forall j \geq I(\J,S',\E).\ \J \not\in\semcapar{\Not S'}{\E}{j})$\\[\NL]
By definition of $\semcapar{\Not S'}{\E}{j}$,
we need to prove  that for any $\J$:\\[\NL]
$(\forall j \geq I(\J,S',\E).\ \J \not\in\semcapar{ S'}{\E}{j})
\Or
(\forall j \geq I(\J,S',\E).\ \J \in\semcapar{ S'}{\E}{j})$\\[\NL]
which holds by induction on $S$, since the term $\J$ is the same and the degree is equal.\\[\NL]

\noindent(iii) Let $S=S'\And S''$. 
In this case, we let\\
 $I(\J,S'\And S'',\E)=max(I(\J,S',\E),I(\J,S'',\E))$.
We want to prove that:
\\[\NL]$
(\forall j \geq max(I(\J,S',\E),I(\J,S'',\E)).\ \J \in\semcapar{S'\And S''}{\E}{j})$\\
$\Or
(\forall j \geq max(I(\J,S',\E),I(\J,S'',\E)).\ \J \not\in\semcapar{S'\And S''}{\E}{j})$
\\[\NL]
This follows immediately from the following two properties, that hold by induction
on $(\J,d_{\E}(S),S)$, since both $S_1$ and $S_2$ have a degree less or equal to $S$, and are strict subterms
of $S$:
\\[\NL]$
(\forall j \geq I(\J,S',\E).\ \J \in\semcapar{S'}{\E}{j})
\Or
(\forall j \geq I(\J,S',\E).\ \J \not\in\semcapar{S'}{\E}{j})$
\\[\NL]$
(\forall j \geq I(\J,S'',\E).\ \J \in\semcapar{S''}{\E}{j})
\Or
(\forall j \geq I(\J,S'',\E).\ \J \not\in\semcapar{S''}{\E}{j})$
\\[\NL]
The same proof holds for the case $S=S'\Or S''$. \\[\NL]


\noindent(iv) Let $S =\PostIte{n}{S'}$. If $\J$ is not an array, then we can take
$I(\J,S,\E)=0$, since $\J$ satisfies $S$ for any index. If $\J=[\J_1, \ldots, \J_m ]$,
then we fix 
$$I([\J_1, \ldots, \J_m ],S,\E) = max_{i\in\SetTo{m}}I(\J_i,S',\E)\qquad\qquad(*)$$
which is well defined by induction, since every $\J_i$ is a strict subterm of $\J$.
Observe that the fact that each $\J_j$ is \emph{strictly} smaller than $\J$, and not
just less-or-equal, is essential since,
in general, the degree of $S'$ may be bigger than the degree of $S$, since $S'$ is in a guarded
position inside $S$.
Consider the semantics of $\PostIte{n}{S'}$:
\\[\NL]$\setst{\J}{ J =  [\J_1, \ldots, \J_m ]  \Rightarrow 
     \forall l\in\SetTo{m} .\ l > n \Rightarrow J_l \in   \semcap{S'}{\E} }$.\\
Now, because of $(*)$,
$\forall j \geq I(\J,S,\E)$, either $\J_l \in\semcapar{S'}{\E}{j}$
or $\J_l \not\in\semcapar{S'}{\E}{j}$, hence 
$(\forall j \geq I(\J,S,\E).\ \J \in\semcapar{\PostIte{n}{S'}}{\E}{j})
\Or
(\forall j \geq I(\J,S,\E).\ \J \not\in\semcapar{\PostIte{n}{S'}}{\E}{j})$
\\[\NL]
Informally, for any $l$ and for any $j\geq max_{i\in\SetTo{m}}I(\J_i,S',\E)$, the question 
``does $\J'$ belong to $\J_l \in\semcapar{S'}{\E}{j}$'' has a fixed answer, hence the question
``does $\J$ belong to $\PostIte{n}{S'}$'' has a fixed answer as well.\\[\NL]

All other {\TO}s can be treated in the same way.

\end{proof}

\begin{theorem}\label{theo:bigone}
For any $\E$ guarded, the following equality holds:
$$ \semca{\E(x)}{\E} = \semca{x}{\E}
$$
Moreover, 
for each equivalence in Figure \ref{tab:corealg-sem},
the equivalence still
holds if we substitute every occurrence of $\semcap{S}{\E}$ with $\semca{S}{\E}$, obtaining for example:\\[\NL]
$\begin{array}{lllll}
 \semca{  \PreIte{l}{S} }{\E} 
 \ =\  \setst{J}{ J =  [\J_1, \ldots, \J_n ]   \Rightarrow 
      n \geq l  \Rightarrow J_l \in   \semca{S}{\E} }\\
\end{array}$ \\[\NL]
from\\[\NL]
$\begin{array}{lllll}
\semcap{  \PreIte{l}{S} }{\E} 
 \ =\  \setst{J}{ J =  [\J_1, \ldots, \J_n ]   \Rightarrow \ 
      n \geq l  \Rightarrow J_l \in   \semcap{S}{\E} }\\
\end{array}$
\end{theorem}

\begin{proof}
This is an immediate consequence of convergence.
Consider any equation such as:\\[\NL]
$\begin{array}{lllll}
 \semcap{  \PreIte{l}{S} }{\E} 
 \ =\  \setst{J}{ J =  [\J_1, \ldots, \J_n ]   \Rightarrow 
      n \geq l  \Rightarrow J_l \in   \semcap{S}{\E} }\\
\end{array}$ \\[\NL]
That is:\\[\NL]
$\begin{array}{lllll}
\J\in \semcap{  \PreIte{l}{S} }{\E} 
\ \Iff\ ( J =  [\J_1, \ldots, \J_n ]   \Rightarrow 
      n \geq l  \Rightarrow J_l \in   \semcap{S}{\E}) \\
\end{array}$ \\[\NL]
If we consider any integer $I$ that is bigger than
$I(\J,\PreIte{l}{S},\E)$ and of every $I(\J_l,S,\E)$, then, if the equation holds
for one index $p\geq I$, then it holds for every such index, hence it holds for the limit.
This is the general idea, and we now present a more formal proof.

We first prove that:
$$
\bigcup_{i\in \mathbb{N}}\bigcap_{j \geq i} \semcapar{x}{\E}{j}=
\bigcup_{i\in \mathbb{N}}\bigcap_{j \geq i} \semcapar{\E(x)}{\E}{j}
$$
Assume that $J\in\bigcup_{i\in \mathbb{N}}\bigcap_{j \geq i} \semcapar{x}{\E}{j}$.
Then, \\$\exists i. \forall j \geq i. J\in  \semcapar{x}{\E}{j}$.
Let $I$ be one $i$ with that property. We have that 
\\$\forall j \geq I. J\in  \semcapar{x}{\E}{j}$, i.e.,
\\$\forall j \geq I. J\in  \semcapar{\E(x)}{\E}{j-1}$, which implies that
\\$\forall j \geq I. J\in  \semcapar{\E(x)}{\E}{j}$, hence
\\$\exists i. \forall j \geq i. J\in  \semcapar{\E(x)}{\E}{j}$.
\\In the other direction, assume 
$J\in\bigcup_{i\in \mathbb{N}}\bigcap_{j \geq i} \semcapar{\E(x)}{\E}{j}$.
Hence, 
\\$\exists i. \forall j \geq i. J\in  \semcapar{\E(x)}{\E}{j}$.
Let $I$ be one $i$ with that property. We have that 
\\$\forall j \geq I. J\in  \semcapar{\E(x)}{\E}{j}$, i.e.,
\\$\forall j \geq I. J\in  \semcapar{x}{\E}{j+1}$, i.e.,
\\$\forall j \geq (I+1). J\in  \semcapar{x}{\E}{j}$, i.e.,
\\$\exists i. \forall j \geq i. J\in  \semcapar{x}{\E}{j}$.

\medskip

For the second property, the crucial case is that for $J\in \semca{\Not S}{\E}$, where we want to prove:
$$J\in\semca{\Not S}{\E}\ \Iff J\not\in\semca{S}{\E}$$.
\\$J\in\semca{\Not S}{\E}\ \Iff$
\\[\NL]$\exists i. \forall j \geq i.\ J\in \semcapar{\Not S}{\E}{j}\ \Iff$
\\[\NL]$\exists i. \forall j \geq i.\ J\not \in \semcapar{S}{\E}{j}\ \Iff (***)$
\\[\NL]$\forall i. \exists j \geq i.\ J\not\in\semcapar{S}{\E}{j}\ \Iff$
\\[\NL]$\Not(\exists i. \forall j \geq i.\ \J\in\semcapar{S}{\E}{j})\ \Iff$
$J\not\in\semca{S}{\E}\ $
\\[\NL] For the crucial $\Iff (***)$ step, 
the direction $\Implies$ is immediate. For the direction $\RevImplies$ we use the convergence Lemma \ref{lem:convergence}:
if we assume that $\forall i. \exists j \geq i.\ J\not\in\semcapar{S}{\E}{j}$, then, 
by considering the case $i=I(\J,S,\E)$, we have that
$\exists j \geq I(\J,S,\E).\ J\not\in\semcapar{S}{\E}{j}$, 
hence, by 
Lemma \ref{lem:convergence},
$\forall j \geq I(\J,S,\E).\ J\not\in\semcapar{S}{\E}{j}$, 
hence $\exists i. \forall j \geq i.\ J\not \in \semcapar{S}{\E}{j}$.

All other cases follow easily from convergence.
Consider for example the case where 
 $\J\in \semca{\CMM{m}{M}{S'}}{\E}$.
 We want to prove:
 $$
\begin{array}{llll}
J\in\semca{\CMM{m}{M}{S'}}{\E}\ \\[\NL]
\Iff
 ( J =  [\J_1, \ldots, \J_n ] 
       \Rightarrow
     m \leq \ |\setst{l}{\J_l\in \semca{S'}{\E}}|\ \leq M)
     \end{array}$$
If $\J$ is not an array, the double implication holds trivially. Consider now the case
$\J=[\J_1, \ldots, \J_n ] $:
\\[\NL]$J\in\semca{\CMM{m}{M}{S'}}{\E}\ \Iff$
\\[\NL]$\exists i. \forall j \geq i.\ J\in \semcapar{\CMM{m}{M}{S'}}{\E}{j}\ \Iff $
\\[\NL]$
\exists i. \forall j \geq i.\  
     m \leq \ |\setst{l}{\J_l\in \semcapar{S'}{\E}{j}}|\ \leq M   \Iff$
\\[\NL]
Here, we choose an $I$ that is greater than
$I(\J,\CMM{m}{M}{S'},\E)$ and is greater than $I(\J_l,S',\E)$ for every $\J_l$ (from the proof
of Lemma \ref{lem:convergence} we know that $I(\J,\CMM{m}{M}{S'},\E)$ as defined in that proof would do the work):
\\[\NL]$
\exists i. \forall j \geq i.\  
     m \leq \ |\setst{l}{\J_l\in \semcapar{S'}{\E}{j}}|\ \leq M   \Iff$
\\[\NL]$
\forall j \geq I.\  
     m \leq \ |\setst{l}{\J_l\in \semcapar{S'}{\E}{j}}|\ \leq M   \Iff$
\\[\NL]
$
     m \leq \ |\setst{l}{\forall j \geq I.\  \J_l\in \semcapar{S'}{\E}{j}}|\ \leq M    \Iff $
\\[\NL]$
     m \leq \ |\setst{l}{\J_l\in \semca{S'}{\E}}|\ \leq M $


\end{proof}

The official {\jsonsch} semantics specifies that $x$ is the same as $\E(x)$ for all schemas where such interpretation
never creates a loop (i.e., for all guarded schemas) and describes, verbally, the equations that we wrote in
the form without the index. Hence, Theorem \ref{theo:bigone} proves that our semantics exactly captures 
the official {\jsonsch} semantics (provided that we wrote the correct equations).
}

\iflong{
\subsection{Semantics of the three extra operators of the positive algebra}

The three operators added in the positive algebra are redundant in presence of negation. They do not correspond to {\jsonsch} operators, but can still be expressed in {\jsonsch}, through the
negation of $\qmof$, $\qpattProps$, and $\qaddIts$.
The semantics of these operators can be easily expressed in the core algebra with negation, as shown
\shortlong{below;}{in Figure \ref{fig:opothers};}
hereafter, we use $S_1 \Implies S_2$ as an abbreviation for $\Not S_1 \Or S_2$:
\ifshort{
\UpdateNL{0.1ex}
}
\iflong{\begin{figure}[htb]}
\[\begin{array}{lll}
%
\NotMof(n) &=&\Type(\Num) \Implies \Not \Mof(n)  \\[\NL]
\PReq(r : S) &=&\Type(\Obj) \Implies \Not \Props(\key{r} : \Not S)  \\[\NL]
\ContAfter{i}{S} &=&\Type(\Arr) \Implies \Not \PostIte{i}{\Not S}  \\[\NL]
\end{array}\]
\iflong{\caption{Semantics of additional operators.}
\label{fig:opothers}
\end{figure}}

\ifshort{\RestoreNL}

\iflong{%
Observe that the semantics of the additional operators is implicative, as for all the others {\ITO}s.

The definition of $\PReq(r : S)$ deserves an explanation. 
The implication $\Type(\Obj) \Implies \ldots$ just describes its implicative nature --- it is satisfied by any
instance that is not an object. Since $\key{r}:\Not S$ means that, if a name matching $r$ is present, then its value satisfies~$\Not S$, any instance that does not satisfy $\key{r}:\Not S$ must possess a member name that matches $r$ and 
whose value does not satisfy~$\Not S$, that is, satisfies $S$. Hence, we exploit here the fact that the negation of an implication
forces the hypothesis to hold.
}
}


\subsection{About regular expressions}\label{sec:regexp}


\ifshort{%
\subsubsection{Mapping {\jsonsch} regular expressions onto standard REs}

Following the example of \cite{DBLP:conf/pods/BourhisRSV17}, we represent 
{\jsonsch} regular expressions (REs) using standard REs.
In practice, in our implementation
we map every {\jsonsch} RE into a standard RE, using a simple incomplete algorithm,\footnote{Dominik Freydenberger suggested this algorithm to us, in personal communication.} and we are currently able to translate more than 97\% of the distinct 
patterns in our corpus. The others mostly contain look-ahead and look-behind.
}

\iflong{%
\subsubsection{Undecidability of {\jsonsch} regular expressions}

{\jsonsch} regular expressions (REs) are ECMA regular expressions. Universality of these REs is undecidable \cite{DBLP:journals/mst/Freydenberger13}, hence the witness generation problem for any sublanguage of
{\jsonsch} that includes $\Not \Pat(r)$ is undecidable.
In our implementation we side-step this problem by mapping every {\jsonsch} RE unto a standard RE, as supported
by the brics library~\cite{brics_automaton}, using a simple incomplete algorithm.\footnote{The rewriting algorithm was suggested to us by Dominik Freydenberger in personal communication.}
When the algorithm fails, we raise a failure. 
This approach allows us to manage the
vast majority of our corpus.\footnote{We are currently able to translate more than 97\% of the unique 
patterns in our corpus. The other ones mostly contain look-ahead and look-behind.}

We limit our complexity analysis to the schemas where our RE translation succeeds, hence, we will hereafter assume that every  {\em {\jsonsch} regexp} that appears in the source schema, can be translated to a standard RE with a linear expansion, similarly to the approach adopted in \cite{DBLP:conf/pods/BourhisRSV17}, where the analysis is restricted to standard REs.
}

\subsubsection{Extending REs with external complement and intersection}

In our algebra, we use a form of \emph{externally extended REs} (EEREs), where the two extra operators
are not
first class RE operators, so that one cannot write $(\CoP{r})*$, but they can be used at the outer level:
$$
r  ::=   \text{\em Any regular expression} \M \CoP{r} \M r_1 \AndP r_2 
$$
This extension does not affect the expressive power of regular expressions
\iflong{, since the set of regular languages is closed under intersection and complement, }%
but affects their succinctness, hence the complexity of problems such as emptiness checking.
We are going to exploit this expressive power in four different ways:
\begin{compactenum}
\item in order to translate $\qaddProps : S$ as \\
   $\Props(\CoPP{r_1|\ldots|r_m}:\Tr{S})$
   \shortlong{(Section \ref{sec:translation});}{, where $\CoP{r}$ is applied to a standard
     RE (Section \ref{sec:translation});}
\ifshort{\item in order to translate $\qpropN : S$ (Section \ref{sec:translation});}
\iflong{\item in order to translate $\qpropN : S$, where a complex boolean combination of $\xpatt$ assertions inside 
   $S$ produces a corresponding complex boolean combination of patterns in the translation (Section \ref{sec:translation});}
\item during not-elimination (Section \ref{sec:notelim}), where $\Pat(\CoP{r})$ is used to rewrite $\Not\Pat(r)$;
\item during object {\crcombination} (Section \ref{sec:objprep}), where we must express the intersection and the difference of patterns
    that appear in $\Props(r:S)$ and $\PReq(r:S)$ operators.
\end{compactenum}

During the final phases of our algorithm (Section \ref{sec:objprepgen}),
we need to solve the following \emph{$i$-enumeration} problem (which generalizes emptiness) for our EEREs: for 
a given EERE $r$ and for a given $i$, either return $i$ words that belong to $\rlan{r}$, 
\ifshort{where $\rlan{r}$ is the language of~$r$,}
or return ``impossible'' if
$|\rlan{r}| < i$. 
It is well-known that emptiness of REs extended (internally) with negation and intersection is non-elementary \cite{StockmeyerPhD}. However,
\ifshort{in the full paper \cite{attouche2022witness} we show that }for our external-only extension $i$-enumeration and emptiness
can be solved in time $O(i^2\times 2^n)$.

\iflong{
\begin{property}\label{pro:circuit}
If $r$ is an EERE, its language can be recognized by a DFA with $O(2^{|r|})$ states,
which can be built in time $O(2^{|r|})$.
\end{property}

\begin{proof}
Let us define a \emph{circuit} of REs to be a term $rr$ generated by the following grammar,
where the graph of dependencies induced by $\Def{x_1}{r_1},\ldots,\Def{x_n}{r_n}$ is acyclic:
$$\begin{array}{llll}
r  &::= &  \text{\em Any regular expression} \M \CoP{r} \M r_1 \AndP r_2  \M x \\[\NL]
rr &::=&  \Doc{r}{\Def{x_1}{r_1},\ldots,\Def{x_n}{r_n}} 
\end{array}$$
The semantics of such a circuit is defined by recursively substituting every $x$ with its definition, which is
guaranteed to terminate because the dependencies are acyclic.
Circuits of $RE$s generalize our EEREs; we prove the desired property for any circuit since this result 
will be useful in Section 
\ref{sec:names}.
We prove that any circuit $rr$ of REs can be simulated by an automaton with $O(2^{|rr|})$ states.
We first transform each basic RE $r_i$ that appears in the circuit into a $DFA$ $A_i$ of size $O(2^{|r_i|})$,
in time $O(2^{|r_i|})$, using standard techniques \cite{DBLP:journals/tocl/GeladeN12}.
We build the product automaton $A_{\Pi}=A_1\times \ldots \times A_n$,
whose states are tuple of states of $A_1\times \ldots \times A_n$ in the standard fashion
\cite{DBLP:books/daglib/0016921}; the states of this automaton grow as $O(2^{|r_1|})\times \ldots \times 2^{|r_n|})$, i.e.\ 
$O(2^{|r_1|+\ldots+|r_n|})$, i.e., $O(2^{|rr|})$.
We associate to each subexpression $r$ in the circuit a set $F(r,rr)$ 
of states of $A_{\Pi}$ that are ``accepting'' for $r$ in the natural way:
for each basic $r_i$, we define $F(r_i,rr)$ to be the states of $A_{\Pi}$ whose $i$-projection is
accepting for $A_i$. We set $F(r\AndP r',rr) = F(r,rr) \cap F(r',rr)$, $F(\CoP{r},rr) = Q\setminus F(\CoP{r},rr)$,
where $Q$ are the states of $A_{\Pi}$, and we set $F(x,\Doc{r}{\E})=F(\E(x),\Doc{r}{\E})$, 
which is terminating since variables form a DAG.
To each subexpression $r$ of $rr$ we associate the automaton $A_r$ whose states and transitions are the same 
as $A_{\Pi}$, and whose final states are $F(r,rr)$.
We define $d_{\E}(r)$ as in the proof of Lemma \ref{lem:convergence}, and
we prove by induction on $(d_{\E}(r),r)$ that $A_r$ recognizes the language of $\Doc{r}{\Def{x_1}{r_1},\ldots,\Def{x_n}{r_n}}$.
When $r=x$, this is true by induction, since $A_{x}=A_{\E(x)}$ and $d_{\E}(x)<d_{\E}(\E(x))$. 
When $r=\CoP{r}$ or $r=r_1 \AndP r_2$, the result follows by induction on $r$.
\end{proof}

\begin{property}\label{pro:igen}
For any extended RE r generated by our grammar starting from standard REs,
the $i$-enumeration problem can be solved in time $O(i^2\times 2^{|r|})$.
\end{property}

\begin{proofsketch}
By Property \ref{pro:circuit},
a DFA $A(r)$ for $r$ with less than  $2^{|r|}$ states can be built in time $O(2^{|r|})$.

Finally, given an automaton of size $2^{|r|}$, it is easy to see that the enumeration of $i$ words can be performed in $O(i^2\times 2^{|r|})$.
\end{proofsketch}
}

\iflong{\section{From {\jsonsch} to the algebra}\label{sec:translation}}
\shortlong{


\subsection{From {\jsonsch} to the Algebra}\label{sec:translation}

\newcommand{\AdPr}{\kw{adPr}}
\newcommand{\PrPr}{\kw{pr}}
\newcommand{\PaPr}{\kw{paPr}}

The translation from {\jsonsch} to the algebra is rather intuitive, and is described in the full paper~\cite{attouche2022witness}.
Essentially, each {\jsonsch} assertion is translated into the corresponding algebraic assertion. However, attention must be paid to certain families of assertions, which must be grouped and translated together:
\begin{compactitem}
\item {\xif}, {\xthen}, {\xelse}, translated using Boolean operators;
\item {\xaddProps}, \xprops, \xpattProps\ 
(here abbreviated as \AdPr, \PrPr, \PaPr):
\PrPr, \PaPr\ correspond to our $\Props$ operator, while
$\QQ{\AdPr}\QQ : S$  associates a schema~$S$ to any name that does not match 
either \PrPr\ or \PaPr\ arguments, and is translated
as $\Props(\CoPP{r_1|\ldots|r_m} : S)$, where
$r_1,\ldots,r_m$ are patterns that represent all arguments 
of all \PrPr\ or \PaPr\  that occur in the same schema;
\item {\xaddIts}, \xit, translated using the algebra assertion $\PostIte{j}{S}$, $\PreIte{l}{S}$;
\item {\xminC}, {\xmaxC}, {\xcont}, which are translated as $\CMM{m}{M}{\ldots}$.
\end{compactitem}
Some redundant operators are mapped to simpler operators:
\begin{compactitem}
\item $\qone : [ S_1,\ldots, S_n]$ requires that a value $\J$ satisfies one of $S_1,\ldots, S_n$ and violates all
the others; it is translated using Boolean operators and variables;
\item $\qpropN : \kw{S}$ requires that every member name satisfies $S$; it is translated as $\Props(\POfS(\Not S):\False)$,
   where $\POfS(\Not S)$ is a pattern that uses $\CoP{r}$ and $\AndP$ in order to encode all strings that
   violate $S$;
\item the \qdeps\ assertion specifies that
if the instance contains a member with name $k_i$, then it must also satisfy some
other assertions; it is translated using $\Req$ and $\Implies$;
\item $\qconst : J$ and $\qenum : [J_1,\ldots,J_n]$, used to restrict a schema to a finite set of values; they are
  translated to structural operators as in \cite{DBLP:conf/issta/HabibSHP21}.
\end{compactitem}

Finally, the definitions-references mechanism of {\jsonsch} (the $\xdref : \key{path}$ operator) is translated into our simpler mechanism, based on variables and environments.

\hide{
\begin{example}\label{ex:run}
\mrevthree{Consider the schema of Figure \ref{fig:running}. This schema is translated into the following algebraic representation.
\[
\begin{array}{llllll}
\Def{r}{\PReq(b : \RRef{x})\Or \Props(\key{a}:\RRef{y})\Or\Props(\key{a.*}:\Not\RRef{r}\Or\RRef{x})  } \\ 
\Def{x}{\TArr} \\
\Def{y}{\TNum}
\end{array}
\]}{3.3}
\end{example}
}

\iflong{
\subsection{How we evaluate complexity}

{\jsonsch} can be translated to the algebra with a polynomial 
size increase.
Later, we show that our algorithm runs in $O(2^{\PN})$ with respect to the size 
$N$ of the input algebra, but with one important caveat:
hereafter, we assume that all $i$ and $j$ constants different from $\Inf$ that appear in $\PreIte{i}{S}$, $\PostIte{i}{S}$,
$\ContAfter{i}{S}$, $\CMM{i}{j}{S}$,  and $\Pro_i^j$, 
are smaller than the input size. We refer to this assumption as the \emph{linear constant assumption}. 
This assumption is justified by the observation that in practical cases these numbers tend to be extremely small in comparison 
to the input size.}
\iflong{Hereafter, whenever a result depends on this assumption, we will say that explicitly.}

}{

\subsection{Structure of the chapter}\label{sec:structure}

A {\jsonsch} schema is a {\json} object whose fields are assertions.
Essentially, the translation $\Tr{S}$ of a schema $S$ applies some simple rules to the single assertions, and
combines them by conjunction, as follows:
$$\begin{array}{llll}
\Tr{ \{ \qkw{a1} : S1, \ldots, \qkw{an} : Sn\}}
&=& \Tr{ \qkw{a1} : S1 } \And \ldots \And \Tr{\qkw{an} : Sn } \\
\Tr{\qmof : q}
&=& \Mof(q)\\
\ldots
\end{array}$$

However, there are some exceptions, that we describe in this chapter.
We first describe how we map the complex referencing mechanism of {\jsonsch} into our simpler
$\Doc{S}{\E}$ construct. We then describe the translation of the redundant operators $\xpropN$, $\xconst$, $\xenum$,
and $\xone$ into the core algebra. Finally, we describe the non-algebraic {\jsonsch} operators, where
a group of related operators must be translated together, and we finish with the easy cases.

\subsection{Representing definitions and references}


{\jsonsch} defines a $\xdref : \key{path}$ operator that allows any subschema of the current schema to be referenced, as well as any subschema of a 
different schema that is reachable through a URI, hence implementing a powerful form of mutual recursion. 
The path $\key{path}$ may navigate through the nodes
of a schema document by traversing its structure, or may retrieve a subdocument on the basis of a special $\xid$, $\xdid$, or 
$\xda$ member ($\xda$ has been added in {\VerEight}), which can be used to associate a name to the 
surrounding schema object. However, according to our collection of JSON schemas, 
the subschemas that are referred are typically 
just those that are collected inside the value of a top-level \xdefs\ member.
Hence, we defined a referencing mechanism that is powerful enough to translate every collection of 
JSON schemas, but that privileges a direct translation of the most commonly used mechanism.



When all references in a {\jsonsch} document refer to a name defined in the \xdefs\ section, we just use  the natural translation:
$$
\begin{array}{lllll}
\Tr{\{ a_1 : S_1, \ldots, a_n : S_n, 
    \xdefs : \{ x_1 : S'_1, \ldots, x_m : S'_m \}
\}} \\
=\ \Doc{\Tr{\{ a_1 : S_1, \ldots, a_n : S_n\}}}
{\Def{x_1} {\Tr{S'_1}} , \ldots, \Def{x_m}{\Tr{S'_m}}}
\end{array}
$$

In the general case, we collect all 
paths that are used in any reference assertion $\xdref : \key{path}$ and that are different from
\xdefs/\kw{k}, we retrieve the referred subschema and copy it inside the \xdefs\ member
where we give it a name \emph{name}, and we substitute all occurrences of $\xdref : \key{path}$ with 
$\xdref : \xdefs/\kw{name}$, until we reach the shape (1) above.
In principle, this may cause a quadratic increase in the size of the schema, in case we have paths 
that refer inside the object that is referenced by another path.
It would be easy to define a more complex mechanism with a linear worst-case size increase, but 
this basic approach does not create any size problem on the schemas we collected.\footnote{When we have a collection of 
documents with mutual references, we first merge the documents together
and then apply the same mechanism, but this functionality has not yet been integrated into our published code.}

\begin{example}
We consider the following {\JS} document

\begin{Verbatim}[fontsize=\small,xleftmargin=5mm]
{ "properties": {
    "Country": { "type": "string" },
    "City":    { "$ref": "#/properties/Country" } }
}
\end{Verbatim}
Definition normalization produces the following, equivalent schema:
\begin{Verbatim}[fontsize=\small,xleftmargin=1mm]
{"properties": {
   "Country": {"type": "string" },
   "City":    {"$ref": "#/definitions/properties_Country"}},
 "definitions": {"properties_Country": {"type": "string" }}
}
\end{Verbatim}
Which is translated as:
$$\begin{array}{llll}
\Props(\key{Country} : \Type(\Str)) \And  \Props(\key{City} : \RRef{properties\_Country})
\\
 \Defs (\RRef{properties\_Country} : \Type(\Str))
\end{array}$$
\end{example}


\subsection{$\qpropN : \kw{S}$  encoded as $\Props(\CoP{r_S}:\False)$} \label{sec:names}

The {\jsonsch} assertion $\qpropN : \kw{S}$ requires that, if the instance is an object, then every member name satisfies $\kw{S}$.
Our translation to the algebra proceeds in two steps. We first translate to a new, redundant, algebraic operator $\Nam(S)$ that has the semantics that we just described:
$$\begin{array}{llll}
 \semca{\Nam(S)}{\E} \\
 =  \setst{ J}{  J= \{ k_1 : J_1,\ldots, k_m : J_m\}  \Rightarrow \forall l\in\SetTo{m}.\ k_{l} \in \semca{S}{\E} }
\end{array}
$$
Hence, $\J\in\semca{\Nam(S)}{\E}$ means that no member name violates $S$.
Hence, if we translate $S$ into a pattern $r=\POfS(S,\E)$ that exactly describes the strings that satisfy~$S$
(whose variables are interpreted by $\E$),
we can translate $\Nam(S)$ into $\CProp{\POfS(\Not S,\E)}{\False}$, which
means:  if the instance is an object, it cannot contain any member whose name does not match $\POfS(S,\E)$.

\newcommand{\EE}{\E}



%

For all the {\ITO}s $S$ whose type is not $\Str$, such as $\Mof(q)$, we define $\POfS(S,\EE) = \TrueP$, since they are satisfied by any string:
$$
\begin{array}{llll}
\POfS(\Mof(a),\EE) \ = \ 
\POfS(\CMM{i}{j}{S},\EE) \ = \ \ldots \ = \  \TrueP
\end{array}
$$

For the other operators, $\POfS(S,\EE)$ is defined as follows.

$$
\begin{array}{llll}
\POfS(\Type(T),\EE) &=&  \FalseP  & \text{if\ } T\neq \Str\\[\NL]
\POfS(\Type(Str),\EE) &=&  \TrueP  \\[\NL]
\POfS(\Pat(r),\EE) &=& r \\[\NL]
\POfS(S_1 \And S_2,\EE) &=&  \multicolumn{2}{l}{\POfS(S_1,\EE) \AndP \POfS(S_2,\EE)}\\[\NL]
\POfS(S_1 \Or S_2,\EE) &=&  \multicolumn{2}{l}{\CoP{\CoP{\POfS(S_1,\EE)} \AndP\CoP{\POfS(S_2,\EE)}}}\\[\NL]
\POfS(\Not S,\EE)  &=&  \multicolumn{2}{l}{\CoP{\POfS(S,\EE)}} \\[\NL]
\POfS(\RRef{x},\EE) &=&  \multicolumn{2}{l}{\POfS(\EE(x),\EE)}  \\[\NL]
\end{array}
$$

Above, while $\POfS(\Mof(q),\EE) = \TrueP$ since $\Mof(q)$ is an Implicative Typed Operator,
$\POfS(\Type(\Num),\EE) = \FalseP$, since $\Type(\Num)$ is not implicative, and is not satisfied by any
string.

Since $\POfS(S,\EE)$ does not depend on the schemas that are guarded by an {\ITO}, 
the above definition is well-founded when recursion is guarded: after a variable $x$ 
has been expanded, $x$ is guarded in the result of any further expansion, hence we 
will not need to expand it again. 

It is easy to prove the following equivalences, which allow us to translate
 $\Nam$, hence $\xpropN$, into the core algebra.

\begin{property}
For any assertion $S$ and for any environment $\E$ guarded and closing for $S$, the following equivalences hold.
$$
\begin{array}{llll}
\semca{\Type(\Str) \And S}{\E} & = & \semca{\Type(\Str) \And \Pat(\POfS(S,\E))}{\E} \\[\NL]
\semca{\Nam(S)}{\E} & = & \semca{\CProp{\POfS(\Not S,\E)}{\False}}{\E}
\end{array}
$$
\end{property}

This translation expands each variable with its definition, hence there exist schemas where 
$\POfS(\Not S,\E)$ is exponential in the size of $(S,\E)$. 
In practice, this is not a problem: in all schemas that we collected, $\qpropN : \kw{S}$ 
(which is quite rare) is invariably used with a very simple $S$, whose expansion
is always small.

To ensure linear-size translation, we should extend regular expressions with a variable mechanism,
for example in the following way, where we would impose a non-cyclic dependencies constraint to variable
environments, so that an expression $rr$ is actually a \emph{Boolean circuit} of regular expressions.
$$\begin{array}{llll}
r  &::= &  \text{\em Any regular expression} \M \CoP{r} \M r_1 \AndP r_2  \M x \\[\NL]
rr &::=&  \Doc{r}{\Def{x_1}{r_1},\ldots,\Def{x_n}{r_n}} 
\end{array}$$
Lifting $\CoP{r}$ and $r \AndP r'$ from EEREs to circuits is very easy. 
We can prove that the complexity of $i$-generation (Section \ref{sec:regexp}) for circuits has the same bound as for 
EEREs, hence this extension would not create complexity problems.
We can now translate an environment 
$$\E=\ldots\Def{x_i}{S_i}\ldots$$
with a pattern environment 
$$\key{patt\_}\E = \ldots\Def{\key{patt\_}x_i}{\POfS(S_i,\E)}\ldots$$
and we can then define 
$$\POfS(\RRef{x},\EE) = \Doc{\key{patt\_}x}{\key{patt\_}\E}.$$
Then, size expansion would be polynomial and not exponential.

Since the problem has, at the moment, no practical relevance, we decided to avoid this complication,
hence we limit our complexity analysis to those schemas that are \xpropN-small, according to the following definition.
If we encounter families of schemas that violate this property, we just need to extend our implementation,
and our analysis, by supporting Boolean circuits of REs.

\begin{definition}[\xpropN-small]
A schema $\Doc{S}{E}$ of the core algebra extended with $\Nam(S)$
is \emph{\xpropN-small} if
$$|\key{PNExpand}(S)| \leq 2\times |\Doc{S}{E}|$$
where \key{PNExpand} is the function that translates all instances of $\Nam(S')$ with 
$\CProp{\POfS(\Not S',\E)}{\False}$.
\end{definition}

Hence, by definition, the translation of \xpropN\ only causes a linear increase in \xpropN-small schemas.

\subsection{Translation of $\xconst$ and $\xenum$}

The assertions $\qconst : J$ and $\qenum : [J_1,\ldots,J_n]$, used to restrict a schema to a finite set of values, 
can be translated by first rewriting them into their algebraic counterparts
$\Enu(J_1,\ldots,J_n)$ and $\Con(J)$, and then by applying the rules in Figure \ref{fig:const}, similar to those
presented in \cite{DBLP:conf/issta/HabibSHP21}.
Hereafter, we use $\keykey{k}$ to denote  a pattern that only matches $\key{k}$;\footnote{Using standard notation,
$\keykey{k}$ would generally coincide with $\key{k}$, unless $\key{k}$ contains special characters,
such as ``.'', ``|'', or ``*'', that need to be escaped.} 
when~$k$ is a string, so that $\qconst : k$ can be translated as
$\Type(\Str) \And \Pat(\keykey{k})$.

\begin{figure*}[t]
\setlength{\NL}{0.2ex}
$$
\begin{array}{lllll}
\Enu(J_1,\ldots,J_n) &=& \Con(J_1) \Or \ldots \Or \Con(J_n) \\[\NL]
\Con(\xnull) & = &  \Type(\Null)   \\[\NL]
\Con(b) & = &  \Type(\Bool) \And \IBT(b) &  b\in \TBool \\[\NL]
\Con(n) & = &  \Type(\Num) \And \Bet_n^n  & n\in \Num 
\\[\NL]
\Con(\key{s)} & = &  \Type(\Str) \And \Pat(\keykey{s})  
\qquad\qquad\qquad\qquad\qquad\qquad\qquad& \key{s}\in \Str 
\\[\NL]
\Con([J_1,\ldots,J_n]) &=&
 \multicolumn{2}{l}{
 \Type(\Arr) \And \CMM{n}{n}{\True} \And \PreIte{1}{\Con(J_1)}\And\ldots\And\PreIte{n}{\Con(J_n)}
 } \\[\NL]
 \multicolumn{4}{l}{
 \Con( \{ \key{k_1} : J_1,\ldots, \key{k_n} : J_n \})
 = \ 
\Type(\Obj) \And \Req(k_1,\ldots,k_n) 
        \And \Pro_0^n \And\ \Props(\keykey{k_1}:\Con(J_1);\True)
                                         \And\ldots\And \Props(\keykey{k_n}:\Con(J_n);\True) }\\[\NL]
\end{array}
$$
\caption{Elimination of $\Enu$ and $\Con$.}
\label{fig:const}
\end{figure*}

\subsection{Translation of $\xone$}\label{sec:oneof}

The assertion $\qone : [ S_1,\ldots, S_n]$ requires that $\J$ satisfies one of $S_1,\ldots, S_n$ and violates all the others.
It can be expressed as follows, where the $x_i$'s are fresh variables, and the $\Defs$ part must actually be added to 
the outermost level:
$$\begin{array}{llll}
\BigOr_{i\in\SetTo{n}}  (\Not x_1 \And\ldots\And \Not x_{i-1}\And\RRef{x_i}\And\Not x_{i+1} \And\ldots\And\Not x_n) \\[\NL]
\Defs\ (\Def{x_1}{\Tr{S_1}}, \ldots,\Def{x_n}{\Tr{S_n}})
\end{array}$$
 
 The definition of the fresh variables is fundamental in order to avoid that a single subschema is copied many times,
 which may cause an exponential size increase.
 The outermost $\BigOr$ has size $O(n^2)$, hence this encoding may still cause a quadratic size increase; 
 this increase can be avoided using a more sophisticated linear 
 encoding that we present in \cite{maybetcs}.\footnote{In our implementation we adopted the basic algorithm, 
having verified that, in our schema corpus, $\xone$ has
on average 2.3 arguments, and, moreover, the quadratic encoding behaves better than the linear one when submitted
to DNF expansion.}

\subsection{The remaining assertions}

While most {\jsonsch} assertions can be translated one by one, as described in Section \ref{sec:structure}, we have four groups of exceptions, that is, four families of assertions whose semantics depends on the occurrence of other assertions of the same family as members of the same schema.
These families are:
\begin{compactenum}
\item {\xif}, {\xthen}, {\xelse};
\item {\xaddProps}, \xprops, \xpattProps;
\item {\xaddIts}, \xit;
\item in {\VerEight}: {\xminC}, {\xmaxC}, {\xcont}.

\end{compactenum}

When translating a schema object, we first partition it into families, we 
complete each family by adding the predefined default value for missing operators
(for example, a missing $\xelse$ becomes $\qelse : \xtrue$),
and we then translate each family as we specify below.
All other assertions are just translated one by one.

The assertion group
$ \qif : S_1, \qthen : S_2, \qelse : S_3 $
is translated as follows, where  $\Def{x}{\Tr{S_1}}$ is inserted in order to avoid duplication of $\Tr{S_1}$, and
is actually lifted at the outermost level, as we do with $\xone$:
$$\begin{array}{llll}
\Doc{ ( (\RRef{x} \And \Tr{S_2}) \Or (\Not\RRef{x} \And \Tr{S_3}))}
      {\Def{x}{\Tr{S_1}}}
\end{array}
$$

The $\xprops$ family is translated as follows, where we  use 
pattern complement $\CoP{r}$ to translate $\xaddProps$,
which associates a schema to any name that does not match either
$\xprops$ or $\xpattProps$ arguments:
$$
\begin{array}{llll}
\BeginTr\qprops : \{ \key{k_1} :  S_1 , \ldots, \key{k_n} :  S_n  \}, \\
\ \ \qpattProps : \{\key{r_1} : PS_1 , \ldots, \key{r_m} :  PS_m \}, \\
\ \ \qaddProps :  S\EndTr  \\[\NL]
= \Props(\keykey{k_1} : \Tr{S_1}) \And \ldots\And \Props(\keykey{k_n} : \Tr{S_n}) \\
\qquad \And\ \Props(\key{r_1} : \Tr{PS_1})\And \ldots\And \Props(\key{r_m}) : \Tr{PS_m}   \\
\qquad \And\ \Props(\CoPP{\keykey{k_1}|\ldots|\keykey{k_n}|r_1|\ldots|r_m} : S)  \\[\NL]
\\[\NL]
\end{array}
$$

{\xit} may have either a schema $S$ or an array $[S_1,\ldots,S_n]$ 
as argument; in the first case, it is equivalent to $\PostIte{0}{S}$, and a co-occurring \xaddIts\ is ignored, while in the second
case it is equivalent to $(\PreIte{1}{S_1}\And\ldots\And\PreIte{n}{S_n})$, and
${\qaddIts : S'}$ means $\PostIte{n}{\Tr{S'}}$.
The family is hence translated as follows.
$$
\begin{array}{llll}
\begin{array}{llll}
\Tr{\qaddIts : S'} \ &=& \PostIte{0}{\Tr{S'}} \\[\NL]
\Tr{\qit : S}  \ 
&=& \PostIte{0}{\Tr{S}} \\[\NL]
\Tr{\qit : S, {\qaddIts : S'}}  \ 
&=& \PostIte{0}{\Tr{S}} \\[\NL]
\end{array}\\
\begin{array}{llll}
\Tr{\qit : [S_1, \ldots, S_n]}  \ 
=\ (\PreIte{1}{\Tr{S_1}}\And\ldots\And\PreIte{n}{\Tr{S_n}})  \\[\NL]
\Tr{\qit : [S_1, \ldots, S_n], {\qaddIts : S'}} \\
\multicolumn{1}{r}{
=\ (\PreIte{1}{\Tr{S_1}}\And\ldots\And\PreIte{n}{\Tr{S_n}}) \And {\PostIte{n}{\Tr{S'}}}
}
\\[\NL]
\end{array}
\end{array}
$$

The  ${\xcont}$ family is translated as follows - a missing lower bound defaults to $1$ (rather than the usual $0$),
and a missing upper bound defaults to $\Inf$:
$$
\begin{array}{llll}
\Tr{\qcont : S, \qminC : m, \qmaxC : M} \qquad  \\[\NL]
\multicolumn{1}{r}{
=\ \CMM{m}{M}{\Tr{S}}
}
\end{array}
$$

Then, we have the \xdeps\ assertion: 
$$\begin{array}{llll}
\QQ\xdeps\QQ : \{  \key{k_1} : [\key{k^1_1}\ldots,\key{k^1_{m_1}}],\ldots,  \key{k_n} : [\key{k^n_1}\ldots,\key{k^n_{m_n}}] \} \\
\QQ\xdeps\QQ :  \{ \key{k_1} : S_1,\ldots,\key{k_n} : S_n \}
\end{array}
$$
The first form specifies that, for each $i\in\SetTo{n}$,
if the instance is an object and if it contains a member with name $k_i$, then it must contain all of 
the member names $\key{k^i_1}\ldots,\key{k^i_{m_i}}$. The second form specifies that, under the same conditions,
the instance must satisfy $S_i$.
Both forms are translated using $\Req$ and $\Implies$: 
$$\begin{array}{llll}
\Tr{\QQ\xdeps\QQ : \{  \key{k_1} : [\key{r^1_1}\ldots,\key{r^1_{m_1}}],\ldots,  \key{k_n} : [\key{r^n_1}\ldots,\key{r^n_{m_n}}] \} } \\
\qquad =
 ( (\TObj\And\Req(k_1)) \Implies \Req( \key{r^1_1}\ldots,\key{r^1_{m_1}}) ) \\
\qquad\quad \And\ldots\And  ((\TObj\And\Req(k_n)) \Implies \Req(\key{r^n_1}\ldots,\key{r^n_{m_n}}))\\[\NL]
\Tr{\QQ\xdeps\QQ : \{ \key{k_1} : S_1,\ldots,\key{k_n} : S_n \}} \\
\qquad =
  ((\TObj\And\Req(k_1)) \Implies \Tr{S_1}) \\
\qquad\quad \And\ldots\And  ( (\TObj\And\Req(k_n)) \Implies \Tr{S_n}  ) \\[\NL]
\end{array}$$

Finally, all the other \jsonsch\ assertions are translated one by one in the natural way, as reported in Table \ref{tab:transl},
where we omit the symmetric cases (e.g.  \qmax : M, \qexmax~: M, etc) that can be easily guessed.

\begin{table}[hbt]
\begin{tabular}{lll}
\Tr{\qmin : m}
&=& $\Bet_m^{\Inf} $\\
\Tr{\qexmin : m}
&=& $\XBet_m^{\Inf} $\\
\Tr{\qmof : n}
&=& $\Mof(n)$\\
\Tr{\qminL : m} 
&=&  $\Pat(\,\hat{}\ .\{m,\}\,\$)$ \\ 
\Tr{\qpatt: r}
 &=&  $\Pat(r)$\\
\Tr{\qminIt : m}
&=&  $\CMM{m}{\Inf}{\True} $\\[\NL]
\end{tabular}
\caption{\label{tab:transl}Translation rules for {\jsonsch}.}
\end{table}


\subsection{How we evaluate complexity}

We have seen that {\jsonsch} can be translated to the algebra with a polynomial (actually, linear) size increase,
and in the rest of the paper we show that our algorithm runs in $O(2^{\PN})$ with respect to the size of the input
algebra, but with one important caveat:
hereafter, we assume that all $i$ and $j$ constants different from $\Inf$ that appear in $\PreIte{i}{S}$, $\PostIte{i}{S}$,
$\ContAfter{i}{S}$, $\CMM{i}{j}{S}$,  and $\Pro_i^j$, 
are smaller than the input size, and we call this assumption the \emph{linear constant assumption}. 
This is a reasonable assumption, since in practical cases these numbers tend to be extremely small when compared
with the input size.
Hereafter, whenever a result depends on this assumption, we will say that explicitly.


}
\iflong{\section{Witness generation}\label{sec:witness}}

\ifshort{\section{The structure of the algorithm}\label{sec:witness}}

\iflong{\subsection{The structure of the algorithm}}

In a recursive algorithm for witness generation,
in order to generate a witness for an {\ITO} such as
$\PReq(r:S)$, one can generate a witness $\J$ for $S$ and use it to build an object with 
a member whose name matches $r$ and whose value is $\J$. The same approach can be
followed for the other {\ITO}s.
For the Boolean operator
$S_1 \Or S_2$, one recursively generates witnesses of $S_1$ and $S_2$.

Negation and conjunction are much less direct: there is no way to generate a witness for $\Not S$ starting from a witness for $S$. Also, given a witness for $S_1$, if it is not a witness for $S_1 \And S_2$, we may
need to try infinitely many others before finding one that satisfies~$S_2$ as 
well.\iflong{\footnote{One may actually
solve the problem by ordered generation of witnesses for $S_1$ and $S_2$ and a merge-sort implementation
of intersection, but the algorithms that we explored with this approach seem far more expensive than ours.}}
We solve this problem as follows.
We first eliminate $\Not$ using not-elimination, then we bring all definitions 
of variables into DNF so that conjunctions are limited to sets of {\ITO}s that regard the
same type (Section \ref{sec:firstphases}).
\mrevmeta{
We then perform a form of \emph{and-elimination} over these homogeneous conjunctions
 (\emph{preparation}), and we finally use these 
``{\crcombined}'' homogeneous conjunctions to generate the witnesses, through a bottom-up
iterative process  (Section \ref{sec:recgen}).}{M.1, 3.3}

\mrevmeta{\emph{Preparation} is the crucial step: here we make all the interactions
between the conjuncted {\ITO}s explicit, which may require the generation of new variables.
This phase is delicate because it is exponentially hard in the general case, and we must organize it
in order to run fast enough in typical case. Moreover, it may generate infinitely many 
new variables, which we avoid with a technique based on ROBDDs, that we define in Section \ref{sec:robdd}.}{M.1, 3.3}
%

\hide{
In greater detail, our algorithm consists of the following steps.

\begin{compactenum}
\item 
\textbf{Transformation in positive, stratified, ground, canonical DNF}:
   we eliminate negations, and we transform the schema into a set of variables each bound to 
   a disjunction of canonical conjunctions, where a conjunction is \emph{canonical}
   when is formed by exactly one assertion~$\Type(T)$ plus a set of {\ITO}s related to
   that type $T$.
 \item \textbf{Object and array {\crcombination}} and \textbf{variable and-comple\-tion}: we rewrite object and array 
    conjunctions 
    into a form that will allow iterative generation of witnesses. This is a form of 
    \emph{and-elimination}, where we
   precompute all variable conjunctions that we may need during the next phase,
   and associate a fresh variable to each one (\emph{and-completion}).
\item \textbf{Iterative generation}: we start from a state where the semantics of every variable is   
     unknown,
      and we repeatedly try and generate a witness for each variable, using the prepared conjunctions,
      until a fix-point is
      reached.   
\end{compactenum}
}




\section{Transformation in positive, stratified, ground, canonical DNF}\label{sec:firstphases}

\mrevmeta{We will illustrate the preliminary phases of our algorithm by exploiting the running example of Figure~\ref{fig:rexample}.}{M.1, 3.3}

\subsection{Premise: ROBDD reduction}\label{sec:robdd}

Two expressions built with variables
and Boolean operators are \emph{Boolean-equivalent} when they can be proved equivalent using 
the laws of the Boolean algebra.
An ROBDD (Reduced Ordered Boolean Decision Diagram) is a data structure that provides the same
representation for two such expressions if, and only if, they are Boolean-equivalent \cite{DBLP:journals/tc/Bryant86}.
Hence, whenever we define a variable $x$ whose body 
$S_x$ is a Boolean combination of variables, in any phase of the algorithm,
we perform the \emph{ROBDD reduction}:
we compute the ROBDD representation of $S_x$,
$\key{robdd}(S_x)$, and we store a 
pair $\Def{x}{\key{robdd}(S_x)}$ in the {\ROBDDT} table, unless a pair
$\Def{y}{\key{robdd}(S_y)}$ with $\key{robdd}(S_x)=\key{robdd}(S_y)$ is already present.
In this case, we substitute every occurrence of $x$ with $y$.
This technique makes the entire algorithm more efficient and, crucially,
it ensures termination of the {\crcombination} phase
(Section~\ref{sec:objprep}). 


%
\newcommand{\Title}[1]{\mbox{\bf #1}}

\begin{figure}

\bcolormeta
\small
\begin{spreadlines}{0ex}
\begin{querybox2l}{(a)}
\bcolormeta
\begin{flalign*}
&\Def{r}{\PReq(b : \RRef{x})\Or \Props(\key{a}:\RRef{y})\Or\Props(\key{a.*}:\Not\RRef{r}\Or\RRef{x}),  } &\\[\NL]
&\Def{x}{\TArr}, 
\qquad
\Def{y}{\TNum}&
\end{flalign*}
\ecolormeta
\end{querybox2l}
\vspace{-0.2cm}
\begin{querybox2l}{(b)}
\bcolormeta
\begin{flalign*}
&\Def{r}{\PReq(b : \RRef{x})\Or \Props(\key{a}:\RRef{y})\Or\Props(\key{a.*}:\NR{r}\Or\RRef{x})  }, &\\[\NL]
&\Def{x}{\TArr}, 
\qquad \Def{y}{\TNum},  &\\[\NL]
&\Def{\NotVar{r}}{\TObj\And\Props(b : \NR{x})\And \PReq(\key{a}:\NR{y})}&\\[\NL]
&\qquad\qquad\qquad\qquad{\And\PReq(\key{a.*}:\RRef{r}\And\NR{x})  }, &\\[\NL]
&\Def{\NotVar{x}}{\TNull\!\Or\!\TBool\!\Or\!\TNum }
{\Or\!\ \TStr\!\Or\!\TObj}, &\\[\NL]
&\Def{\NotVar{y}}{\TNull\!\Or\TBool\!\Or\TStr } 
{\Or\!\ \TObj\!\Or\!\TArr} &
\end{flalign*}
\ecolormeta
\end{querybox2l}
\vspace{-0.2cm}
\begin{querybox2l}{(c)}
\bcolormeta
\begin{flalign*}
&\Def{r}{\PReq(b : \RRef{x})\Or \Props(\key{a}:\RRef{y})\Or\Props(\key{a.*}:\RRef{\textit{crx}})  }, &\\[\NL]
&\Def{\NotVar{r}}{\TObj\And\Props(b : \NR{x})\And \PReq(\key{a}:\NR{y})} &\\[\NL]
&\qquad\qquad\qquad\qquad{\And\PReq(\key{a.*}:\RRef{\textit{rcx}})  }, &\\[\NL]
&\Def{\textit{crx}}{\NR{r}\Or\RRef{x}}, 
\qquad \Def{\textit{rcx}}{\RRef{r}\And\NR{x}} &
\end{flalign*}
\ecolormeta
\end{querybox2l}
\vspace{-0.2cm}
\begin{querybox2l}{(d)}
\bcolormeta
\begin{flalign*}
&\Def{\textit{crx}}{\{\TObj,\Props(b : \NR{x}), \PReq(\key{a}:\NR{y}),} &\\[\NL]
&\qquad{ \ \ \ \PReq(\key{a.*}:\RRef{\textit{rcx}})\}\ \ \  \Or\ \ \  \{\TArr\} }, &\\[\NL]
&\Def{\textit{rcx}}{\{ (\PReq(b : \RRef{x}),\TNull \}  \Or \{ (\PReq(b : \RRef{x}),\TBool \}  } &\\[\NL]
&\qquad\Or\{ (\PReq(b : \RRef{x}),\TNum \}  \Or \{ (\PReq(b : \RRef{x}),\TStr \}  &\\[\NL]
&\qquad\Or\{ (\PReq(b : \RRef{x}),\TObj \} &\\[\NL]
&\qquad\Or\{ \Props(\key{a}:\RRef{y}),\TNull \}  \Or \{ \Props(\key{a}:\RRef{y}),\TBool \}  &\\[\NL]
&\qquad\Or\{ \Props(\key{a}:\RRef{y}),\TNum \}  \Or \{ \Props(\key{a}:\RRef{y}),\TStr \}  &\\[\NL]
&\qquad\Or\{ \Props(\key{a}:\RRef{y}),\TObj \} &\\[\NL]
&\qquad\Or\{ \Props(\key{a.*}:\RRef{\text{crx}}),\TNull \}  \Or \ldots &\\
&\qquad\Or\{ \Props(\key{a.*}:\RRef{\text{crx}}),\TObj \} &
\end{flalign*}
\ecolormeta
\end{querybox2l}
\vspace{-0.2cm}
\begin{querybox2l}{(e)}
\bcolormeta
\begin{flalign*}
&\Def{r}{\{\TObj,\PReq(b : \RRef{x})\}\Or 
            \{\TObj,\Props(\key{a}:\RRef{y})\}  } &\\[\NL]
&            \qquad{ \Or
            \{\TObj,\Props(\key{a.*}:\RRef{\textit{crx}}\} \Or \CTNull 
            } &\\[\NL]
&            \qquad{\Or \CTBool\Or\CTNum\Or\CTStr\Or\CTArr  },&\\[\NL]
&\Def{\textit{rcx}}{ \{\TObj,\PReq(b : \RRef{x})\} \Or \{\TObj,\Props(\key{a}:\RRef{y})\}} &\\[\NL]
&\qquad{\Or  \{\TObj,\Props(\key{a.*}:\RRef{\textit{crx}})\} \Or \CTNull 
            } &\\[\NL]
&            \qquad{\Or \CTBool\Or\CTNum\Or\CTStr\Or\CTArr  } &
\end{flalign*}
\ecolormeta
\end{querybox2l}
\end{spreadlines}

\RestoreNL
\ecolormeta

\caption{
(a)~Original term. (b)~After not-elimination. (c)~After stratification, omitting unaffected variables. (d)~After transformation to GDNF.
(e)~After canonicalization.}
\label{fig:rexample}

\end{figure}

\subsection{Not-elimination}

\label{sec:notelim}
\comment
{
\begin{theorem}
The positive core algebra cannot express $\NotUni$.
\end{theorem}

\begin{proof}
Assume, for a contradiction, that we can express the following type.
$$
\Ite(;\Type(\Int)) \And \NotUni
$$
Hence, there exists a schema $S$ written in the positive core algebra such that, for any number $n$, the sequence
$1,\ldots,n,n$ is allowed while $1,\ldots,n$ is not. Let $J$ be the biggest finite $j$ such that either $i:j : S'$ or $j:\Inf;S$
is present in $S$ and consider the sequences $1,\ldots,J+1$ and $1,\ldots,J+1,J+1$\ldots
To be completed keeping into account also $\Ex_i^j S$. 
\end{proof}
}

\iflong{%
Not-elimination, described in detail in our technical report~\cite{maybetcs}, proceeds in two phases.
\begin{compactenum}
\item Not-completion of variables: for every variable $\Def{x_n}{S_n}$ we define a 
corresponding $\Def{not\_x_n}{\Not S_n}$.\footnote{We do this, unless a variable
whose body is Boolean-equivalent to $\Not S_n$ already exists,
in which case that variable is used through ROBDD reduction}
\item Not-rewriting: we rewrite every expression $\Not S$ into an expression where the negation has been pushed inside.
\end{compactenum}


\paragraph*{Not-completion of variables}\label{par:notcompletion}

Not-completion of variables is the operation that adds a variable
$\key{not\_x}$ for every variable $\key{x}$ as follows:
$$\begin{array}{lllll}
\text{not-completion}(\Def{x_0}{S_0} , \ldots, \Def{x_n}{S_n}) = \\[0.8ex]
\quad\Def{x_0}{S_0},
, \ldots, \Def{x_n}{S_n}, \\[0.8ex]
\quad\Def{not\_x_0}{\Not S_0} , \ldots, 
\Def{not\_x_n}{\Not S_n}
\end{array}$$

After not-completion, every variable has a complement variable
$\NotVar{x_i}=not\_x_i$ and $\NotVar{not\_ x_i}=x_i$.
The complement $\NotVar{x}$ is used for not-elimination\ifshort{.}
 (and also
in the {\crcombination} phase).

} 

\ifshort{%
Not-elimination, described in detail in \cite{maybetcs}, proceeds in two phases: \emph{not-completion} of variables and \emph{not-rewriting}. 

During not-completion of variables, for every variable $\Def{x_n}{S_n}$ we define a 
corresponding $\Def{not\_x_n}{\Not S_n}$.  After not-completion, every variable has a complement variable
$\NotVar{x_i}=not\_x_i$ and $\NotVar{not\_ x_i}=x_i$.
The complement $\NotVar{x}$ is used for not-elimination.

During not-rewriting, we rewrite every expression $\Not S$ into an expression where the negation has been pushed inside,
by applying the rules
reported in~\cite{maybetcs}, which  include the following \mrevmeta{(Fig.\ \ref{fig:rexample}(b))}{}:
} 
\iflong{\paragraph*{Not-rewriting}\label{par:notcrewrite}}
\ifshort{%
\UpdateNL{0.1ex}
{\small
{
\[
\begin{array}{llll}
\Not(\Pat(r)) &=& \Type(\Str) \And \Pat(\CoP{r}) \\[\NL]
\Not(\Props(\key{r} : S))
 & = &
\Type(\Obj)  \And  \PReq(\key{r} : \Not S)
\\[\NL]
\Not(\PReq(r:S)) & = & \TObj \And \Props(r:\Not S) \\[\NL]
\Not(\PreIte{l}{S}) &=&
\Type(\Arr) \And \PreIte{l}{\Not S_i} \And \CMM{l}{\Inf}{\True}\\[\NL]
\Not(\PostIte{i}{S}) &=&
\Type(\Arr) \And \ContAfter{i}{\Not S}\\[\NL]
\Not(\ContAfter{i}{S}) & = & \TArr \And \PostIte{i}{\Not S} \\[\NL]
\Not(\RRef{x}) &=& \NR{x} 
\end{array}
\]}
}
\RestoreNL%
Not-elimination of a schema of size $N$ produces an equivalent sche\-ma 
of size $O(N)$~\cite{maybetcs};
hereafter, we use $N$ 
to indicate the size of the abstract syntax tree of the 
original schema, where numbers and strings are represented in binary notation,
assuming that the $i$ and $j$ constants different from $\Inf$ that appear in $\PreIte{i}{S}$, $\PostIte{i}{S}$,
$\ContAfter{i}{S}$, $\CMM{i}{j}{S}$,  and $\Pro_i^j$, 
are smaller than the input size. 
This assumption is justified by the observation that in practical cases these numbers tend to be extremely small in comparison 
to the input size.
}
\iflong{%
We rewrite $\Req(k)$ as $\PReq(\keykey{k}:\True)$, and then we inductively apply the rules
in Figure \ref{fig:notelimvar}.
It is easy to prove that not-elimination can be performed in linear time and increases the schema size of a linear
factor.
We report here the following result from \cite{maybetcs}.
%
%
\begin{property}
For any system where recursion is guarded,
not elimination preserves the semantics of every variable.
\end{property}
}
\iflong{From now on, every other phase of the algorithm will only produce schemas that belong
to the positive algebra.}

\iflong{
\newcommand{\EQ}{\ =\ }
\renewcommand{\EQ}{& = &}
\newcommand{\SEP}{\ ;\ }

\StoreNL
\setlength{\NL}{0.6ex}

\begin{figure}[!tb]
\small
\[
\begin{array}{llll}
\Not(\IBT(\xtrue)) &=& \Type(\Bool) \And \IBT(\xfalse) \\[\NL]
\Not(\IBT(\xfalse)) &=& \Type(\Bool) \And \IBT(\xtrue) \\[\NL]
\Not(\Pat(r)) &=& \Type(\Str) \And \Pat(\CoP{r}) \\[\NL]
\Not(\Bet_{m}^{M}) & = &
\Type(\Num) \And (\XBet_{-\Inf}^{m} \Or \XBet_{M}^{\Inf}) \\[\NL]
\Not(\XBet_{m}^{M}) & = &
\Type(\Num) \And (\Bet_{-\Inf}^{m} \Or \Bet_{M}^{\Inf}) \\[\NL]
\Not(\Mof(q)) & = &
\Type(\Num) \And \NotMof(q)  \\[\NL]
\Not(\NotMof(q))  & = & \TNum \And  \Mof(q) \\[\NL]
\Not(\Props(\key{r} : S))
 & = &
\Type(\Obj)  \And  \PReq(\key{r} : \Not S)
\\[\NL]
\Not(\PReq(r:S)) & = & \TObj \And \Props(r:\Not S) \\[\NL]
\Not(\Pro_{i}^{j}) & = & 
\Type(\Obj) \And (\Pro_{0}^{i-1} \Or \Pro_{j+1}^{\Inf}) \\[\NL]
\Not(\PreIte{l}{S}) &=&
\Type(\Arr) \And \PreIte{l}{\Not S_i} \And \CMM{l}{\Inf}{\True}\\[\NL]
\Not(\PostIte{i}{S}) &=&
\Type(\Arr) \And \ContAfter{i}{\Not S}\\[\NL]
\Not(\ContAfter{i}{S}) & = & \TArr \And \PostIte{i}{\Not S} \\[\NL]
\Not(\CMM{i}{j}{S}) & = &
\Type(\Arr) \And (\CMM{0}{i-1}{S} \Or \CMM{j+1}{\Inf}{S}) \\[\NL]
\Not(\Type(T)) & = & \BigOr( \Type(T') \M T' \neq T ) \\[\NL]
\Not(\RRef{x}) &=& \NR{x}\\[\NL]
\Not (S_1 \And S_2) \EQ  (\Not S_1) \Or (\Not S_2) \\[\NL]
\Not (S_1 \Or S_2) \EQ  (\Not S_1) \And (\Not S_2) \\[\NL]
\Not (\Not S) \EQ  S 
\\[\NL]
\end{array}
\]
\caption{Not-pushing rules --- unsatisfiable disjuncts, such as $\Pro_{0}^{-1}$ or $\Pro_{\Inf}^{\Inf}$,  are generated
as $\False$.} 
\label{fig:notelimvar}
\end{figure}

\RestoreNL

}

\iflong{\subsection{Stratification}\label{sec:varnorm}}
\ifshort{\subsection{Stratification and Transformation in Canonical Guarded DNF}\label{sec:varnorm}\label{sec:DNF}}

\ifshort{\subsubsection*{Stratification}}
We say that a schema is \emph{stratified} when every schema argument of every {\ITO} is a variable,
so that 
$ \PReq{(a:\RRef{x}\And\RRef{y})}$ is not stratified
while $ \PReq{(a:\RRef{w})}$ is stratified.

\iflong{

Stratification makes it easy to build a witness for a typed group such as
$$
\TG{\Ob, \PReq(\PPP{a}:\RRef{x}), \PReq(\PPP{b}:\RRef{y})}
$$
after a witness for each involved variable has been built.

In this phase, for every {\ITO} that has a subschema $S$ in its syntax, 
such as $\CMM{i}{j}{S}$, 
when $S$ is not a variable,
we create a new variable $\Def{x}{S}$, and we substitute 
$S$ with $\RRef{x}$.  
For every variable $\Def{x}{S}$ that we define, we must also define its complement
 $\Def{not\_x}{\Not S}$, and perform not-elimination and stratification on $\Not S$
 --- see Figure \ref{fig:rexample}(c).
As specified in Section \ref{sec:robdd}, we apply ROBDD reduction to $\Def{x}{S}$
and $\Def{not\_x}{\Not S}$.

}
\ifshort{During stratification, for every {\ITO} that has a subschema $S$ in its syntax, 
such as $\CMM{i}{j}{S}$, 
when $S$ is not a variable,
we create a new variable $\Def{x}{S}$,  we substitute 
$S$ with $\RRef{x}$, and we apply not-completion, not-elimination and ROBDD reduction
to $\Def{x}{S}$ \crevmeta{(Figure \ref{fig:rexample}(c))}.

}

\ifshort{Stratification of a schema of size $N$ produces an equivalent sche\-ma of size $O(N)$
 \cite{attouche2022witness}.}

\iflong{
\begin{property}
Stratification transforms a schema $\Doc{S}{\E}$ into a schema $\Doc{S'}{\E'}$
such that $\semca{S}{E} = \semca{S'}{E'}$.
\end{property}

\begin{property}
Stratification transforms a schema $\Doc{S}{\E}$ into a schema $\Doc{S'}{\E'}$
such that $|\Doc{S'}{\E'}|$ is in $O(N)$, where $N = |\Doc{S}{\E}|$.
\end{property}

\begin{proof}
Assume that stratification is performed bottom up, so that  $\CMM{i}{j}{ \CMM{l}{k}{S}}$ is first transformed into 
$\CMM{i}{j}{ \CMM{l}{k}{x}}$ with  $\Def{x}{S}$ and $\Def{not\_x}{\Not S}$, and then in
$\CMM{i}{j}{y}$ with  $\Def{y}{\CMM{l}{k}{x}}$ and $\Def{not\_y}{\Not \CMM{l}{k}{x}}$. 
In this way, every $S$ that is moved
to the environment is only copied twice (once below negation), and each such operation generates two instances of $x$ and one of $\Not x$.
Hence, each node in the original tree corresponds to a constant number of nodes in the stratified tree - in the worst case, it generates three variables, one negation, and two copies of the original node. At this point we apply not-elimination,
and this step is linear as well.
\end{proof}

}

\iflong{\subsection{Transformation in Canonical GDNF}\label{sec:DNF}}

\newcommand{\phn}{^{n}}
\newcommand{\phe}{^{G}}
\newcommand{\Ts}{{\mathcal{T}}}
\newcommand{\Sch}[2]{({#1},{#2})}

\subsubsection*{Guarded DNF}
A schema is in Guarded Disjunctive Normal Form (GDNF) if it 
has the shape 
$
\Or(\And(S_{1,1},\ldots,S_{1,n_1}),\ldots,\And(S_{l,1},\ldots,S_{l,n_l}))
$
and every $S_{i,j}$ is a {\TE}. 
\iflong{Every conjunction may be trivial ($n_i = 1$), and so may be the disjunction ($l=1$).}
\ifshort{To transform an environment $\E$ into a corresponding one $\E\phe$ in  GDNF, we just substitute every 
variable that is not in the scope of an {\ITO}
with its body, a process that is guaranteed to terminate, thanks to the guardedness condition on 
$\E$, and we bring the result in DNF by distributivity of $\And$ and $\Or$ \crevmeta{(Figure~\ref{fig:rexample}(d))};
hereafter, for brevity, we use $\TG{S_1,\ldots,S_n}$ to indicate a conjunction
$S_1\And\ldots\And S_n$. 
Reduction to GDNF can lead to an exponential explosion, and is actually a very expensive phase, 
according to our experiments (Section ~\ref{sec:exp}):
\crevmeta{observe, in Figure~\ref{fig:rexample}(d), how the size of \key{rcx} is the product of the
sizes of $r$ and that of $\NotVar{x}$.}
}

\iflong{
To produce a new environment $E\phe$ in GDNF starting from a positive and stratified 
environment $E$, we first define an ordered enumeration $\Set{x_1,\ldots,x_o}$
of the variables in $\Vars(\E)$ such that when $x_i$ directly depends of $x_j$ (as defined
in Section \ref{sec:syntax}) then $j<i$. We know that such enumeration exists because 
recursion is guarded.
We now compute $E\phe(x_i)$ starting from $x_1$ and going onward, so that, when
we compute $E\phe(x_i)$, $E\phe(x_j)$ has already been computed for each $j<i$.

Let $\Ts$ denote the set of all {\TE}s that appear in $E$ as subterms of $\E(y)$ for any $y$,
 so that,
if $$\E\ \ =\ \ \Def{x}{(\TNum \And \PReq(\PPP{a}:x))\Or \Mof(3)}$$ then 
$\Ts = \Set{\TNum,\ \PReq(\PPP{a}:x),\ \Mof(3)}$.
As we will show, reduction in GDNF does not create any new typed expression, hence every 
term in GDNF corresponds to a set $DC$ (\emph{Disjunction of Conjunctions}) of subsets of $\Ts$  as follows.
$$ {\E\phe(x)} =
{\BigOr_{C \in DC_x}  \BigAnd_{S\in C}   S\ \ \ \text{where}\ \ \ \  DC_x \in \Parts(\Parts(\Ts))}$$

To compute this set-of-sets representation $g(E(x))$ of the GDNF of
the body $\E(x)$ of every  $x$ defined in $\E$, we apply the following rules:
$$\begin{array}{llll}
g(S) &=& \Set{\Set{S}} \qquad\qquad \mbox{\ if $S$ is a {\TE}} \\[\NL]
g(y) &=& E\phe(y)  \\[\NL]
g(S_1  \Or S_2) &=& g(S_1) \cup g(S_2)  \\[\NL]
g(S_1  \And S_2)  &=& \bigcup_{(C_1,C_2) \in g(S_1)\times g(S_2)}  (C_1\cup C_2) 
\end{array}$$

When $S$ is a typed expression, it is translated into a trivial GDNF. Each variable $y$ inside $\E(x)$ had its body
already transformed. The rule for $\Or$ is trivial, while the rule for $\And$ is Boolean algebra distributivity:
for each conjunction $\BigAnd_{S\in C_1}S$ of $S_1$ and for each conjunction $\BigAnd_{S\in C_2}$ of $S_2$, the
conjunction $\BigAnd_{S\in C_1}S \And \BigAnd_{S\in C_1}S = \BigAnd_{S\in C_1\cup C_2}S$ is inserted in the result.


Reduction to GDNF can lead to an exponential explosion, and it is actually the most expensive phase
of our algorithm, according to our measures (Section \ref{sec:exp}).

\begin{property}
For a given schema $\Doc{x}{E}$, such that $n = |\Doc{x}{E}|$, the size of $\Doc{x}{E\phe}$ is in $O(2^n)$,
and it can be build in time $O(2^n)$.
\end{property}

\begin{proof}
The schema $\Doc{x}{E\phe}$ has $O(n)$ variables.
The body of each variable can be represented as a set $DC$ belonging to $\Parts(\Parts(\Ts))$
The set $\Parts(\Ts)$ has size $O(2^n)$, hence every set of sets $DC$
contain at most $O(2^n)$ sets, and each of these sets can be represented using $n$ bits.
This yields a total upper bound of $O(n)\times O(n)\times O(2^n)$ for $\Doc{x}{E\phe}$.
As for the construction time, the most expensive part is the computation of 
$\bigcup_{(C_1,C_2) \in g(S_1)\times g(S_2)}  (C_1\cup C_2)$, that may take place once
for each variable. The size of $g(S_1)\times g(S_2)$ is in $O(2^n)$, the size of $C_1$ and $C_2$
is in $O(n)$, hence this computation is in $O(2^n)$.
\end{proof}
}


\subsubsection*{Canonicalization}


\newcommand{\Filter}[2]{\key{filter}({#1},{#2})}

\iflong{Canonicalization is a process defined along the lines of \cite{DBLP:conf/issta/HabibSHP21}.}
We say that a conjunction that contains exactly one assertion $\Type(T)$ and a set of {\ITO}s of that
same type $T$ is a \emph{typed group} of type $T$; canonicalization splits every conjunct of the 
GDNF into a set of  \emph{typed groups} \mrevmeta{(Figure \ref{fig:rexample}(e),
where we also applied elementary equivalences, such as idempotence of $\Or$).}.

\hide{%
For example, a trivial conjunction $\{\Mof(3.5)\}$ is not a typed group, since it is missing a $\Type(T)$
assertion, and is transformed in the following disjunction of six groups:
$$
 \{\TNum, \Mof(3.5)\}  \Or \{\TNull\}  \Or \{\TBool\} \Or \ldots
$$
}

\iflong{%
In order to transform a conjunction $C$ of a GDNF $DC$ into a typed group, we first repeatedly apply 
the following rewriting rules, which preserve the meaning of the conjunction. In the third rule,
$\key{{\ITO}}(T')$ are the {ITO}s associated to type $T'$, which are trivially satisfied when 
in conjunction with a $\Type(T)$ with $T \neq T'$:
$$\begin{array}{llllll}
\Type(T), \Type(T) &\To & \Type(T) & \\[\NL]
\Type(T), \Type(T') &\To & \False & T \neq T' \\[\NL]
\False, S &\To & \False &\\[\NL]
\Type(T), S &\To & \Type(T) & S \in \key{{\ITO}}(T'), \ T' \neq T \\[\NL]
\end{array}
$$
The first three rules ensure that the result is either $\False$, which is then deleted from the disjunction,
or has exactly one $\Type(T)$ assertion, or has none. If it has exactly one $\Type(T)$ assertion,
then the fourth rule ensures that all the $\ITO$s refer to type $T$.
If it has no $\Type(T)$ assertion, we transform it in the following equivalent disjunction, where
$\Filter{\TG{S_1,\ldots,S_n}}{T}$ is the conjunction of those {\ITO}s in $\TG{S_1,\ldots,S_n}$ 
whose type is $T$:
$$\begin{array}{llllll}
(\TNull) & \Or (\TBool \And \Filter{C}{\Bo}) \\
& \Or (\TStr \And  \Filter{C}{\St})\ldots
\end{array}
$$
so that every $C\in DC$ denotes a set of values of the same type.
}

\mrevmeta{
By construction, every phase described in this section transforms  a {\JS} document into
an equivalent one.
}{M.3, 2.8}
\mrevmeta{
\begin{property}[Equivalence]\label{pro:preliminarycc}
The phases of not-elimination, stratification, transformation 
into Canonical GDNF, transform a {\JS} document into
an equivalent one.
\end{property}
}{M.3, 2.8}


\section{{\Crcombination} and \iflong{witness }generation}\label{sec:recgen}

\renewcommand{\Next}{\kw{next}}
\renewcommand{\EOV}{\kw{EOV}}
\renewcommand{\DOV}{\kw{DOV}}
\renewcommand{\AA}{A}
\newcommand{\cf}{\key{cf}}
\newcommand{\Meets}{\akey{\#}}
\renewcommand{\Meets}{\cap}
\newcommand{\JJOfD}[1]{\mathit{J}^{\mathit{#1}}}
\newcommand{\MeetsD}[1]{\cap\JJOfD{#1}}

\newcommand{\AAE}[1]{{\mathcal A}^{#1}_{\E}}
\newcommand{\AAI}{{\mathcal A}^{\Inf}_{\E}} 

\subsection{Assignments and bottom-up semantics}

\renewcommand{\Sch}{\Doc}

\ifshort{
We define an assignment $\AA$ for an environment $\E$ as a function mapping each variable of $\E$ to 
a set of {\json} values.
An assignment is sound when it maps each variable $y$ to
a subset of its semantics $\semca{y}{\E}$, which denotes the
set of {\json} values that satisfy $y$ in $\E$.}

\iflong{
Let us define an assignment $\AA$ for an environment $\E$ as a function mapping each variable of $\E$ to 
a set of {\json} values.
An assignment is sound when it maps each variable to
a subset of its semantics.
We order assignments by variable-wise inclusion.}

\newcommand{\asEval}{assignment-evaluation}

\iflong{
\begin{definition}[Assignments, Soundness, Order]
}
\ifshort{
\begin{definition}[Assignments, Soundness]
}
An assignment $\AA$ for an environment $\E$ is a function mapping each variable of $E$ to a set of {\json}
values.
\iflong{ }
An assignment $\AA$ for $\E$ is sound iff for all $y\in \Vars(E)$:
$\AA(\RRef{y}) \subseteq \semca{y}{\E}$.
\iflong{ 
We say that $\AA \leq \AA'$ iff $\forall y.\ \AA(y)\subseteq \AA'(y)$.}
\end{definition}

\ifshort{
\mrevthree{
Given a schema $\Sch{S}{\E}$ and an assignment $\AA$ for $\E$, we can evaluate
$S$ using $\AA$ to interpret any variable in $S$,  by applying the rules exemplified in Figure~\ref{tab:corealg-sem}
(see \cite{attouche2022witness} for a complete list),
where universal quantification on an empty set is true, the set $\SetTo{0}$ is empty,
and $\semt(\Bool)$ is the set of {\json} values of type $\Bool$.
For example, if $\AA(x)=\Set{\J}$, and 
if $S={\{\TArr,\PostIte{0}{x},\CMM{1}{2}{\True}\}}$, 
then $\semas{S}{\AA}=\Set{[\J],[\J,\J]}$.
Intuitively,  $\semas{S}{\AA}$ uses the witnesses collected by
$\AA$ in order to build bigger witnesses
for $S$.
}{3.1}

}
\ifshort{
\begin{figure}[tb]
\UpdateNL{0.5ex}
{\small{
\[
\begin{array}{lcl}
\semas{\RRef{x}}{\AA}  &= & \AA(x) \\[\NL]
   \semas{\IBT(b)}{\AA} & = & \setst{J}{J\in\semt(\Bool) \Rightarrow  J=b }\\
\semas{\Props(\key{r} : S)  }{\AA} & = & \SetOpen\, {J} \,\mid\, J = \{(k_1:J_1),\ldots,(k_n:J_n)\} \Rightarrow  \\
     &&
   \forall i\in \SetTo{n}.\ k_i\in\rlan{r} \ \Rightarrow J_i \in \semas{S}{\AA} \,\SetClose \\
 \semas{  \PreIte{l}{S} }{\AA} & = & \SetOpen\, {J} \,\mid\, J =  [\J_1, \ldots, \J_n ]   \Rightarrow \\
     &&
      n \geq l  \Rightarrow J_l \in   \semas{S}{\AA}  \,\SetClose  \\
 \semas{  \PostIte{i}{S} }{\AA} & = & \setst{J}{ J =  [\J_1, \ldots, \J_n ]  \Rightarrow \\
     &&
     \forall j\in\SetTo{n} .\ j > i \Rightarrow J_j \in   \semas{S}{\AA} }\\
\semas{  \CMM{i}{j}{S}  }{\AA} & = & \setst{J}{ J =  [\J_1, \ldots, \J_n ] 
       \Rightarrow \\
     &&
      i \leq \ |\setst{l}{\J_l\in \semas{S}{\AA}}|\ \leq j }\\
 \semas{S_1 \And S_2}{\AA} & = &  \semas{S_1 }{\AA}  \cap   \semas{ S_2}{\AA} \\
 \semas{S_1 \Or S_2}{\AA} & = &  \semas{S_1 }{\AA}  \cup   \semas{ S_2}{\AA} \\
 \semas{\PReq(r:S)}{\AA} & = & \setst{J}{J = \Set{(k_1:J_1),\ldots,(k_n:J_n)}   \Rightarrow \\
     &&  \exists i \in \SetTo{n} .\  k_i\in\rlan{r}) \ \And J_i \in \semas{S}{\AA} } \\
 \semas{\ContAfter{i}{S}}{\AA} & = & \setst{J}{ J =  [\J_1, \ldots, \J_n ] \Rightarrow  \\
     &&   \exists l.\ l > i \ \And \J_l \in \semas{ S}{\AA} }
 \end{array}
\]}}
\RestoreNL
\caption{Rules for assignment evaluation.}
\label{tab:corealg-sem}
\end{figure}
}


\iflong{
Given a schema $\Sch{S}{\E}$, an assignment $\AA$ for $\E$ defines an \emph{\asEval}
for $S$ by applying the rules in Figure \ref{tab:corealg-set}, 
which are the same rules that define environment-based
semantics $\semca{S}{\E}$, with the only difference that a variable $\RRef{x}$ is not interpreted
by interpreting the schema $\E(x)$, but directly as the set of values $\AA(x)$
(we always assume that every schema $\Sch{S}{\E}$ is closed and guarded).

For all schemas  not containing subschemas, such as $\IBT(b)$, we just define
$\semas{\IBT(b)}{\AA} = \semca{\IBT(b)}{\E}$, and neither $\AA$ nor $\E$ play any role in the definition
\begin{figure}[ht]

\[
\begin{array}{lcl}
\semas{\RRef{x}}{\AA}  &= & \AA(x) \\[\NL]
   \semas{\IBT(b)}{\AA} & = & \setst{J}{J\in\semt(\Bool) \Rightarrow  J=b }\\[\NL]  
  \semas{\Props(\key{r} : S)}{\AA} & = & \setst{J}{J = \{(k_1:J_1),\ldots,(k_n:J_n)\} \Rightarrow  \\
     &&
   \forall i\in \SetTo{n}.\ k_i\in\rlan{r} \ \Rightarrow J_i \in \semas{S}{\AA} } \\[\NL]
\iflong{
\semas{\PreIte{l}{S} }{\AA} & = & \setst{J}{ J =  [\J_1, \ldots, \J_n ] \Rightarrow  \\
     &&
          n \geq l \Rightarrow J_l \in   \semas{  S }{\AA} }\\[\NL]        
\semas{  \CMM{i}{j}{S}  }{\AA} & = & \setst{J}{ J =  [\J_1, \ldots, \J_n ] 
       \Rightarrow \\
     &&
      i \leq \ |\setst{l}{\J_l\in \semas{S}{\AA}}|\ \leq j }\\[\NL]
}
 \semas{S_1 \And S_2}{\AA} & = &  \semas{S_1 }{\AA}  \cap   \semas{ S_2}{\AA} \\[\NL]
\iflong{
 \semas{S_1 \Or S_2}{\AA} & = &  \semas{S_1 }{\AA}  \cup   \semas{ S_2}{\AA} \\[\NL]
}
 \ldots
 \end{array}
\]
\caption{Rules for {\asEval}.}
\label{tab:corealg-set}
\end{figure}

For schemas in the positive algebra, iterated
{\asEval} yields an alternative notion of semantics, as follows.
}

\ifshort{
\mrevthree{
Hence, the repeated application of these rules, 
starting from an empty assignment  $\AAE{0}$, 
defines a sequence of assignments $\AAE{i}$ containing more and more
witnesses, whose limit $\AAI$ defines a bottom-up semantics for
{\jsonsch}.
}{3.1}

\begin{definition}[$\AAE{i}$, $\AAI$]
For a given positive environment $\E$, the sequence of assignments $\AAE{i}$ is defined 
as\iflong{ follows}:
$$
\begin{array}{llllllllll}
\forall y\in\Vars(\E):&  &\AAE{0}(y) = \emptyset 
 \\[10\NL] \forall y\in\Vars(\E):
 && \AAE{i+1}(y) =  \semas{\E(y)}{\AAE{i}} \\
\end{array}
$$

The assignment $\AAI$ is defined as $\bigcup_{i \in \Nat}\AAE{i}$.
\end{definition}

}

\iflong{
\begin{definition}
For a given positive environment ${\E}$, the corresponding assignment transformation $T_{\E}(\_)$ is the function from assignments to assignments defined as follows:
$$\begin{array}{llll}
\forall y\in\Vars(\E).\ T_{\E}(\AA)(y) = \semas{\E(y)}{\AA}
\end{array}$$
\end{definition}

Intuitively, if $\AA$ collects witnesses for the variables in $\E$,
then $T_{\E}(\AA)$ uses $\E$ in order to build new witnesses
starting from those in $\AA$.
For example, if $\E$ contains $\Def{y}{\{\TArr,\PostIte{0}{x},\CMM{1}{3}{\True}\}}$, 
if $\AA(x)=\Set{\J}$, then $T_{\E}(\AA)(y)=\Set{[\J],[\J,\J],[\J,\J,\J]}$.

For any positive environment $\E$, the corresponding assignment transformation is monotone in $\AA$, by positivity
of $\E$, hence $T_{\E}$ has a minimal fix-point, 
that is the limit $\AAI$ of the sequence $\AAE{i}$ defined accordingly
to Tarski theorem, starting from the empty assignment and then reapplying $T_{\E}$.

\begin{definition}[$\AAE{i}$, $\AAI$]
For a given positive environment $\E$, the sequence of assignments $\AAE{i}$ is defined 
as\iflong{ follows}:
$$
\begin{array}{llllllllll}
& \forall y\in\Vars(\E).\ \AAE{0}(y) = \emptyset 
\shortlong{&&}{\\[\NL]}
 & \AAE{i+1} = T_{\E}(\AAE{i}) \\
\end{array}
$$

The assignment $\AAI$ is defined as $\bigcup_{i \in \Nat}\AAE{i}$.
\end{definition}

\begin{property}
For any positive ${\E}$, the assignment $\AAI$ is the minimal fix-point
of the assignment transformation $T_{\E}$.
\end{property}
}

\ifshort{
\mrevthree{
The function $\semca{S}{\E}$ described in the full paper \cite{attouche2022witness} corresponds to the official semantics, 
and is based on the top-down substitution of variables with their definitions during the validation of a JSON value.
In \cite{attouche2022witness}, we show that, on positive schemas, the bottom-up interpretation
$\semas{S}{\AAI}$ corresponds to $\semca{S}{\E}$.
}{3.1}
}

\iflong{
In Section \ref{sec:semantics}, we adopted the official top-down semantics for {\json} schema in order to follow the standard and
because it also applies to negative operators. However, on positive schemas, the top-down semantics and the
bottom-up fix-point coincide.

\begin{property}
For any positive schema $\Sch{S}{\E}$, 
the following equality holds:
$$
\semca{S}{\E} = \semas{S}{\AAI}
$$
\end{property}
}

\iflong{
\begin{proofsketch}
We prove, by induction on $i$ and, when $i$ is equal, on $S$, that for all $i$, and for any positive 
assertion $S$ that is closed 
wrt $\E$, the following holds:
$$\semcai{S}{\E} = \semas {S}{\AAE{i}}
$$
For the inductive step $i+1$, if $S$ is an operator that contains no schema subterm, the equality
$$\semcapar{S}{\E}{i+1} = \semas {S}{\AAE{i+1}}
$$ is immediate.
If $S$ is a variable, we have, by definition,
$\semcapar{y}{\E}{i+1}=\semcapar{\E(y)}{\E}{i}$
and
$\semas{y}{\AAE{i+1}} = (\AAE{i+1})(y) = \semas{\E(y)}{\AAE{i}}$;
we can conclude since
$\semcapar{\E(y)}{\E}{i} = \semas{\E(y)}{\AAE{i}}$ holds by induction on $i$.
For  $S=S_1 \And S_2$ we reason by induction on $S$ as follows:  
$$\begin{array}{llll}
\semcapar{S_1 \And S_2}{\E}{i+1}
 = \semcapar{S_1}{\E}{i+1}\cap \semcapar{ S_2}{\E}{i+1}\\[\NL]
 = \semas{S_1}{\AAE{i+1}}\cap\semas{S_2}{\AAE{i+1}}
 =\semas{S_1\And S_2}{\AAE{i+1}}
\end{array}$$
For all other operators we reason in the same way.

Finally, the base case $i = 0$. 
When $S=x$, then both $\semcapar{x}{\E}{0}$ and $\semas{x}{\AAE{0}}$ are the empty set.
In all other cases, we reason as in case $i>0$.

Now, since $\semcap{S}{\E}$ coincides with $\semas{S}{\AAE{p}}$ for any $p$, 
then $\semcap{S}{\E}$ is a succession of sets that grows with $p$, 
hence $\bigcap_{p \geq i}\semcapar{S}{\E}{p} = \semas{S}{\AAE{i}}$,
hence $\bigcup_{i\in N}\bigcap_{p \geq i}\semcapar{S}{\E}{p}
= \bigcup_{i\in\Nat} \semas{S}{\AAE{i}}
= \semas{S}{\AAI}$.

\end{proofsketch}
}

Any {\json} value $\J$ has a \emph{depth} $\Dep(\J)$, that is the number of levels of its tree
representation, formally defined as follows.

\begin{definition}[Depth $\Dep(\J)$, $\JJOfD{\key{d}}$]
The depth of a {\json} value $\J$, $\Dep(\J)$, is defined as follows, where $\max(\Set{\ })$ 
is defined to be 0:
$$\begin{array}{lllllllll}
\mbox{$\J$ belongs to a base type}: & \Dep(\J) = 1  \\[\NL]
\J = [\J_1,\ldots,\J_n]:                     & \Dep(\J) = 1 + \max(\Set{\Dep(\J_1),\ldots,\Dep(\J_n)}) \\[\NL]
\J = \{ \ \key{a_1}: \J_1,\ldots, \key{a_n}: \J_n\ \}:      & \Dep(\J) = 1 + \max(\Set{\Dep(\J_1),\ldots,\Dep(\J_n)}) \\[\NL]
\end{array}$$

$\JJOfD{\key{d}}$ is the set of all {\json} values $\J$ with $\Dep(\J)\leq d$.

\end{definition}

The assignment $\AAE{i}$ includes all witnesses of depth $i$:
for any depth $i$, it can be proved that $(\semca{y}{\E}\MeetsD{i})\subseteq \AAE{i}(y)$.

\iflong{%
Bottom-up semantics is the basis of bottom-up witness generation: we will compute a witness for $\Sch{S}{\E}$
by approximating the sequence $\AAE{i}$.
}

\subsection{Bottom-up iterative witness generation}\label{sec:bottomup}
\ifshort{\label{sec:implementation}}

\newcommand{\Gen}{\ensuremath{\text{\tt Gen}}}

Since $\Sch{S}{\E}$ is equivalent to $\Sch{x}{\Def{x}{S},\E}$, we will discuss here, for
simplicity, generation for the $\Sch{x}{\E}$ case.

Our algorithm for bottom-up iterative witness generation for a schema $\Sch{x}{\E}$
produces a sequence of finite assignments $\AA^i$, each approximating the assignment $\AAE{i}$,
until we reach either a witness for $x$ or an ``unsatisfiability fix-point'', 
which is a notion that we will introduce shortly.


\mrevthree{
$\AA^i$ is built as follows: $\AA^0=\AAE{0}$;
then, at step $i$, for each $y\in\Vars(\E)$, we compute a set of new values for $y$ based
on the current assignment $\AA^i$ using a generation algorithm
$\Gen(\E(y),\AA^i)$ that computes a subset of $\semas{\E(y)}{\AA^i}$; formally,
$\AA^{i+1}(y)=\Gen(\E(y),\AA^i)$. 
Our specific \Gen\ algorithm is defined in the next section, but we show
now that any generic algorithm $g$ can be used 
to approximate $\semas{\E(y)}{\AA^i}$,
provided that $g$ is 
\emph{sound} and \emph{generative}. 
}{3.1}

We first 
introduce a notion of \emph{$i$-witnessed assignment} $\AA$:
if a variable $y$ has a witness $\J$ with $\Dep(\J)\leq i$, then $y$ has a witness 
in an \emph{$i$-witnessed assignment} $\AA$.

\begin{definition}[$i$-witnessed]
For a given environment $\E$, and an
assignment $\AA$ for $\E$, we say that $\AA$ is $i$-witnessed if:
$$\forall y\in\Vars(\E). \ (\semca{y}{\E}\MeetsD{i})\neq \emptyset\ 
\Implies\ \AA(y)\neq\emptyset
$$
\end{definition}

\mrevthree{Generativity of $g$
means 
that, if 
$\AA$ is $i$-wit\-nessed, then the assignment computed using
$g$ is ($i$+1)-witnessed, so that, by repeated application of $g$ starting from
$\AA^0$, every non-empty variable  will be 
eventually ``witnessed'' (Property \ref{pro:coandco}).
}{3.1}

Hereafter, we say that a triple $(S,\E,\AA)$ is coherent if $\E$ is guarded and closing
for $S$, and if $\Vars(\E)=\Vars(\AA)$.

\begin{definition}[Soundness of $g$]
A function
$g(\_,\_)$ mapping each pair assertion-assignment to a set of {\json} values is \emph{sound}
iff, for every coherent $(S,\E,\AA)$,
if $\AA$ is sound for ${\E}$, then $g(S,\AA)\subseteq \semca{S}{\E}$.
\end{definition}

\begin{definition}[Generativity of $g$]\label{def:gen}
A function
$g(\_,\_)$ mapping each pair assertion-assignment to a set of {\json} values is \emph{generative}
for an assertion  $S$ iff for any $\E$ and $\AA$ such that $(S,E,A)$ is coherent:
\begin{compactenum}
\item if $(\semca{S}{\E}\MeetsD{1}) \neq \emptyset$, then $g(S,\AA) \neq \emptyset$;
\item for any $i\geq 1$, if $\AA$ is $i$-witnessed, and if
   $(\semca{S}{\E}\MeetsD{i+1}) \neq \emptyset$,   then $g(S,\AA) \neq \emptyset$.
\end{compactenum}
$g$ is \emph{generative for $\E$} if it is generative for $\E(y)$ for each \ifshort{$y$}\iflong{variable $y\in \Vars(\E)$}.
\end{definition}

\iflong{
Soundness of \Gen\ inductively implies that every assignment in every $\AA^i$ is sound.
Generativity implies that each $\AA_i$ computed by the $i$-th pass of the algorithm is
$i$-witnessed, so that, if a variable has a witness $\J$ of depth $d$, then
$\AA^{i}\neq \emptyset$ for every $i \geq d$.}

We can now define our bottom-up algorithm (Algorithm 1)\iflong{
as follows}.
\RestyleAlgo{ruled}
\begin{algorithm}\label{alg:bottomup}
\footnotesize
\caption{Bottom-up witness generation}
\SetKwProg{Fn}{}{}{end}
\SetKw{kwWhere}{where}
\SetKwFunction{BotUp}{BottomUpGenerate}
\SetKwFunction{Prepare}{Prepare}
\SetKwFunction{Eval}{Eval}
\SetKwFunction{Gen}{Gen}

\Fn{\BotUp{x,E}}{ 
  \Prepare(E)\;
  $\forall y.\ A[y] := \text{nextA}[y] := \emptyset$ \; 
  \While{A[x] == $\emptyset$}{  
    \For{y in vars(E) \kwWhere A[y] == $\emptyset$}{
        $\text{nextA}[y] :=  \Gen{\E(y),A}\;$}
     \lIf{($\forall y.\ $nextA[y] == A[y])} {\Return (unsatisfiable)} 
     \Else{$\forall y.\ A[y] := \text{nextA}[y]$\;}
   } 
   \Return ($A[x]$)\;
} 
\end{algorithm}

\emph{Prepare(E)} rewrites $\E$ and prepares all the extra variables needed for generation,
as explained later.
Then, we initialize $\AA^0$ as the empty assignment $\lambda y.\ \emptyset$.
We repeatedly execute a pass that
sets $\AA^{i}(y) =  \Gen(\E(y),\AA^{i-1})$ for any $y$ such 
that  $\AA^{i-1}(y) =\emptyset$ ---
we call it ``pass $i$''.
We say that a pass~$i$ is \emph{useful} if there exists $y$ such that 
$\AA^{i}(y)\neq\emptyset$ while $\AA^{i-1}(y)=\emptyset$, and we say that pass
$i$ was \emph{useless} otherwise.
Before each pass $i$, if $\semas{x}{\AA^{i-1}}\neq\emptyset$, then the algorithm stops with success. 
After pass $i$, if the pass was useless, the algorithm stops with ``unsatisfiable''.

\iflong{%
We can now prove that this algorithm is correct and complete, as follows.
}

\begin{property}[Correctness and completeness]\leavevmode\label{pro:coandco}
If \Gen\ is sound and is generative for $\E$ after preparation, then Algorithm 1 enjoys the following properties.
\begin{compactenum}
\item If the algorithm terminates with success after step $i$, then 
$A^i(x)$ is not empty and
is a subset of $\semca{x}{\E}$.
\item If the algorithm terminates with ``unsat.'', then $\semca{x}{\E} = \emptyset$.
\item The algorithm terminates after at most $|\Vars(\E)|+1$ passes.
\end{compactenum}
\end{property} 

\ifshort{
\bcolormeta
\begin{proofsketch}
\mrevmeta{
Property (1) is immediate.}{M.3, 2.3, 2.8}
\crevmeta{For (2), we first prove the following property: if the algorithm terminates 
with ``unsatisfiable'' after step $j$, then, for every variable $y$:
$\AA^{j}(y)=\emptyset \ \Implies\ \semca{y}{\E}= \emptyset.$
Assume, towards a contradiction, that there is a non empty set of variables $Y$ such that
$y\in Y\ \Implies\ (\AA^{j}(y)=\emptyset  \ \And\   \semca{y}{\E}\neq\emptyset).$
Let $d$ be the minimum depth of $\bigcup_{y\in Y}\semca{y}{\E}$, and let
$w$ be a variable in $Y$ and such that $d$ is the minimum depth of the
values in $\semca{w}{\E}$.
Minimality of $d$ implies that every variable $z$ with a value in $\semca{z}{\E}$ 
whose depth is less than
$d-1$ has a witness in $\AA^{j}$, hence, since the step $j$ was useless,
every such $z$ has a witness in $\AA^{j-1}$,
hence  $\AA^{j-1}$ is $(d-1)$-witnessed, hence, by generativity, $w$ should have a witness 
generated during step $j$, which contradicts the hypothesis.}

\crevmeta{If the algorithm terminates with ``unsatisfiable'', this means that $\semas{x}{\AA^{j-1}}=\emptyset$,
hence $\semas{x}{\AA^{j}}=\emptyset$ since the step $j$ was useless, hence $\semca{x}{\E}=\emptyset$,
since we proved that 
$\AA^{j}(y)=\emptyset \Implies \semca{y}{\E}=\emptyset.$}

\crevmeta{Property (3) holds since at every useful pass the number of variables such that $\AA^i(y)\neq\emptyset$
diminishes by at least 1.}
\end{proofsketch}
\ecolormeta
}

\iflong{
\begin{proof}
Property (1) is immediate:
by induction and by soundness of $\Gen$, we have that 
${\AA^i}$ is sound for any $i$, that is,
$\semas{S}{\AA^i} \subseteq \semca{S}{\E}$.

For (2), we first prove the following property: if the algorithm terminates 
with ``unsatisfiable'' after step $j$, then, for every variable $y$:
$$\AA^{j}(y)=\emptyset \ \Implies\ \semca{y}{\E}= \emptyset.$$
Assume, towards a contradiction, that there is a non empty set of variables $Y$ such that
$$y\in Y\ \Implies\ (\AA^{j}(y)=\emptyset  \ \And\   \semca{y}{\E}\neq\emptyset).$$
Let $d$ be the minimum depth of $\bigcup_{y\in Y}\semca{y}{\E}$, and let
$w$ be a variable in $Y$ and such that $d$ is the minimum depth of the
values in $\semca{w}{\E}$.
Minimality of $d$ implies that every variable $z$ with a value in $\semca{z}{\E}$ 
whose depth is less than
$d-1$ has a witness in $\AA^{j}$, hence, since the step $j$ was useless,
every such $z$ has a witness in $\AA^{j-1}$,
hence  $\AA^{j-1}$ is $(d-1)$-witnessed, hence, by generativity, $w$ should have a witness 
generated during step $j$, which contradicts the hypothesis.

If the algorithm terminates with ``unsatisfiable'', this means that $\semas{x}{\AA^{j-1}}=\emptyset$,
hence $\semas{x}{\AA^{j}}=\emptyset$ since the step $j$ was useless, hence $\semca{x}{\E}=\emptyset$,
since we proved that 
$$\AA^{j}(y)=\emptyset \Implies \semca{y}{\E}=\emptyset.$$

Property (3) is immediate: at every useful pass the number of variables such that $\AA^i(y)\neq\emptyset$
diminishes by at least 1, hence we can have at most $|\Vars(\E)|$ useful passes plus one useless pass.
\end{proof}
}

\hide{
Not every schema $S$ has a sound function that is generative for it.   
Definition \ref{def:gen} (2) implies that \Gen\ is able to verify whether 
$(\semca{S}{\E}\MeetsD{i+1})\neq \emptyset$
only knowing \emph{one} arbitrary value of  $\semca{x}{\E}$ for any variable such that
$(\semca{x}{\E}\MeetsD{i})\neq \emptyset$.
This requires a certain degree of ``independence'' of the different assertions inside $S$. 
Consider for example an array group
$$S=\TG{\TArr, \CMM{1}{1}{\True},  \CMM{1}{\Inf}{\RRef{x}}, \CMM{1}{\Inf}{\RRef{y}} }$$
Let $\E = \Def{x}{\True}, \Def{y}{\True}$. 
The group is 2-witnessed in $\E$ by any array $[\J]$.
Hence, a function generative for $S$ must be able to generate a witness starting from any $1$-witnessed assignment,
such as $\AA(x)=\Set{12}$ and $\AA(y)=\Set{34}$, but this is impossible, since the assignment does not
provide any value that satisfies both $x$ and $y$, and a sound function cannot attribute to $x$ or $y$ any
value that is not included in the assignment.
If we want to generate a witness for $S$ with our approach, we must first rewrite $S$ in a form that makes
the interaction of the different assertions explicit, for example by adding to $\E$ a fresh variable
$z$ equivalent to $x\And y$ and by rewriting $S$ as
$\TG{\TArr, \CMM{1}{1}{\True},  \CMM{1}{\Inf}{\RRef{z}} }$.

For this reason, our approach first makes every interaction between assertions explicit (\emph{\crcombination}), at the price of introducing new variables, in order to  produce interaction-free schemas where the ability to build a witness for $S$ from witnesses for its free variables only depends on which of these variables have a witness, but not on the specific witness provided for them.
}


\iflong{%
We can finally describe the phases of {\crcombination} and generation for all typed groups.
}

\ifshort{%
We finally describe the phases of {\crcombination} and generation for object groups, corresponding
respectively to the functions {\tt Prepare} and {\tt Gen} of Algorithm 1. 
For reasons of space we leave the description of preparation and generation for arrays in the full paper \cite{attouche2022witness}, where we also detail generation for strings and numbers.
}

\mrevthree{
Preparation is a crucial phase, where we make explicit the interactions between different object or array operators\ifshort{,}
\iflong{found in a same typed group, }
and we create new variables to manage these interactions.
}{3.1}


\subsection{Object group {\crcombination} and generation} \label{sec:objprepgen}

\subsubsection{Constraints and requirements} 

We say that an assertion $S=\Props(r:\RRef{x})$ or $S=\Pro_0^M$ is a
\emph{constraint}. A \emph{constraint}  has the following features: (a) 
$\{\ \} \in\semca{S}{\E}$ and 
(b) $\{k_1:\J_1,\ldots,k_n:\J_n,k_{n+1}:\J_{n+1}\}\in\semca{S}{\E}\Implies
\{k_1:\J_1,\ldots,k_n:\J_n\}\in\semca{S}{\E}$ --- constraints can prevent the addition of 
members, but they never require the presence of a member,
similarly to a \emph{for all fields} quantifier.

We say that an assertion $S=\PReq(r:\RRef{x})$ or $S=\Pro_m^\Inf$ with $m>0$ is a
\emph{requirement}. A \emph{requirement} $S$ has the following features: (a)
$\{\ \} \not\in\semca{S}{\E}$ and 
(b) $\{k_1:\J_1,\ldots,k_n:\J_n\}\in\semca{S}{\E}\Implies
\{k_1:\J_1,\ldots,k_n:\J_n,k_{n+1}:\J_{n+1}\}\in\semca{S}{\E}$ --- requirements can require the addition of 
a member, but they never prevent adding a member, 
similarly to an \emph{exists field} quantifier.

\iflong{As a consequence, a possible algorithm to build an object is: start from the empty object, add one member
at a time until all requirements are satisfied, but, whenever you add a member to satisfy \emph{some} requirements,
verify that it satisfies \emph{all} constraints too.}

\subsubsection{{\Crcombination} and generation}

For a typical object group, where every pattern is trivial and where each type in each
$\PReq$ is just~$\XTrue$%
\ifshort{ (which we use to indicate the only variable whose body is $\True$)}, 
object generation is very easy.
Consider the following group: 
\[
\{\ \TObj, \Props(\qkey{a} : \RRef{x}), 
    \PReq(\qkey{a} : \XTrue) , 
   \PReq(\qkey{c} : \XTrue) \ 
\}
\]

In order to generate a witness, we just need to generate a member $\key{k} : \J$
for each required key, respecting the corresponding $\Props$ constraint if present. Hence,
here we generate a member $\qkey{a} : \J$ where $\J\in\AA^i(\RRef{x})$,
and a member $\qkey{c} : \J'$, where $\J'$ is arbitrary.

\mrevthree{
Unfortunately, in the general case where we have non-trivial patterns and
where the $\PReq$ operator specifies a non-trivial schema for the required member, 
the situation is much more complex, and we must keep into account the following issues:
\begin{compactenum}
\item need to compute the intersections between patterns of different assertions;
\item need to generate new variables when patterns intersect;
\item possibility for one member to satisfy many requirements.
\end{compactenum}
}{3.1}

\bcolorthree
To exemplify the first two problems, consider the following object group:
$
\{\ \TObj, \Props(\key{p} : \RRef{x}), 
   \PReq(\key{r} : \RRef{y}) , 
   \Pro_1^1
\ \}
$.

There are two distinct ways of producing a witness $\{\ \key{k} : \J \ \}$
for the object above: either we generate a $k$ that matches $\key{r} \AndP \CoP{\key{p}}$,
and a witness $\J$ for $\RRef{y}$, or we generate a
$k$ that matches $\key{r} \AndP \key{p}$, and a witness~$\J$ for $\RRef{x} \And\RRef{y}$.
This exemplifies the first two issues above:
\ecolor
\begin{compactenum}
\item patterns: we need to compute which of the combinations $\key{r} \AndP \CoP{\key{p}}$
   and $\key{r} \AndP \key{p}$ have a non-empty language, in order to know which approaches
   are viable w.r.t.\ to pattern combination;
\item new variables: we need a new variable whose body is $\RRef{x} \And\RRef{y}$,
   in order to generate a witness for this conjunctive schema.
\end{compactenum}
Let us say that a member $k: \J$ has shape $r : S$ when $k\in\rlan{r}$ and~$\J$ is a witness for $S$. Then, we can rephrase the example above by saying that an object $\{\ \key{k} : \J\ \}$ satisfies that object group  
iff $\key{k} : \J$ either has shape $(\key{r} \AndP \CoP{\key{p}} : \RRef{y})$
or $(\key{r} \AndP \key{p} : \RRef{x} \And \RRef{y})$.

To exemplify the last problem --- one member possibly satisfying many requirements ---
consider the following object group:
\[
\{\ \TObj, \PReq(\key{r_1} : \RRef{y_1}), 
   \PReq(\key{r_2} : \RRef{y_2}), 
   \Pro_{min}^{Max}
\}
\]

In order to satisfy both requirements, we have two possibilities:
\begin{compactenum}
\item  producing just one member with shape
$\key{r_1} \AndP \key{r_2} : \RRef{y_1} \And \RRef{y_2}$;
\item  producing two members, with shapes
$\key{r_1} : \RRef{y_1} $ and $\key{r_2} : \RRef{y_2} $.
\end{compactenum}

In order to explore all possible ways of generating a witness, we need to consider
both possibilities. But, in order to consider the first possibility, we need a new variable
whose body is \iflong{equivalent to }$\RRef{y_1} \And \RRef{y_2}$.

We solve all these issues by transforming, during the  {\crcombination} phase,
every object into a form where all possible interactions between assertions are made explicit,
and we create a fresh new variable for every conjunction of variables that is 
relevant for witness generation. 
\iflong{The generative witness-generation function that is 
used during bottom-up evaluation, and that will be described in the Section \ref{sec:witobj},
will be applied to this {\crcombined} form. }

\subsubsection{Object group {\crcombination}} \label{sec:objprep}

Consider a generic object group 
\[
\begin{array}{lllll}
\{\ \TObj, & \Props(\key{p_1} : \RRef{x}_1),\ldots, \Props(\key{p_m} : \RRef{x}_m), \\
& \PReq(\key{r_1} : \RRef{y}_1) ,\ldots,\PReq(\key{r_n} : \RRef{y}_n) , \Pro_{min}^{Max}\ \}
\end{array}
\]

\newcommand{\No}[1]{\overline{(\qkey{#1})}}
\newcommand{\Yes}[1]{\qkey{#1}}
\renewcommand{\No}[1]{\CoP{#1}}
\renewcommand{\Yes}[1]{{#1}}

\mrevthree{
We use $CP$ (\emph{constraining part}) to denote the set of $\Props$ assertions
$\Set{ \Props(\key{p_i} : \RRef{x}_i) \M i \in 1..m }$ 
and  $RP$ (\emph{requiring part}) to denote the set of $\PReq$ assertions.
Any witness for this object group is a collection of fields $(k,\J)$ where every field
satisfies every constraint  $\Props(\key{p_i} : \RRef{x}_i)$ such that $k\in\rlan{p_i}$,
and such that every requirement $\PReq(\key{r_j} : \RRef{y}_j)$ is satisfied by
a matching field. Hence, every field is associated to a set $CP'\subseteq CP$
of constraints
and to a set $RP'\subseteq RP$ of requirements.
Only some pairs of sets $(CP',RP')$ make sense, because of pattern compatibility.
Object preparation generates all, and only, the pairs (actually, the \emph{triples},
as we will see) that will be useful to the task of exploring all 
ways of generating a witness.
}{3.1}

Formally, to every pair $(CP',RP')$, where $CP'\subseteq CP$ and $RP'\subseteq RP$,
we associate a \emph{characteristic pattern} $cp(CP',RP')$ that describes all
strings (maybe none) that match every pattern in $(CP',RP')$ and no pattern
in $(CP\setminus CP',RP\setminus RP')$, as follows.

\begin{definition}[Characteristic pattern]
Given an object group \\
$\{ \TObj, CP, RP, \Pro_{min}^{Max} \}$ and
two subsets $CP'\subseteq CP$ and $RP'\subseteq RP$,
the characteristic pattern $cp(CP',RP')$ is defined as follows:
\[\begin{array}{llll}
      \multicolumn{3}{l}{cp(CP',RP')} \\[\NL]
      &=&(\,\BigAndP_{\Props(\key{p} :\_)\in CP'}{p}\,) 
                                 \AndP (\, \BigAndP_{\Props(\key{p} :\_)\in (CP\setminus CP')}{\No{p}}\,) \\[\NL]
                       &  &     \AndP\, (\,\BigAndP_{(\PReq(\key{r} :\_)\in RP'}{r}\,) 
                                 \AndP (\,\BigAndP_{(\PReq(\key{r} :\_)\in (RP\setminus RP')}{\No{r}}\,) 
\end{array}
\] 
\end{definition}

Consider for example the following object group, \mrevmeta{corresponding, modulo variable names, to a fragment of our running example (Figure \ref{fig:rexample}(d))}{M.1, 3.3}: 
$$\begin{array}{llll}
\{ \TObj, \CProps{\qkey{b} : \RRef{x}}, 
 \PReq(\qkey{a} : y1) , \PReq(\qkey{a.*} :\RRef{y2}) 
            \}
\end{array}$$

For space reason, we adopt the following abbreviations for the assertions that
belong to $CP$ and $RP$: 
$$\begin{array}{llllllllll}
pb = \CProps{\qkey{b} : \RRef{x}},\ \ \ \ \ \ 
ra = \PReq(\qkey{a} : \RRef{y1}),\\
ras = \PReq(\qkey{a.*} :\RRef{y2})
\end{array}$$

Here we have $2^3$ pairs $(CP',RP')$ that are elementwise included in
$(CP,RP)$, each pair defining its own characteristic pattern; for each
pattern we indicate an equivalent extended regular expression
(``.+'' stands for any non-empty string)
or $\emptyset$ when the pattern has an empty language:
$$\begin{array}{llll}
cp(\Set{},\Set{}) & = & \No{b} \AndP \No{a}
                                            \AndP \No{a.*}\!\!\! & \equiv  \No{b} \AndP \No{a.*}\\[\NL]
cp(\Set{},\Set{ra}) & = & \No{b} \AndP \Yes{a}
                                            \AndP \No{a.*} & \equiv \emptyset   \\[\NL]
cp(\Set{},\Set{ras}) & = & \No{{b}} \AndP \No{a}
                                            \AndP \Yes{a.*} & \equiv \Yes{a.+}  \\[\NL]
cp(\Set{},\Set{ra,ras}) & = & \No{b} \AndP \Yes{a}
                                            \AndP \Yes{a.*} & \equiv \Yes{a}   \\[\NL]
cp(\Set{pb},\Set{}) & = & \Yes{b} \AndP \No{a}
                                            \AndP \No{a.*} & \equiv \Yes{b}  \\[\NL]
cp(\Set{pb},\Set{ra}) & = & \Yes{b} \AndP \Yes{a}
                                            \AndP \No{a.*} & \equiv \emptyset \\[\NL]
cp(\Set{pb},\Set{ras}) & = & \Yes{{b}} \AndP \No{a}
                                            \AndP \Yes{a.*} &  \equiv \emptyset  \\[\NL]
cp(\Set{pb},\Set{ra,ras}) & = & \Yes{b} \AndP \Yes{a}
                                            \AndP \Yes{a.*} & \equiv \emptyset  \\[\NL]
\end{array}$$

All different pairs $(CP',RP')$ define languages that are mutually disjoint by 
construction,
but many of these are empty, as in this example.
The non-empty languages cover all strings, by construction, hence they always define a partition
of the set of all strings.

Consider now a member $\key{k}: J$ which we may use to build a witness of the 
object group. 
The key $k$ matches 
exactly one non-empty characteristic pattern
$cp(CP',RP')$, hence $J$ must be a witness for all variables $\RRef{x_i}$ such that
$\Props(\key{p_i} : \RRef{x}_i)\in CP'$,
\iflong{ since each relevant constraint must be satisfied,} 
but, as far as the assertions 
$\PReq(\key{r_j} : \RRef{y}_j)\in RP'$
are concerned, there is much more choice. If $J$ is a witness for every such $\RRef{y}_j$,
then this member satisfies all requirements in $RP'$. But it may be the case that some of these
$\RRef{y}_j$'s are mutually exclusive, hence we must choose which ones will be satisfied by $J$.
Or, maybe, none of the $\RRef{y}_j$ is satisfied by~$J$, but we may still use $\key{k}: J$ in
order to satisfy a $\Pro_{m}^{\Inf}$ requirement with $m\neq 0$.
Hence, in order to explore all different ways of generating a member $(\key{k}:\J)$ 
for a witness of the object
group, we must choose a pattern $cp(CP',RP')$, and a subset $RP''$ of $RP'$
that we require $\J$ to satisfy. Hence, we define a \emph{choice} to be a 
triple $(CP',RP',RP'')$, with $RP''\subseteq RP'$.
The $(CP',RP',\_)$ part specifies the pattern that is satisfied by $k$, while the
$(CP',\_,RP'')$ part, with $RP''\subseteq RP'$, specifies the variables that $\J$ must satisfy.
 
We also distinguish \emph{R-choices}, where $RP''$ is not empty, hence they are useful
in order to satisfy some requirements in $RP$, and \emph{non-R-choices}, 
where $RP''$ is empty, hence they can only be used to satisfy a $\Pro_{m}^{\Inf}$
requirement.
The only choices that may describe a member are those where \iflong{the set of strings}
$\rlan{cp(CP',RP')}$ is not empty;
we call them \emph{non-cp-empty choices}.

\begin{definition}[Choice, R-Choice,  cp-empty choice]
Given an object group 
$\{\ \TObj, CP, 
RP, \Pro_{m}^{M}\ \ \}$
with constraining part $CP = \Set{ \Props(\key{p_i} : \RRef{x}_i) \M i \in 1..m }$
and 
$RP= \Set{ \PReq(\key{r_j} : \RRef{y}_j) \M j \in 1..n }$,
a choice is a triple $(CP',RP',RP'')$ such that $CP'\subseteq CP$,
$RP''\subseteq RP'\subseteq RP$.
The \emph{characteristic pattern} $cp(CP',RP',RP'')$ of the choice is defined by its
first two components, as follows:
$$cp(CP',RP',RP'')=cp(CP',RP')$$
The \emph{schema} of the choice $s(CP',RP',RP'')$ is defined by the first and the third component,
as follows:
\[
s(CP',RP',RP'') \ =\  \BigAnd_{\Props(\key{p} :\RRef{x})\in CP'}{\RRef{x}} 
                             \And  \BigAnd_{\PReq(\key{r} :\RRef{y})\in RP''}{\RRef{y}} 
\]
A choice is  \emph{cp-empty} if $\rlan{cp(CP',RP',RP'')}$ is empty, is \emph{non-cp-empty}
otherwise.

A choice is an \emph{R-choice} if $RP''\neq\Set{ }$, is a \emph{non-R-choice}
otherwise.
\end{definition}

In the object group of our previous example we have 4 non-cp-empty pairs,
$(\Set{},\Set{})$, $(\Set{pb},\Set{})$, $(\Set{},\Set{ras})$,
$(\Set{},\Set{ra,ras})$, which correspond to the
following 8 non-cp-empty choices -- for each, we indicate the corresponding schema.
{
$$\begin{array}{lllllll}
s(\Set{},\Set{},\Set{}) &\!\!=\!\!& \RRef{\XTrue} & \text{non-R-choice}\\
s(\Set{pb},\Set{},\Set{})&\!\!=\!\!& \RRef{x} & \text{non-R-choice}\\
s(\Set{},\Set{ras},\Set{})&\!\!=\!\!& \RRef{\XTrue} & \text{non-R-choice}\\
s(\Set{},\Set{ras},\Set{ras})&\!\!=\!\!& \RRef{y2} & \text{R-choice}\\
s(\Set{},\Set{ra,ras},\Set{})&\!\!=\!\!& \RRef{\XTrue} & \text{non-R-choice}\\
s(\Set{},\Set{ra,ras},\Set{ra})&\!\!=\!\!& \RRef{y1} & \text{R-choice}\\
s(\Set{},\Set{ra,ras},\Set{ras})&\!\!=\!\!& \RRef{y2} & \text{R-choice}\\
s(\Set{},\Set{ra,ras},\Set{ra,ras})&\!\!=\!\!& 
\RRef{y1}\And \RRef{y2}  
           & \text{R-choice}\\
\end{array}$$
}

The schema of a choice is always a conjunction of variables,
say $x_1\And\ldots\And x_n$. 
During bottom-up generation, we need to know which non-cp-empty choices have a witness
in the current assignment $\AA^i$,
hence we need to associate every non-cp-empty  choice with just one variable, not with a conjunction.
\mrevthree{
Hence, we need to create a new variable $y$ for each conjunction
$x_1\And\ldots\And x_n$ that we have never seen
before, then we execute GDNF normalization over $x_1\And\ldots\And x_n$, transforming
it into a guarded disjunction of typed groups $S$, then we add $\Def{y}{S}$ to the current
environment and we apply \emph{preparation} again to this new variable;
we call this process \emph{and-completion}.
In the example above, this may be the case for $y1\And y2$, unless 
$y1\And y2$ is Boolean-equivalent to some variable that already exists.
}{3.1}


Preparation can be regarded as a sophisticated form of and-elimi\-nation. Here,
\emph{and-comple\-tion} plays the same role that 
{not-comple\-tion} plays for not-elimination: it creates the new variables that we need in order
to push conjunction through the object group operators.
But, crucially, and-completion is \emph{lazy}: we do not 
pre-compute every possible conjunction, but only those that are really needed by some specific
non-cp-empty choice. This laziness is 
crucial for the practical feasibility of the algorithm: when different constraints, or requirements,
are associated to disjoint patterns, we have very few non-cp-empty choices, and in most cases they do not
need any fresh variable, as in the example.
\mrevthree{
Despite laziness, this prepare-generate-normalize-prepare
loop can still generate a huge number of variables.
We keep their number under control using the {\ROBDDT} data structure that we
introduced in Section \ref{sec:robdd}, which allows us to create a new variable only when
none of the existing variables is boolean-equivalent to its body; this crucial
optimization also ensures that this phase can never generate an infinite loop.
}{3.1}

Hence, object {\crcombination} proceeds as follows:
\begin{compactenum}
\item determine the set of non-cp-empty pairs $(CP',RP')$,
   that is the pairs such that $cp(CP',RP')$ is not empty;
\item for each non-cp-empty pair  $(CP',RP')$ compute the corresponding choices
   $(CP',RP',RP'')$
   and, if the variable intersection $s(CP',RP',RP'')$ has no equivalent
   variable in the environment, add a new variable
   $\Def{x}{s(CP',RP',RP'')}$ to the environment,
    apply GDNF reduction to $s(CP',RP',RP'')$, apply {\crcombination} to the GDNF-reduced 
    conjunction.
\end{compactenum}

\iflong{When we describe object generation, we will show how the set of all prepared
choices
can be used in order to enumerate all possible ways of generating a witness for an object
group.}

Step (1) has, in the worst case, an exponential cost, but in practice it is much cheaper:
in the common case where every pattern matches a single string, a set of $n$ properties and 
requirements generates at most $n+1$ non-empty pairs (one for each string plus one for the
complement of the string set), $n$ R-choices,
and $n+1$ non-R-choices.
Since before preparation we have at most $O(N)$ distinct variables (where $N$ is the input size),
step (2) may generate at most $O(2^{N})$ new variables, each of which has a body which
can be {\crcombined} in time $O(2^{\PN})$.
Hence, the global cost of this phase is still $O(2^{\PN})$. 
\ifshort{In our implementation we use an algorithm, sketched in the full paper \cite{attouche2022witness}, that runs in polynomial time 
in the common case when the number of non-cp-empty pairs is actually polynomial
in the size of the object group, and our experiments
show that this cost is, for most real-world schemas, tolerable.
}

\iflong{Our experiments
show that this cost is, for most real-world schemas, tolerable.}

\begin{property}
Object {\crcombination} can be performed in  $O(2^{\PN})$ time.
\end{property}

%

\hide{
\begin{remark}
In our implementation, the generation of all non-cp-empty pairs is not performed by brute 
force enumeration, but using an algorithm, sketched in the full paper \cite{attouche2022witness}, that runs in polynomial time 
in the common case when the number of non-cp-empty pairs is actually polynomial
in the size of the object group.
\end{remark}
}

\iflong{
\begin{remark}
In our implementation, the generation of all non-cp-empty pairs is not performed by brute 
force enumeration, but using an algorithm based on the following schema:
it matches every pair of patterns $r1$ and $r2$
coming from either $CP$ and $RP$ and, in case the two are neither equal nor disjoint,
splits them into three patterns $r1\AndP\CoP{r2}$, $r1\AndP r2$ and $\CoP{r1}\AndP r2$.
This algorithm has a cost that is quadratic in the number of non-empty pairs that are generated.
Hence, it is $O(2^n)$ in the worst case but is just quadratic in the typical case, the
one where the number of non-empty pairs is linear in the size of the object group.
\end{remark}
}
\newcommand{\CC}{\ensuremath{\mathbb{C}}}
\newcommand{\CCP}{\ensuremath{\mathbb{C'}}}

\subsubsection{Witness generation from a {\crcombined} object group}\label{sec:witobj}

After the object group has been {\crcombined} once for all, at each pass of bottom-up
witness generation we use the following sound and generative algorithm,
listed  as Algorithm \ref{alg:owg},
to compute a witness for the prepared object group starting from the current assignment $\AA^i$.

\mrevthree{
In a nutshell, we (1) pick a list of choices that contains enough R-choices to satisfy all requirements
--- each choice will correspond to one field in the generated object, and vice versa; 
(2) we verify that the list is \emph{pattern-viable}, i.e., that it does not require two fields with the same name; 
(3) to satisfy any unfulfilled $\Pro_{m}^{\Inf}$ requirement, we add some
non-R-choices, still keeping the choice list \emph{pattern-viable}, as defined above.
In order to keep the search space in $O(2^{\PN})$, we limit ourselves to the subset of the \emph{disjoint}
solutions, and we prove that it is big enough to have a complete algorithm.
}{3.1}

\bcolorthree
In greater detail, consider a generic object group 
with the 
form $\{\ \TObj, CP, RP, \Pro_{m}^{M}\ \}$
and assume that the corresponding non-cp-empty choices have been prepared.

To generate an object, we first choose a list of choices that satisfies all of $RP$.
To reduce the search space, we first observe that
a single object can be described by many different choice lists.
\ecolor
For example,
assume that `1' belongs to both $\semca{x}{\E}$ and $\semca{y}{\E}$ and assume that:
$$
\begin{array}{lllll}
rx = \PReq(\qkey{a|b}: \RRef{x}) &&  \\
ry = \PReq(\qkey{a|b}: \RRef{y})  \\
RP = \{\ rx, ry\ \}   \\
\end{array}
$$
then $\{\ \qkey{a}:1, \qkey{b}:1\ \}$ is described 
by each the following four choice lists (and by others),
where every choice  could be used to generate/describe each of the two members:
$$\begin{array}{llllllll}
CL_1 =  \ [\ (\Set{},\Set{rx,ry},\Set{rx}), &(\Set{},\Set{rx,ry},\Set{ry}) &]\\
CL_2 =  \ [\ (\Set{},\Set{rx,ry},\Set{rx,ry}),&(\Set{},\Set{rx,ry},\Set{}) &]\\
CL_3 =  \ [\ (\Set{},\Set{rx,ry},\Set{rx,ry}),&(\Set{},\Set{rx,ry},\Set{rx,ry}) &]\\
CL_4 =  \ [\ (\Set{},\Set{rx,ry},\Set{rx,ry}),&(\Set{},\Set{rx,ry},\Set{rx}) &]
\end{array}$$
This example shows that we do not need to explore any possible choice list,
but just \emph{enough} choice lists to generate \emph{all} witnesses.
To this aim, we focus on \emph{disjoint solutions}, defined as follows,
whose completeness will be proved in Theorem \ref{the:sogen}.

\begin{definition}[Disjoint solution, Minimal disjoint solution]
Fixed a set \iflong{of requirements}
$RP$, a size limit $M$, and a set of choices $\CC$,
a multiset 
$\CCP=\Set{(C_l,R'_l,R''_l) \M l \in L}$ with elements in $\CC$
is a \emph{solution} \iflong{(for the fixed $RP$ and $M$) }
iff:
$$
\bigcup_{l \in L}R''_l = RP 
 \ \mbox{\ and\ \ }  |\CCP| \leq M
$$

The solution is  \emph{disjoint} if:
$
 i\neq j \ \Implies\ R''_i \cap R''_j = \emptyset.
$

The solution is \emph{minimal} if every choice in $\CCP$ is an R-choice.
\end{definition}

In the previous example, only $CL_1$ and $CL_2$ are disjoint, and only $CL_1$ is disjoint and minimal.

\iflong{
Every object described by a solution for an object group is a witness for the
that group.

\begin{definition}[describes-in-$\AA$]\label{def:solution}
A choice $C=(CP',RP',RP'')$ for a prepared object group
\emph{describes} \emph{in an assignment $\AA$} a field $k:\J$, iff 
$k\in\rlan{cp(C)}$ and $\J\in\AA(var(C))$.
A choice list $\CC$ \emph{describes}  in $\AA$ an object $\J$ if there is a bijection mapping each
field $k:\J'$ in $\J$ to a choice $C$ in $\CC$ such that $C$ describes $k:\J'$.
\end{definition}

\begin{property}\label{pro:solution}
For any prepared object group $$S=\TG{\TObj, CP, RP, \Pro_{m}^{M}}$$
with the corresponding environment $E$ and choices $\CC$,
if $\CC'$ is a choice list over $\CC$ with $m \leq |\CC'| \leq M$ that is a solution for
$RP$, if $\AA$ is sound for $\E$, and if $\J$ is described in  $\AA$  by $\CC$,
then $\J\in\semca{S}{\E}$.
\end{property}
}

%
%

\ifshort{
\renewcommand{\Pop}{\emph{Witnessed}}
\renewcommand{\Open}{\emph{NonWitnessed}}
}

{Object generation depends on the current assignment $\AA^{i}$.
We say that a variable $x$ is \Pop\ (in $\AA^{i}$) when $\AA^{i}(x)\neq \emptyset$,
and is \Open\ otherwise. We say that a choice is \Pop, or \Open,
when its schema variable is \Pop, or is \Open.}
In order to generate a witness, we first generate a \emph{disjoint minimal solution}
for $RP$ with bound $M$, only using R-choices that are \Pop.
Then, in order to deal with the constraint that all names in an object are distinct, we check
that the solution is \emph{pattern-viable}. Informally, pattern-viability ensures that,
if we have $n$ choices in the solution with the same characteristic pattern $cp$, then the 
language of $cp$ has at least $n$ different strings, which can be used to build $n$ different members corresponding
to those $n$ choices.
We will exemplify the issue after the definition.

\begin{definition}[Pattern-viable]
A set of choices $\CC$ is 
pattern-viable iff for every pair $(CP',RP')$, the number of choices in $\CC$ with shape
$(CP',RP',\_)$ is smaller than the number of words in $\rlan{cp(CP',RP')}$:
$$
\begin{array}{llll}
\forall CP', RP'.\ \ \\
|\, \Set{(CP',RP',RP'') \M (CP',RP',RP'')\in\CC}\,| \ \leq\  |\,\rlan{cp(CP',RP')}\,|
\end{array}
$$
\end{definition}

For example, the following choice list $\CC$ is not viable since it describes an object with two members
that  share the same characteristic pattern $\qkey{a}$ that only contains one string:
$$\begin{array}{llllll}
 \multicolumn{4}{l}{rx = \PReq(\qkey{a}:\RRef{x}), \ ry = \PReq(\qkey{a}:\RRef{y})} \\[\NL]
\CC = \ [ &(\Set{},\Set{rx, ry},\Set{rx}),  
      &(\Set{},\Set{rx, ry},\Set{ry}) & ]
\end{array}
$$
But it would be viable if the pattern $\qkey{a}$ were substituted by $\qkey{a|b}$.
 
Finally, for each viable disjoint solution, we check whether it also satisfies the 
$\Pro_{m}^{\Inf}$ requirement (line 6 of Algorithm \ref{alg:owg}). 
If it does not, we try and extend the solution by adding some \Pop\ non-R-choices (line 7).
Observe that the disjoint solution contains each R-choice 
$(CP',RP',RP'')$ at most once, 
because of disjointness; however,
we can add the same non-R-choice as many times as we need in order to reach $m$ members.
A non-R-choice $C$ can only be added if the result remains
viable;
hence, a minimal disjoint solution $\CC$ may have a viable extension $\CCP$ of length $m$, 
obtained by adding a multiset of non-R-choices
(lines 6-13), or it may not have such a viable extension, and then we need to start from a different
minimal solution.
If no viable disjoint solution admits a viable extension of length at least $m$, then the algorithm returns \shortlong{``no witness''}{``Open''}
(according to the current assignment).
Otherwise, we use the extended solution $\CCP$ to build a witness: 
for each choice $C\in\CCP$, we generate a name $k$ satisfying $cp(C)$,
we pick a value $\J$ from $\AA^{i}(var(C))$, and the set of members $k:\J$ that we obtain is a witness
for the object group. When $n$ different choices inside $\CCP$ have the same characteristic pattern, we generate $n$  different names,
which is always possible since the solution is viable\iflong{ --- this is the $n$-enumeration 
problem for EEREs 
that we introduced in Section \ref{sec:regexp}}.

\RestyleAlgo{ruled}
\begin{algorithm}
\footnotesize
\caption{Object witness generation\label{alg:owg}
}
\SetKwProg{Fn}{}{}{end}
\SetKw{kwWhere}{where}
\SetKwFunction{obi}{Gen}
\SetKwFunction{minSol}{minDisjointSols}

\Fn{\obi{RPart, WitRChoices,WitNonRChoices, min, Max,}}{ 
\For{Solution in \minSol(WitRChoices,RPart,Max)}{
    \If{(viable(Solution))}{
       missing := min --- size(Solution)\; 
       nonViableChoices := $\emptyset$\;
        \While{(missing > 0 and nonViableChoices !=WitNonRChoices)}{ 
            choose NRC from (WitNonRChoices-nonViableChoices)\;
             \If{(viable([NRC]++Solution))}{
                Solution := [NRC]++Solution\;
                missing := missing-1\;
             }
             \lElse{nonViableChoices := [NRC]++nonViableChoices}
        }
         \If{(missing == 0)}{
            \Return (``\shortlong{Witnessed}{Populated}'', WitnessFrom(Solution))\;
         }
    }
}
\Return (``\shortlong{NonWitnessed}{Open}'')\; 
} 
\end{algorithm}

\begin{theox}[Soundness and generativity]\label{the:sogen}
Algorithm \Gen\ is sound and generative.
\end{theox}

\ifshort{
\crevmeta{
\begin{proofsketch}
\mrevmeta{
Our algorithm is sound by construction.
For generativity, assume that the group
$S = \{\ \TObj, CP, RP, \Pro_{min}^{Max}\ \}
$
has a witness of depth $d+1$ in $\semca{S}{\E}$.}{M.3, 2.3, 2.8}
Assume that $\AA$ is $d$-witnessed for $\E$.
We want to prove that \Gen, applied to 
$S$ and $\AA$, will generate at least one witness. 
Let 
$\J=\{ \key{a_1}:\J_1, \ldots, \key{a_l}:\J_l \}$
be a witness for $S$ in $\E$ with depth $d+1$.
We can extract from  the fields of $\J$
a multiset of choices $\CC=(C'_i,R'_i,R''_i)$ with $i\in\SetTo{l}$, that describes these fields,
as detailed in \cite{attouche2022witness}.
We prove that all these choices are \Pop\ in $\AA$, by exploiting the fact that 
 $\J$ has depth $d+1$, hence every $\J_i$ that appears in the witness has depth $d$
 at most, and $\AA$ is $d$-witnessed.
Finally, we prove that our algorithm would generate at least one
solution for the group.
To this aim, we first remove every non-R-choice from $\CC$, hence obtaining a minimal
disjoint solution, and we then add non-R-choices back if required by a $\Pro_m^{\Inf}$ requirement, and we observe that this is
a viable solution, hence our algorithm
would find it.
\end{proofsketch}}
}

\begin{appendixproof}
Our algorithm is sound by construction.
For generativity, assume that the object group
$$S = \{\ \TObj, CP, RP, \Pro_{min}^{Max}\ \}
$$
has a witness of depth $d+1$ in $\semca{S}{\E}$.
Assume that $\AA$ is $d$-witnessed for $\E$.
We want to prove that \Gen, applied to 
$S$ and $\AA$, will generate at least one witness. 
Let 
$$\J=\{ \key{a_1}:\J_1, \ldots, \key{a_l}:\J_l \}$$
be a witness for $S$ in $\E$ with depth $d+1$.
We can now extract from  $$\{ \key{a_1}:\J_1, \ldots, \key{a_l}:\J_l \}$$
a set of choices $(C'_i,R'_i,R''_i)$ with $i\in\SetTo{l}$, as follows.
$C'_i$ and $R'_i$ are defined by the only pair $(C'_i,R'_i)$ whose language includes
$a_i$. 
In order to define $R''_i$, we observe that, since $\J$ satisfies $RP$,
then, we can associate to each $S$ in $RP$ one member $i$ such that 
$\key{a_i}:\J_i$ satisfies $S$ --- if many such members exist, we just choose one. 
The inverse of this relation associates to each member $i$ a subset $R''_i$ of $R'_i$.
The collection of choices $\CC=\setst{(C'_i,R'_i,R''_i)}{i\in\SetTo{l}}$ that we have defined is actually a multiset,
since a non-R-choice may appear more than once, and is a disjoint
solution since, by
construction, $\bigcup_{i\in\SetTo{l}}R''_i = RP$, $l\leq Max$,
and $1\leq i < j \leq l\ \Implies \ R''_i\cap R''_j = \emptyset$, since every requirement is
mapped to exactly one member.
We now prove that all these choices are \Pop\ in $\AA$.
To this aim, consider a choice $C=(C'_i,R'_i,R''_i)$ in $\CC$ and
the field $\key{a_i}:\J_i$ that we used to define it.
By construction, the schema $s(C)$ is the conjunction of the variables of all constraints $C'_i$
that must be satisfied by and $\J$ that is associated to $a_i$ in any witness of
$S$, plus the variables a set of requirements $R''_i$ whose variables are satisfied by
$\J_i$, hence $\J_i\in\semca{s(C)}{\E}$, hence, by definition of $var(C)$,
$\J_i\in\semca{var(C)}{\E}$.
Since $\J$ has depth $d+1$, then
$\Dep(\J_i)\leq d$, hence $\AA(var(C))\neq\emptyset$ since $\AA$ is $d$-witnessed,
hence every choice in $\CC$ is \Pop\ in $\AA$.

Now we prove that our algorithm would generate at least one subsequence of $\CC$ that is a solution,
unless it stops since it is able to generate a different solution; in both cases, our algorithm
generates a solution for the group.

To prove this, we remove every non-R-choice from $\CC$, and so we get a collection $\CCP$ that is a minimal
disjoint solution. 
If $min>|\CCP|$, then we choose $min-|\CCP|$ non-R-choices out of $\CC$ and add them
to $\CCP$. Being a subset of $\CC$, the result is viable and, by construction, is an extension of a minimal disjoint solution $\CCP$ with a multiset of non-R-choices. Our algorithm scans every such extension of every minimal disjoint solution, hence,
if it is not stopped because it finds a different solution, it finds this one, and it generates a
corresponding witness.
\end{appendixproof}

\begin{property}[Complexity]
Given a schema of size $N$, each run of the \Gen\ 
algorithm has a complexity in $O(2^{\PN})$.
\end{property}

\iflong{
\begin{proof}
Let $N$ we the size of the original schema.
Let us first focus on a single, arbitrary, group.
For any object group, $RP$ has at most $N$ elements, and any choice has a size that is $O(N)$.
Let $M$ be an upper bound for the number of non-empty choices for an arbitrary object group.
Since every minimal disjoint solution contains at most $|RP|\leq N$ choices, we can generate all minimal disjoint solutions
by scanning the list of all $N$-tuples of choices, which can be done in time
$O(M^N)$. We then need to scan the list of all non-R-choices for at most $min$ times, which adds another
$O(M^N)$ factor, since $min\leq N$ by the linear constants assumption, hence we arrive at $O(M^{2N})$
solutions.
For every solution that contains $i$ choices, we need to solve at most $i$ times the 
$i$-enumeration problem, with $i\leq N$, in order to verify viability and to generate the witness when a witness exists. 
The pattern expression $cp(C)$ of each choice $C$ of the solution has a size that is in $O(\PN)$, hence running $i$ times the $i$-enumeration problem has a cost that is $O(2^{\PN})$, hence we can examine 
$O(M^{2N})$ solutions in time $O(M^{2N}\cdot\PN\cdot 2^{\PN})$. 
Since $M$ is in $O(2^{\PN})$, each pass of object generation is in $O(2^{\PN})$ for each {\crcombined} object
group.
Since we have less then $O(2^{\PN})$ groups, each pass  of object generation is in $O(2^{\PN})$.
\end{proof}
}

\renewcommand{\RR}{R_{RP}}

\ifshort{%
\hide{\subsection{Array group {\crcombination} and generation}\label{subsec:arraygrprep}

In theory, arrays and objects are almost identical, since they are both finite mappings from labels to
values. However, arrays introduce additional issues:
\begin{compactenum}
\item Arrays have a domain downward closure constraint, that specifies that, when a value is associated to a label $n+1$, then a value is associated to $n$ as well, for every $n \geq 1$;
\item The $\CMM{i}{j}{\RRef{x}}$ operator specifies an upper bound, and requires counting,
while $\PReq(\key{a}:\RRef{x})$ only specifies the existence of at least one member matching $a$ with 
schema $\RRef{x}$, with no upper bound and no counting ability.
\end{compactenum}


For more details on array preparation and generation, see \cite{attouche2022witness}.}
}

\iflong
{%
\subsection{Array group {\crcombination} and generation}\label{subsec:arraygrprep}

\subsubsection{Constraints and requirements} 

As with objects, we say that an assertion $S=\ContAfter{i}{\RRef{x}}$  or $S=\CMM{i}{\Inf}{\RRef{x}}$ with $i>0$ is a
\emph{requirement}, since it is not satisfied by $[\,]$ and, if $\J^+$ extends $\J$, then 
$\J\in\semca{S}{\E}\Implies\J^+\in\semca{S}{\E}$.

We say that an assertion $S=\PreIte{l}{\RRef{x}}$, $S=\PostIte{i}{\RRef{x}}$,  
or $S=\CMM{0}{j}{\RRef{x}}$ is a
\emph{constraint}, since it is satisfied by $[\,]$ and  if $\J^+$ extends $\J$, then 
$\J^+\in\semca{S}{\E}\Implies\J\in\semca{S}{\E}$.

An assertion $S=\CMM{i}{j}{\RRef{x}}$ with $i\neq 0$ and $j\neq \Inf$ combines a requirement and a constraint.

\subsubsection{Array group {\crcombination}}

\renewcommand{\IJ}[2]{[{#1},{#2}]}
\newcommand{\Inter}{\key{In}}
\newcommand{\HL}{\key{HL}}
\newcommand{\hel}{head-length}

An array group is a set of assertions with the following shape:
$$\{\ \TArr, IP, AP, KP\ \}
$$
Here, $IP$ is a set of \emph{item} constraints $\PreIte{l}{\RRef{x}}$ and $\PostIte{i}{\RRef{x}}$,
$AP$ is a set of \emph{contains-after} requirements with shape $\ContAfter{l}{\RRef{x}}$,
$KP$ is a set of \emph{counting} assertions $\CMM{i}{j}{\RRef{x}}$, where every assertions combines
a requirement $\CMM{i}{\Inf}{\RRef{x}}$ and a constraint $\CMM{0}{j}{\RRef{x}}$.%
\footnote{For the sake of simplicity, in our formal treatment we do not distinguish
$\CMM{i}{j}{\XTrue}$ from the other counting assertions, where $\XTrue$ here indicates the variable whose body is $\True$,
although in the implementation we actually exploit its special properties for efficiency reasons.}

In theory, arrays and objects are almost identical, since they are both finite mappings from labels to
values, but arrays have some extra issues:
\begin{compactenum}
\item Arrays have a domain downward closure constraint, that specifies that, when a value is associated to a label $n+1$, then a value is associated to $n$ as well, for every $n \geq 1$; objects do not have anything similar.
\item The $\CMM{i}{j}{\RRef{x}}$ operator specifies an upper bound, and requires counting,
while $\PReq(\key{a}:\RRef{x})$ only specifies the existence of at least one member matching $a$ with 
schema $\RRef{x}$, with no upper bound and no counting ability.
\end{compactenum}

Consider for example the following array group.
$$ \{\ \TArr, \PreIte{2}{\RRef{x}}, \ContAfter{0}{\RRef{y}}, \CMM{1}{1}{\RRef{z}}, 
\CMM{2}{2}{\XTrue} \}
$$

It describes an array of exactly two elements. The one at position 2 must satisfy $\RRef{x}$.
At least one of the two elements must satisfy $\RRef{y}$.
One, but only one, of the two elements must satisfy $\RRef{z}$.

Let us say that an array has shape $[S_1,\ldots,S_k]$ if it contains exactly $k$ items
$[\J_1,\ldots,\J_k]$, and if each item $\J_i$ satisfies $S_i$.
Then, the group above is satisfied by arrays with one of the following four shapes:
$$\begin{array}{lllllllllllll}
\ [\ \RRef{y} \And \RRef{z},\ &\RRef{x} \And \NR{z}\ ], &
\ [\ \RRef{y} \And \NR{z} ,\ &\RRef{x} \And \RRef{z}\ ],  \\
\ [\ \RRef{z} ,\ &\RRef{x} \And \RRef{y} \And \NR{z}\ ], &
\ [\ \NR{z}, \ &\RRef{x} \And \RRef{y} \And \RRef{z}\ ] 
\end{array}$$

We recognize the two problems that we have seen with objects: interaction between constraints and requirements,
resulting in conjunctions of $\RRef{x}$ with other variables in position 2,
and the possibility of one element to satisfy two requirements, resulting in $\RRef{y} \And \RRef{z}$
conjunctions, but we have the extra problem of the upper bound, that results in the presence of 
the dual variable $\NR{z}$ in some positions.

Hence, our algorithm to {\crcombine} arrays and to generate the corresponding witnesses is somehow different
from that of objects, although similar in spirit. It obviously differs in the presence of dual variables like $\NR{z}$, motivated
by upper bounds, but also differs in the strategy that we use to explore the space of witnesses. Instead of starting
the exploration from the requirements, hence from the ``first choices'', here we are guided by the 
domain closure constraint, hence we start the exploration from the first position of the array.

We need to define some terminology.
We first define a notion of head-length for an array group $S$ (Definition \ref{def:head}):
intuitively, when the head-length of $S$ is $h$, then, for any witness $\J$ of $S$, if the elements of $J$ from position $h+1$ onwards --- which constitute the \emph{tail} of $\J$ --- are permuted, then 
$\J$ is still a witness; the elements in positions $1$ to $h$ constitute the \emph{head}, and their
position may matter.
For example, an array group $\{\TArr, \PreIte{3}{x}\}$ has head-length 3.
The head-length $n$ may be $0$, and actually this is the most common {\hel} that we
encounter in practice.
The interval of an assertion $\Inter(S)$ is the interval of positions of the array 
that the assertion describes,
which may belong to the head of the group, to the tail, or may cross both. 

\begin{definition}[$\IJ{i}{j}$, $\HL(S)$, $\Inter(S)$]\label{def:head}
$\IJ{i}{j}$, with $i\in\Nat, j\in\Nat^{\Inf}$, denotes the interval between $i$ and $j$, which is infinite when $j=\Inf$, and is empty
when $i>j$.
The head-length $\HL(S)$ and the interval $\Inter(S)$ of an array {\ITO} $S$,
and of an array group, 
are  defined as follows:
$$\begin{array}{llllllll}
\IJ{i}{j} & = & \setst{l}{ l\in\Nat,\  i \leq  l \leq j} \\[\NL]
\HL(\PreIte{l}{S}) & = & l & \\[\NL]
\HL(\PostIte{i}{S}) & = & i & \\[\NL]
\HL(\ContAfter{l}{S}) & = & l \\[\NL]
\HL(\CMM{i}{j}{S}) & = & 0 \\[\NL]
\HL((\{ \TArr, IP, AP, KP \}) & = & \max_{S\in IP\cup AP}(\HL(S))  \\[\NL]
\Inter(\PreIte{l}{S}) & = & \IJ{l}{l} & \\[\NL]
\Inter(\PostIte{i}{S}) & = & \IJ{i+1}{\Inf} & \\[\NL]
\Inter(\ContAfter{l}{S}) & = & \IJ{l+1}{\Inf} \\[\NL]
\Inter(\CMM{i}{j}{S}) & = & \IJ{1}{\Inf} \\[\NL]
\end{array}
$$
\end{definition}

\begin{property}[Irrelevance of tail position]
If $S$ is an array typed group, $J=[\J_1,\ldots,\J_n]\in\semca{S}{\E}$, 
for all $i, j$ with
 $HL(S) < i \leq j \leq n$, if $\J'$ is obtained from $\J$
by exchanging $\J_i$ with $\J_j$, then $J'\in\semca{S}{\E}$.
\end{property}

In order to define a choice we need a last definition: for a set of assertions 
$\mathcal{S}$, we define its restriction to $\IJ{i}{j}$, denoted by
$\mathcal{S} \Meets \IJ{i}{j}$, as the subset of $\mathcal{S}$ containing the
assertions whose interval intersects $\IJ{i}{j}$.

\begin{definition}[$\mathcal{S} \Meets \IJ{i}{j}$]
$$\mathcal{S} \Meets \IJ{i}{j} = \setst{S}{S\in \mathcal{S}, \ (\IJ{i}{j} \cap \Inter(S)) \neq \emptyset}$$
\end{definition}

Now, we define a choice for an array group $IP$, $AP$, $KP$ with $h=\HL(IP\cup AP)$,
as a quintuple $(\IJ{i}{j},IP',AP',KP^{+},KP^{-})$ 
where:
\begin{compactenum}
\item either $i=j \leq h$ or $i=h+1$ and $j=\Inf$, hence a choice describes either
     a single element $\IJ{i}{i}$ in the head of the array group, or an element in the tail
     interval $\IJ{h+1}{\Inf}$;
\item $IP'$ is equal to $IP \Meets \IJ{i}{j}$;
\item $AP'$ is a subset of $AP \Meets \IJ{i}{j}$;
\item $KP^{+}$ is a subset of $KP$;
\item $KP^{-}$ is a subset of $KP \setminus KP^{+}$.
\end{compactenum}

Hence, for each interval $\IJ{i}{j}$, the element $IP'$ is fixed, but we may still have many choices for 
$AP'$, $KP^{+}$ and $KP^{-}$.
Intuitively, a choice $(\IJ{i}{j},IP',AP',KP^{+},KP^{-})$ describes an element in a position that
belongs to $\IJ{i}{j}$,
that satisfies all the constraints in $IP\Meets \IJ{i}{j}$, that satisfies 
the assertions in $AP'$ and in $KP^{+}$, and does not satisfy any assertion in $KP^{-}$.
With respect to object choices, here the label is not represented by a pair of sets of assertions 
$(CP',RP')$, but just
by an interval $\IJ{i}{j}$, while the schema is a bit more complex since it has three positive components $IP'$,
$AP'$ and
$KP^{+}$, playing the roles of $CP'$ and $RP''$,
but also a negative component $KP^{-}$. 
Observe that, while $IP'$ and $AP'$ are restricted to the assertions that apply to $\IJ{i}{j}$,
we do not have this restriction for $KP$, since every counting assertion analyzes all positions of the 
array.
Hence, the schema of a choice is defined as follows.

\begin{definition}[$s(\IJ{i}{j},IP',AP',KP^{+},KP^{-})$]\label{def:schemaac}
$$
\begin{array}{lll}
\multicolumn{2}{l}{
s(\IJ{i}{j},IP',AP',KP^{+},KP^{-})
} \\[\NL]
\qquad = &(\BigAnd_{(\PreIte{l}{\RRef{x}})\in IP'} {\RRef{x}}) 
\And
(\BigAnd_{(\PostIte{i}{\RRef{x}})\in IP'} {\RRef{x}}) \\[\NL]
&\And\ 
(\BigAnd_{(\ContAfter{l}{ \RRef{x}})\in AP'} {\RRef{x}})  \\[\NL]
&\And\ 
(\BigAnd_{(\CMM{i}{j}{\RRef{x}})\in KP^{+}} {\RRef{x}})
\And
(\BigAnd_{(\CMM{i}{j}{\RRef{x}})\in KP^{-}} {\NR{x}} )
\end{array}
$$
\end{definition}

As with object groups, a generative exploration of the space of all possible solutions does not
require the generation of all possible choices, and different strategies are possible.
In our implementation, we limit ourselves to the choices where 
$KP^{-} = KP \setminus KP^{+}$, which we 
call here the co-maximal choices. 
We prove later that this strategy ensures the generativity property that we need.
More optimized strategies would be possible, but we believe that they are not 
worth the effort, since in practice the array types that we have to deal with are usually quite simple.

Hence, array {\crcombination} consists of the following steps.

\begin{compactenum}
\item compute $h = \HL(IP,AP)$;
\item for each interval $\IJ{i}{i}$ corresponding to an $i\in \IJ{1}{h}$, and for each subset $AP'$ of $AP$
and $KP'$ of $KP$
produce the corresponding co-maximal choice: 
$$
(\IJ{i}{i},IP \Meets \IJ{i}{i},AP',KP',KP \setminus KP')
$$
and check whether the variable intersection that corresponds to the schema of that choice
is equivalent to some existing variable, and, if not, create a new variable that will become the schema
of that choice, and apply {\crcombination} to the body of this new variable, as in the case of 
object {\crcombination};
\item do the same for the interval $\IJ{h+1}{\Inf}$, and for each subset $AP'$ of $AP$
and $KP'$ of $KP$.

\end{compactenum}

As happens with object {\crcombination}, also array {\crcombination} has an exponential cost
that is quite low in practice, since in the vast majority of cases the head-length of array groups is
zero or one, and the set $AP\cup KP$ is either empty or a singleton. For this reason, we did not put any
special effort into the optimization of this phase.

\begin{property}
Array {\crcombination} can be performed in time $O(2^N)$, where $N$ is the size of the input schema.
\end{property}

%

\subsubsection{Witness generation from a {\crcombined} array group}\label{sec:witarrj}

Array {\crcombination} applied to an array group
$\{ \ \TArr, IP, AP, KP\ \}$ with head-length $h$ produces a set of co-maximal 
choices, each characterized by an interval $\IJ{i}{j}$ with shape $\IJ{i}{i}$ when $i\leq h$, or $\IJ{h+1}{\Inf}$ otherwise, 
and by two subsets $AP', KP'$ of $AP, KP$.
We indicate with $C(i,AP',KP')$ the co-maximal choice that is characterized by these three parameters, and with
$s(i,AP',KP')$ and $s(i,AP',KP')$  its schema and the associated variable, as follows:
$$\begin{array}{llll}
\multicolumn{3}{l}{
C(i,AP',KP')\ \ \ \mbox{with\ }   1 \leq i \leq h
}\\
& \!= & \!(\IJ{i}{i},IP\Meets\IJ{i}{i},AP',KP',KP \setminus KP')  \ \ \\[\NL]
\multicolumn{3}{l}{
C(h+1,AP',KP')
}\\
& \!= & \!(\IJ{h+1}{\Inf},IP\Meets\IJ{h+1}{\Inf},AP',KP',KP \setminus KP')   \\[\NL]
\multicolumn{3}{l}{
s(i,AP',KP') \ = \ s(C(i,AP',KP'))
}\\[\NL]
\multicolumn{3}{l}{
var(i,AP',KP') \ = \ var(C(i,AP',KP'))
}
\end{array}$$

A choice $C(i,AP',KP')$ is a \emph{head choice} when $i\leq h$, and is a \emph{tail choice} when $i=h+1$. 
At any pass of the generation algorithm, a choice is \Pop\ or \Open, depending on its schema variable.

Given a list of choices $\CC$ and a set of contains-after and counting assertions $\Set{AP,KP}$
(where $\Set{AP',KP'}$ abbreviates $AP'\cup KP'$),
we define the \emph{incidence} of $\CC$ over $\Set{AP,KP}$ as a function that maps each $S\in \Set{AP,KP}$
to the number of elements of $\CC$ that are guaranteed to satisfy $S$, as follows:
$$\begin{array}{llll}
\mbox{if }S\notin (AP'\cup KP') :\ & I_{C(i,AP',KP')}(S) = 0 \\[\NL]
\mbox{if }S\in (AP'\cup KP') :\ & I_{C(i,AP',KP')}(S) = 1 \\[\NL]
\multicolumn{2}{l}{
I_{[C_1,\ldots,C_n]}(S) =  \sum_{i \in \SetTo{n}} I_{C_i}(S) 
}
\end{array}
$$

We say that a list of choices $\CC$ is a solution for $\Set{AP,KP}$
when the incidence of the list satisfies all requirements and does not violate any constraint, as follows.

\begin{definition}[Well formed list, Solution] 
A list of choices for an array group is well-formed for head-length $h$ iff 
\begin{compactdesc}
\item[(1)] every choice in the 
list has either an interval $\IJ{i}{i}$ with $i\leq h$ or the interval  $\IJ{h+1}{\Inf}$; 

\item[(2)] if two consecutive choices in the list have intervals $\IJ{i}{\_}$ and $\IJ{j}{\_}$, then either
$j=i+1$ or $j=i=h+1$.
\end{compactdesc}
\end{definition}

For example, $[\,([3,3],\ldots), [4,4],\ldots), ([5,\Inf],\ldots), ([5,\Inf],\ldots)\,]$,
$[\,([5,\Inf],\ldots)\,]$, 
and $[\ ]$ are well formed
for head-length 4.

\begin{definition}[Solution] 
Fixed an array group
$\TG{ \ \TArr, IP, AP, KP\ }$ with head-length $h$,   a choice list  $\CC$
is a solution for the array group iff all the following hold:
\begin{compactdesc}

\item[(1)] it is well formed for $h$; 

\item[(2)] either $\CC$ is empty or the first choice has
interval $\IJ{1}{\_}$; 

\item[(3)] for every assertion $\CMM{m}{M}{x}\in KP$ we have $I_{\CC}(S) \leq M$;

\item[(4)] for every assertion $\CMM{m}{M}{x}\in KP$ we have $I_{\CC}(S) \geq m$;

\item[(5)] for every requirement $S\in AP$ we have $I_{\CC}(S) > 0$.

\end{compactdesc}

\end{definition}

Observe that an incidence $I_{\CC}(S)=n$ guarantees that an array described by $\CC$ 
has exactly $n$ elements that satisfy $S$ if $S\in KP$, and \emph{at least} $n$ elements that satisfy $S$ if $S\in AP$.
This happens by design, and is sufficient to guarantee the essential property that every array described by a solution is a witness
for the corresponding group.

\begin{definition}[describes-in-$\AA$]\label{def:arrdescr}
A choice $C=(\IJ{i}{j},\ldots)$ for a prepared array group
\emph{describes} \emph{in an assignment $\AA$} an element $J_l$  of an array $[\J_1,\ldots,\J_n]$, iff 
$l\in \IJ{i}{j}$ and $\J\in\AA(var(C))$.
A choice list $[C_1,\ldots,C_n]$ \emph{describes} in $\AA$  an array $\J= [\J_1,\ldots,\J_n]$ if every 
$C_l$ describes in $\AA$ the element $J_l$.
\end{definition}

\begin{property}\label{pro:arrsolution}
For any prepared array group $$S=\TG{\TArr, IP, AP, KP}$$
with the corresponding environment $E$ and choices $\CC$, if 
$\AA$ is sound for $\E$,
if the choice list $\CC'$ over $\CC$ is a solution for $S$, and if $\J$ is described in  
$\AA$ by $\CC$,
then $\J\in\semca{S}{\E}$.
\end{property}

\begin{proof}
Consider a prepared group $S=\TG{\TArr, IP, AP, KP}$ and the corresponding choices
$\CC$ and environment $\E$.
Let $\AA$ be sound for $\E$ and assume that $\CC'=[C_1,\ldots,C_n]$ describes $\J= [\J_1,\ldots,\J_n]$.

By definition of $I_{\CC'}(S)$, for any $S=\ContAfter{i}{x}\in AP$, if $I_{\CC'}(S)=k$, then there are exactly 
$k$ choices $C$ in $\CC'$ such that $C=C(l,AP',KP')$, and $S\in AP'$.
By definition of $s(C)$ and $var(C)$, 
for all of these 
choices we have that $s(C)$ is a conjunction of $x$ with other variables,
hence $\semca{var(C)}{\E}\subseteq \semca{x}{\E}$. 
For all of these choices, the corresponding $J_l$ belongs to $\AA(var(C))$, since $\CC'$ describes in $\AA$ $J$.
Since $\AA$ is sound for $\E$, we conclude that, for these choices, we have that $J_l\in  \semca{x}{\E}$.
Hence, if $I_{\CC'}(S)>0$ with $S=\ContAfter{i}{x}$, we have at least one element of $J$ which satisfies $x$.
We must now prove that the position of that elements is greater than $i$.
By definition of choice, every choice that includes $\ContAfter{i}{x}$ has an interval that intersects $\IJ{i+1}{\Inf}$.
Since the head-length of the object group is at least $i$, every choice whose interval intersects $\IJ{i+1}{\Inf}$
is either a head-choice with interval $\IJ{j}{j}$ and $j > i$ or a tail choice with interval $\IJ{h+1}{\Inf}$ and
$h \geq i$. In both cases, every position described by that choice is strictly greater than $i$.

By definition of $I_{\CC'}(S)$, for any $S=\CMM{m}{M}{x}\in KP$, if $I_{\CC'}(S)=k$, this implies that there are exactly 
$k$ choices $C$ in $\CC'$ such that $C=C(l,AP',KP')$, and $S\in KP'$, and, as in the previous case, for all of these 
choices we have that $\semca{var(C)}{\E}\subseteq \semca{x}{\E}$. 
Since we only consider co-maximal choices, for all the other $n-k$ choices we have that $S\in KP^{-}$,
hence for the other choices we have that $s(C)$ is a conjunction of $\NR{x}$ with other variables,
hence
$\semca{var(C)}{\E}\cap \semca{x}{\E}=\emptyset$.
Since $\AA$ is sound for $\E$, and $\CC'$ describes in $\AA$ $J$, we conclude that \emph{exactly} $k$
elements of $\J$ belong to $\semca{x}{\E}$. Since $m \leq I_{\CC'}(S) \leq M$, we conclude that $\J$ satisfies
$\CMM{m}{M}{x}$.

Consider any $S=\PreIte{l}{x}\in IP$ and any choice $C$ whose interval intersects $\IJ{l}{l}$.
By construction, $\semca{var(C)}{\E}\subseteq \semca{x}{\E}$, hence, by soundness of $\AA$, the 
element described by $C$ satisfies $S$.

Consider any $S=\PostIte{i}{x}\in IP$ and any choice $C$ whose interval intersects $\IJ{i+1}{\Inf}$.
By construction, $\semca{var(C)}{\E}\subseteq \semca{x}{\E}$, hence, by soundness of $\AA$, the 
element described by $C$ satisfies $S$.

Hence, every assertion in $\TG{\TArr, IP, AP, KP}$  is satisfied by $\J$.
\end{proof}

We finally need a notion of \emph{useful choices}, which is similar in spirit to the 
\emph{R-choices} that we defined for the object case, and which will be crucial to ensure the termination
of the algorithm: a choice $C$ is \emph{useful} for a set $\Set{AP,KP}$ iff
some assertion in $\Set{AP,KP}$ is affected by $C$.

\begin{definition}[useful choice]
A choice $C(i,AP', KP')$ 
is \emph{useful} for a set of assertions $\Set{AP'',KP''}$ 
iff $$(\Set{AP',KP'} \cap \Set{AP'',KP''}) \neq \emptyset.$$

\end{definition}

We can now describe our algorithm.

Our algorithm \emph{cList(hLen, aList, fLen, fInc, pChoices)} recursively solves the following generalized problem:
assume you have a list of assertions \emph{aList} and you already have a choice list \emph{firstC} of length
\emph{fLen}, whose incidence 
on  \emph{aList} is \emph{fInc};
find the rest of the list --- that is, find a well formed choice list $\CC$ such that the concatenation of \emph{firstC} with
$\CC$ is a solution for \emph{aList}.

If \emph{aList} is already satisfied by \emph{fInc}, then \emph{cList} returns the empty choice list (line 2).
Otherwise, for each $C$ in \emph{pChoices} that can describe position \emph{fLen+1}, 
we try to solve
the subproblem \emph{cList(hLen, aList, fLen+1, fInc', pChoices')}, where
\emph{fInc'} is the incidence updated after \emph{C}, and, when the position \emph{fLen+1} belongs to the tail, \emph{pChoice'}
only contains the elements of \emph{pChoice} that are still useful to solve \emph{aList} after a \emph{CLFirst}
with incidence \emph{fInc}
--- this reduction of \emph{pChoice} will be commented later on.
If such a C exists, and $\CC$ is a solution for  \emph{cList(hLen, aList, fLen+1, fInc', pChoices')}, then we return 
$[C]\mbox{++}\CC$ (lines 9-11).
If \emph{pChoices} contains no choice $C$ such that  \emph{cList(hLen, aList, fLen+1, fInc', pChoices')} has a solution,
then we return ``unsatisfiable''.

Hence, at each pass, we start from an assignment $\AA$, 
we collect all choices that are \Pop\ wrt $\AA$ in a list \emph{pChoices}, and we invoke the algorithm
 \emph{cList(\hel,0,\Set{AP,KP},allZeroes,pChoices)}.
Termination is ensured by the fact that, once we arrive to the tail, we only keep the useful choices,
hence every choice that is selected either 
(a) increments to one  the incidence over an assertion $\ContAfter{i}{\RRef{x}}$ whose incidence was zero, or
(b) increments by one the incidence over an assertion $\CMM{m}{M}{\RRef{x}}$ 
whose incidence was still below $m$,
hence the algorithm stops after not more than $\key{MaxSteps}$ steps:
$$ \key{MaxSteps}\ = \ h + |AP| + \Sigma_{\CMM{m}{M}{\RRef{x}}\in KP}\ m
$$
Here, $h$ is the {\hel}, $|AP|$ is an upper bound for the (a) steps, 
and $\Sigma_{\ldots}\ m$ is an upper bound for the steps of type (b).
If the algorithm returns a solution, we use it to generate a witness by substituting
each choice with a witness from the corresponding \Pop\ schema.

\RestyleAlgo{ruled}
\begin{algorithm}
\caption{Pseudo-code for array solution generation}
\label{algo:array}
\footnotesize

\SetKw{kwIn}{in}
\SetKw{kwIsNot}{is not}
\SetKw{kwExists}{exists}
\SetKw{kwWhere}{where}
\SetKw{kwAnd}{and}

\SetKwFunction{FcList}{cList}
\SetKwFunction{FemptySat}{emptyListSatisfies}
\SetKwFunction{FtailUsefulChoices}{tailUsefulChoices}
\SetKwFunction{FpChoices}{pChoices}
\SetKwFunction{FaddChoiceToInc}{updateIncAfterChoice}
\SetKwFunction{FmaxViolated}{maxViolated}
\SetKwFunction{FinInterval}{inInterval}

\SetKwProg{Fn}{}{}{end}

\Fn{\FcList{hLen, aList, fLen, fInc, pChoices}} {
    \lIf{\FemptySat{aList, fInc}}{return [ ]}
    \If{fLen >= hLen} {
        pChoices $\gets$ \FtailUsefulChoices{pChoices, aList, fInc, hLen}\;
    }
    \For{C in \FpChoices \kwWhere \FinInterval{{hLen}+1,C}} {
        newFInc $\gets$ \FaddChoiceToInc{aList, fInc, C}\;
        \lIf{\FmaxViolated{aList, newFInc}}{continue}
        \Else{
            restSolution = \FcList{hLen, aList, fLen+1, newFInc, pChoices}\;
            \lIf{restSolution \kwIsNot null}{\Return ([C] ++ restSolution)}
            \lElse{continue}
        }
    }
    \Return null\;
}

\Fn{\FtailUsefulChoices{choices, aList, fInc, hLen}}{
    result = [ ]\;
    \For{C in choices \kwWhere start(C)=={hLen+1}}{
        \If{\kwExists ContAftInC \kwIn APPrimeOf(C) \\
            \kwWhere fInc(ContAftInC)=0}{add C to result\;}
        \If{\kwExists MinMaxInC \kwIn KPPrimeOf(C) \\
            \kwWhere min(MinMaxInC) $>$ fInc(MinMax)}{add C to result\;}
    }
    \Return results\;
}
\end{algorithm}

\nop{
\begin{figure}[h]
\begin{tabbing}
aa\=aa\=aa\=aa\=aa\=aaa\=aaaa\=aaaa\=aaaa\=aaaa\=aaaa\=aaaa\=aaaa\=\kill
cList(hLen,from,r,pChoices) \+\\
  if (emptyListSatisfies(r)) return []; \\
  if (from == hLen+1) \+\\
           pChoices = tailUsefulChoices(pChoices,r,hLen); \-\\
   for C in pChoices(from) \{ \+ \\
     r = residuate(r,C);      \\
     if (False in r)  continue;  //go to next C \\
    else \{ \+ \\
       if (from <= hLen)   from = from+1; \\
       restSolution =  cList(hLen, from, r, pChoices) \\
       if (restSolution is not null) \+ \\
            return  ([C] ++ restSolution); \- \\
       else continue; // go to next C \- \\
     \} \- \\
   \} \\
return null; \- // every pChoice has been tried \\
\\

aa\=aa\=aa\=aaa\=aa\=aaa\=aaaa\=aaaa\=aaaa\=aaaa\=aaaa\=aaaa\=aaaa\=\kill

// r is a list [(or,res),...,(or,res)] of original/residuated assertions \\
tailUsefulChoices(choices,r,hLen) \+\\
  result = []; \\
  for C in choices \+ \\
  if (index(C) == hLen+1)  \+ \\
     if  (\>exists ContAft in APPrimeOf(C), (or,res) in r  \+ \\
             where or==ContAft \ )\+ \\
              add C to result; \-\- \\\
     if  (\>exists MinMax in KPPrimeOf(C), (or,res) in r  \+ \\
             where or==MinMax  and  min(res) > 0) \+ \\
              add C to result; \-\-\-\- \\\
  return(results) \- 
\end{tabbing}
\caption{Pseudo-code for array solution generation}
\label{fig:arraygeneration}
\end{figure}
}

This algorithm is sound and generative.

\begin{property}[Soundness and generativity]
The algorithm \emph{cList} is sound and generative.
\end{property}

\begin{proof}
Assume that an array group $S = \{\ \TArr, IP, AP,KP\ \}$ with {\hel} $h$ has a witness with depth
$d+1$, and consider such a witness $\J=[ \J_1, \ldots, \J_o ]$. 
For every $i$ of $\SetTo{o}$, we define 
$$\begin{array}{llll}
A(i) & = & \setst{S}{S=\ContAfter{l}{\RRef{x}},\ S\in AP,\ i > l,\ \J_i \in \semca{x}{\E}} \\[\NL]
K(i) & = & \setst{S}{S=\CMM{m}{M}{\RRef{x}},\ S\in KP,\ \J_i \in \semca{x}{\E}}
\end{array}
$$
Now we build a choice list $\CC$ that is derived from $\J$, as follows.

We define an index $i$, initialized to 1, and a \emph{cumulative incidence} function $in$, that maps every assertion to 0.
If the function $in$ satisfies already both $AP$ and $KP$, then $\CC=[]$.
Otherwise, we consider the choice $C(i,A(i),K(i))$.
We say that a choice is useful for $\Set{AP,KP}$ ``after a list of choices described by $in$'', if the choice contains
some requirements from $\Set{AP,KP}$ that are not yet satisfied by an array that is described by a list of choices 
whose incidence is $in$, which can be verified as described by function \emph{tailUsefulChoices} in the algorithm.
If $i\geq h+1$ and  $C(i,A(i),K(i))$ is not a useful choice for $\Set{AP,KP}$ after a list of choices described by $in$, then
we can remove $\J_i$ from
the array and what we obtain is still a witness: all requirements are already satisfied by the part of the array with
incidence $in$, and the fact that all elements after $\J_i$ decrease their position by 1 is irrelevant since we
are in the tail.
If we are not in the tail, or we are in the tail and $C(i,A(i),K(i))$ is a useful choice, 
then we leave $J_i$ in the array witness, we put $C(min(h+1,i),A(i),K(i))$ in $\CC$,
we update the cumulative incidence function $in$, we increment $i$, and we continue.

At the end of this process, we have a new witness $\J'$, obtained by deleting some elements from
the tail of $\J$, and a  choice list $\CC$ that describes $\J'$. 
By the definition of $A(i)$ and $K(i)$, every $\J'_i$ in $\J'$ belongs to $\semca{x}{\E}$ for all variables $x$ that appear positively in 
$s(C(i,AP',KP'))$ and 
does not belong to $\semca{x}{\E}$ for all variables $x$ that appear complemented in $s(C(i,AP',KP'))$,
hence it belongs to  $\semca{\NR{x}}{\E}$ for all these variables. Since $\J'$ is a witness for $S$, then
$\J'_i$ also satisfies all applicable constraints in $IP$,
hence it belongs to $\semca{s(C(i,AP',KP'))}{\E}$, hence it belongs to $\semca{var(C(i,AP',KP'))}{\E}$.
If we assume that $\J$ has depth $d+1$, then every $\J'_i$ has a depth smaller than $d$, hence, for any $\AA$
that is $d$-witnessed, every variable $var(C(i,AP',KP'))$ in the list $\CC$ is populated.
Hence, the choice list $\CC$ is a list of choices that are populated, such that every tail choice $C$ is useful
after the choices that have been chosen before $C$, hence the choice list $\CC$
would be generated by our algorithm unless a different solution were generated, hence our algorithm is 
generative.

Soundness of the algorithm is immediate.
\end{proof}

\begin{property}[Complexity]
For any array group whose size is in $O(N)$, each pass of algorithm \emph{cList}
has a complexity in $O(2^{\PN})$.
\end{property}

\begin{proof}
The \emph{cList} algorithm explores at most $O(2^{\PN})$ choices at each step, and the total number of steps is at most: 
$$ \key{MaxSteps}\ = \ h + |AP| + \Sigma_{\CMM{m}{M}{\RRef{x}}\in KP}\ m
$$
By the linear constants assumption, $\key{MaxSteps}$ is in $O(N^2)$, hence the algorithm explores at most
$O({(2^{\PN})}^{N^2})=O(2^{\PN*N^2})$ tuples, and the operation that must be executed for each tuple can be performed
in time $O(2^{\PN})$.
\end{proof}
}

%
%

\iflong{
\subsection{Witness Generation from Base Typed Groups}


Witness generation for groups with a base type needs no {\crcombination}, is fully accomplished during the first
pass, and is not difficult, as detailed below.

\subsubsection{Witness generation from a canonical schema of type $\Null$ or $\Bool$}

A canonical group of type \Null\ has the shape $\{ \TNull \}$ and generates
$\xnull$.

A group of type $\Bool$ that does not contain any $\IBT(b)$ operator will generate either $\xtrue$ or $\xfalse$.
If it contains a collection of $\IBT(\xtrue)$ operators, it will only generate $\xtrue$, and similarly 
for $\IBT(\xfalse)$. If it contains both, it is not satisfiable, and will return ``unsatisfiable''.

\subsubsection{Witness generation from a canonical schema of type \Str}

A canonical group of type $\Str$ is just the conjunction of zero or more extended regular expressions,
which we reduce to one by computing their intersection, whose size is linear in the size of the input regular expressions.
At this point, we generate a witness for this regular expression, which can be done in time $O(2^\PN)$
(Section \ref{sec:regexp}).

%

\subsubsection{Witness generation from a canonical schema of type \Num} 

For a canonical schema of type $\Num$, we can first merge all intervals into one and
all $\Mof(m)$ operators into one, let us call it $\Mof(M)$;
if the group contains an assertion $\NotMof(n)$ with $M=n\times i$ for any integer $i$, then the group 
returns ``unsatisfiable''.
Otherwise, we obtain one interval (if none is present, we add $\Bet_{-\Inf}^{\Inf}$),
a set of zero or many $\NotMof(n)$ constraints, and one optional $\Mof(m)$
with $m\neq n\times i$ for every $i\in\IntSet$ and for every $\NotMof(n)$.
At this point, to simplify some operations, we substitute any
negative argument $n$ of $\Mof(n)$ or $\NotMof(n)$ with its opposite. 
The interval may be open at both extremes,  closed at both, or mixed. We distinguish
five cases. In the last three cases we describe an open interval $\XBet_{min}^{Max}$, but the reasoning when one extreme, or both, are included, is essentially the same.
\begin{compactenum}
\item Empty interval: we return ``unsatisfiable''.
\item One-point interval $\Bet_{m}^{m}$: if $m$ satisfies all  $\NotMof$ and  $\Mof$
     assertions we return $m$, otherwise we return ``unsatisfiable''.
\item No $\Mof(m)$, i.e.,\ many-points interval $\XBet_{min}^{Max}$ 
     with no $\Mof(m)$ constraint and $l$ $\NotMof(n_j)$ constraints: choose
      $\epsilon$ such that
      $$0 < \epsilon \leq \frac{min((Max-min),n_1,\ldots,n_l)}{l+2}$$
     If we consider the set $B = \setst{min+i\times \epsilon}{i\in \SetTo{(l+1)}}$, then 
     every value in $B$ satisfies $\XBet_{min}^{Max}$, and no assertion $\NotMof(n_j)$
     can be violated by two distinct values in $B$, hence at least one value in $B$ is a witness.
\item Finite $Max-min$ and $\Mof$, i.e.,\ interval $\XBet_{min}^{Max}$ with a $\Mof(m)$ constraint 
   and finite values for both $min$ and $Max$: we list 
   all multiples of $m$ starting from $min$ (excluded in case of $\XBet$) until we find one 
   that satisfies all $\NotMof$ assertions, or until we go over $Max$ (excluded or included depending on the interval), in which case we return ``unsatisfiable''.
\item Infinite $Max-min$ and $\Mof$, i.e.,\ interval $\XBet_{min}^{Max}$ where either $min$ or
   $Max$ is not finite, and with a $\Mof(m)$ constraint: 
   bring all arguments of $\Mof(m)$ and $\NotMof(n)$ into a fractional form where they share the same
   denominator $d$, as in $\Mof(M/d)$, $\NotMof(n_j/d)$. Select any prime number $p$ that is strictly 
   bigger than every $n_j$ and such that either $p\times M/d$ or its
   opposite belongs to the interval. Such a number clearly exists, and
   it is easy to prove that primality of $p$ and the fact that $(M/d)\neq (n_j/d)\times i$ for every $i\in\IntSet$ 
   and for every $\NotMof(n_j/d)$, imply that $p\times M/d$ satisfies all $\NotMof$ assertions.
 \end{compactenum}

\begin{property}
If a group of type $\Num$ has a witness, then the above algorithm will return a witness.
\end{property}

\begin{proof}
The only difficult case is case (5).
  Assume, towards a contradiction, that exists $n_j/d$ and an integer $i$ with 
  $p\times M/d = i\times (n_j/d) $, that is $p\times M = i\times n_j$.
Since $p$ is prime and is bigger than $n_j$, then $p$ is prime wrt $n_j$.
Since $p$ is a factor of $i\times n_j$ and is prime wrt $n_j$,
then $p$ is a factor of $i$,
hence there exists an integer $i'$ such that $i=i'\times p$,
that is,
$p\times M = i'\times p \times n_j$,
that is, $M = i' \times n_j$, which is impossible.
\end{proof}

\begin{property}
If a group of type $\Num$ has a witness, one can be generated in time $O(2^{\PN})$,
where $N$ is the size of the input schema.
If a group of type $\Num$ has a witness, this fact can be proved in time $O(2^{\PN})$.
\end{property}

\begin{proof}
Here we do not need the linear constant assumption over any of the involved parameters.
Let $N$ be the size of the input schema.
In case (3), we try $O(N)$ witnesses.
In case (4), we must try at most $(Max-min)/m$ possible witnesses, which is in $O(2^N)$, because of
binary notation.
In case (5), we exploit the fact that the numbers are decimal, hence the number of digits of $d$ is 
linear in $N$, hence the size of every $n_j$ is still limited by $N$. 
We must also assure that either $p\times M/d$ or its opposite belongs to the interval.
For example, when $min$ is finite, $p$ must satisfy 
$(p\times M)/d > min$ hence  $p > min\times d / M$, and again
all the constants have a bitmap representation linear in $N$. 
A prime number greater than $k$ can be generated in time
that is polynomial in $k$, hence we are still in $O(2^{\PN})$.
\end{proof}
}

\ifshort{
\bcolormeta
\subsection{Completeness and correctness}

\mrevmeta{The algorithm described in this paper  is correct and complete.}{M.3, 2.3, 2.8}

\begin{theorem}[Correctness and completeness]
The witness generation algorithm is correct and complete: it returns a witness if, and only if,
the schema admits a witness, and otherwise it indicates that the schema is not satisfiable
\end{theorem}

\begin{proofsketch}
This follows from Property \ref{pro:preliminarycc},
Property \ref{pro:coandco}, and Theorem \ref{the:sogen} (more details in \cite{attouche2022witness}).


\hide{By Property \ref{pro:coandco},
the bottom-up algorithm described in Section \ref{sec:bottomup} terminates with success if, and only if,
the schema admits a witness, and otherwise indicates that the schema is not satisfiable,
provided that the algorithm used for preparation and generation is sound and generative,
which holds by Theorem \ref{the:sogen}.}

\end{proofsketch}

\ecolormeta
}

\section{Experimental analysis}\label{sec:exp}


\subsection{Implementation and experimental setup}\label{subsec:impl}
We implemented our \ifshort{algorithm}
\iflong{witness generation algorithm for {\JS} \VerSix}
in Java~11, using the Brics library~\cite{brics_automaton} to generate witnesses from patterns, and the~\emph{jdd} library~\cite{jdd} for \mbox{ROBDDs}. 
Our experiments were run on a Precision~7550 laptop with a 12-core Intel~i7 2.70GHz CPU, 32 GB of RAM%
\iflong{
, running Ubuntu~21.10}. 
\iflong
{We set the JVM heap size to 10 GB.  }
Witnesses were validated by an external tool~\cite{jsvalidator1} (version~1.0.65), and additionally by hand, since the external tool reported false negatives in a few cases. 
\iflong{
}
Each schema is processed by a single thread, and all reported times are measured for a single run. 
Our reproduction package~\cite{repro_package} can be used to confirm our results.

\subsection{Tools for comparative experiments}


Due to the lack of equivalent tools, we compare our tool against a Data Generator and a Containment Checker.

\smallskip
\emph{Data generator (DG).}
We use an open source test data generator for JSON Schema~\cite{jsongen} (version~0.4.6).
This Java implementation pursues a try-and-fail approach: an example is first generated, then validated against the schema, and potentially refined if validation fails, exploiting the error message. This tool lends itself to a comparison although it is not able to detect schema emptiness: given an unsatisfiable schema, it will always return an (invalid) instance.

\smallskip
\emph{Containment checker (CC).}
We compare our tool against the containment checker by Habib et al.~\cite{jsonsubschema} (version~0.0.5), 
described in~\cite{DBLP:conf/issta/HabibSHP21}, and
designed to check interoperability of data transformation operators~\cite{baudart_et_al_2020-automl_kdd}.
\iflong{Typically, these schemas do not contain negation or recursion.}  
\iflong{The \enquote{CC~tool} only supports \VerFour\ schemas, a limitation that we consider when comparing against this tool.}



\subsection{Schema collections}

\ifshort{We conduct experiments with different schema collections.  
Table~\ref{tab:results} states the respective numbers of satisfiable/unsatisfiable schemas.}

\iflong{We conduct experiments with \crevmeta{six} different schema collections: four real-world and two synthetic.
Table~\ref{tab:results} states their origin, the number of schemas, broken down into satisfiable and unsatisfiable sche\-mas, and the average and 
maximal size of schemas.}



\smallskip
\emph{Real-world schemas.}
\ifshort{For the GitHub collection,
we retrieved virtually all files from GitHub that present the features of 
{\JS}, based on a BigQuery search on the GitHub public dataset.
We downloaded the 80K identified schemas (shared online~\cite{schema_corpus}).
We performed duplicate-elimination and data cleaning (see \cite{attouche2022witness}), arriving at
6,427 schemas,
40 of which are unsatisfiable (according to our tool and
 confirmed by direct inspection). We renamed all occurrences of {\xuniqIt}, 
treating it as a user-defined keyword.
}
\iflong{The largest of the real-world schemas collection was obtained from GitHub.
We retrieved virtually every accessible, open source-li\-censed {\json} file from GitHub that presents the features of 
a schema, based on a BigQuery search on the GitHub public dataset;
Google hosts a snapshot of all open source-licensed on GitHub,
refreshed on a regular basis. 
The schemas were downloaded in July 2020, and are shared online~\cite{schema_corpus}.
We obtained over 80K schemas. 
As can be expected, 
we encountered a multitude of problems in processing these non-curated, raw files: files with syntactic errors, files which do not comply to any JSON Schema draft, and files with references that we are unable to resolve.
Notably, there is a large share of duplicate schemas, with small variations in syntax and semantics. We rigorously removed such files, eliminating schemas with the same occurrences of keywords, condensing the corpus down to 7,046. 
We further excluded 619 schemas which are either ill-formed, or use specialized types (audio, video) that we do not support, or use an old draft with a different syntax, or employ patterns not supported by the third-party automaton library, or use unguarded recursion.
More precisely, we excluded 17 ill-formed schemas, 105 schemas with specialized types, 355 schemas expressed in Draft-3,  61 schemas whose patterns contain negative lookahead, 68 schemas using unreachable references or references to fragments  expressed inside specific keywords (like \emph{properties}) that our tool does not yet correctly handle, and 13 schemas using unguarded recursion.
Of the remaining 6,427 schemas, 40 are well-formed but unsatisfiable. We identified these schemas using our tool,
and then performed a manual verification on all of them.

}

\mrevmeta{
The three remaining real-world collections correspond
to specifications of standards for deploying applications (Kubernetes~\cite{kuber}), 
ruling interactions within a specific system (Snowplow~\cite{snow}),
and describing data produced by content management systems (Washington Post~\cite{wp}).
To increase the number of processable schemas, we inlined references to external schemas.
An earlier version of these collections where already used in~\cite{DBLP:conf/issta/HabibSHP21} to check inclusion.
Almost all schemas are satisfiable, except 5 from Kubernetes.
}{M.2, 1.2, 1.4, 3.2}

\paragraph{Hand-written schemas}
\mrevmeta{Real-world schemas reflect real usage and can be 
quite big, but they focus on the commonest operators and combination of 
operators. Hence, for stress-testing, we inserted in our reproduction packages
233 handwritten schemas that are small but have been crafted to
exemplify complex interactions between the language operators. }{}
\mrevmeta{To illustrate such an interaction, consider the following schema.
}{M.2, 1.2, 3.2, 3.4}

\smallskip
\crevmeta{\begin{tabular}{l}
\{ $\Def{r}{\Props(a : \RRef{x})\And \Props(\key{a.*}:\RRef{y}) \And  \Req(\key{a})},$\\
\ \ $\Def{x}{\TStr \And  \Pat(a(c|e))}$,\\
\ \ $\Def{y}{\TStr \And  \Pat(a(b|c))}$ \}
\end{tabular}
}{}

\smallskip

\crevmeta{Here we have an interaction between two $\Props$ and a $\Req$ with overlapping patterns,
and associated with two different variables $x$ and $y$ whose schema present non-trivial overlapping.
}{}

\crevmeta{Array operators also present interactions, as in the following example.
}{}

\smallskip
\crevmeta{
\begin{tabular}{l}
\{ $\Def{r}{\PreIte{1}{x} \And  \CMM{1}{1}{y}},$\\
\ \ $\Def{x}{\TArr \And \CMM{2}{\infty}{t}}$, \\
\ \ $\Def{y}{\CMM{1}{\infty}{\Type(\Num) \And \Mof{(3)}}}$\}
\end{tabular}
}{}

\smallskip
\crevmeta{
This example describes an array with schema $r$ that contains another array with schema $x\And y$, this one having at least two elements
 (because of $\CMM{2}{\infty}{t}$), one of which is multiple of 3.
}{}

\crevmeta{The collection has been built by systematically considering  operators for objects, arrays, strings and numbers,
following software-engineering principles for testing complex programs. Ultimately, this collection has proved particularly helpful in debugging.
}{}

\iflong{
\crevmeta{More precisely, we considered the following combinations of typed operators by involving boolean operators with the goal of testing virtually all non-trivial interactions.}{}
\begin{itemize}
\item \crevmeta{
for objects, we test interactions among $\Props$ (as in the previous example) and between $\Props$ and $\Pro_{i}^{j}$ by setting one bound at a time than both the lower and the upper bounds, }{}
\item \crevmeta{
for arrays, we test the interactions among $\PreIte{l}{S}$ and $\PostIte{i}{S}$, but also  between these operators and $\CMM{i}{j}{S}$,}{}
\item \crevmeta{
for strings, we basically test the interaction between patterns ($\Pat$) and the lower/upper-bound for the length of string, which, in our algebra is captured in the pattern itself,}{}
\item
\crevmeta{
 for numbers, we test the interaction among $\Bet_{m}^{M}$ and $\XBet_{m}^{M}$, $\Mof(q)$, than any combination thereof.}{}
\end{itemize}
}

\hide{
which describes nested arrays such that the inner array has at least two elements and the constant 42.
This schema uses both $\xmaxIt$ and $\xcont$ to capture constraints about the array size and about its content, and happens to be a good candidate for testing the ability of any descent tool to unify between such constraints.  

\begin{center}
{\small 
\begin{tabular}{l}
\{"type" :"array",\\
\ \   "items" :    \{ "type" : "array", "minItems": 2\},\\
\ \   "contains" : \{ "contains" :\{"const" :  42 \}\}\\
\}
\end{tabular}
}
\end{center}

}

\hide{ 
this is an example that exemplifies subtle interactions
{
  "allOf": [
    { "patternProperties" : {"^(ba|ab)" : {"$ref": "#/definitions/var1"},
                             "(ab|ba)$" : {"$ref": "#/definitions/var2"}
                            },
      "additionalProperties" : false
    },
    { "not" : {"patternProperties" : {"^a" : false }}},
    { "not" : {"propertyNames" : {"not" : {"pattern" : "b" }}}},
    {"propertyNames": {"pattern" : "^..$"}},
    {"maxProperties": 1}
   ]
  ,
  "definitions": {
    "var1": { "not" : {"propertyNames" : {"not" : {"pattern" : "a"}}}},
    "var2": { "not" : {"propertyNames" : {"not" : {"pattern" : "b" }}}}
  }
}
}

\paragraph{Synthesized schemas} 
We include schemas that are neither real-world nor hand-written, but they are \emph{synthesized},
that is, they are generated
 from the reference test suite for {\jsonsch} validation~\cite{json-schema-test-suite}, designed to cover all language operators.
The derivation is described in~\cite{DBLP:conf/er/AttoucheBCDFGSS21,json-schema-containment-test-suite}, and yields triples $(S_1,S_2,b)$ 
where the Boolean~$b$ specifies whether $S_1\subseteq S_2$ holds for schemas~$S_1, S_2$. 
Here, we restrict ourselves to schemas in  \VerFour, since the CC-tool is restricted to this version. 
We excluded selected schemas that contain features that we do not yet support, such as the {\xformat} keyword (a mere technicality)
or references to external files.

We check a containment $S_1 \subseteq S_2$ by trying to generate a witness for the schema $S_1\wedge \neg S_2$, which is
unsatisfiable if, and only if, $S_1 \subseteq S_2$  holds; we thus obtain both satisfiable and unsatisfiable schemas.
The CC~tool  accepts two schemas as input and does not need this encoding.
We also test the DG~tool, where comparison is only meaningful
for pairs where $S_1\wedge \neg S_2$ is satisfiable, since the DG tool cannot recognize unsatisfiable schemas.

\subsection{Research hypotheses}

We test the following hypotheses: 
(H1)~\emph{correctness} of our implementation, that we test with the help of an external tool that verifies the generated witnesses;
(H2)~\emph{completeness} of our implementation, that we test by using an ample and diverse test-set;
(H3)~it can be used to fulfill some specific tasks better than existing tools;
(H4)~it can be implemented to run in \emph{acceptable time} on sizable real-world schemas, despite its asymptotic complexity.
We test the latest hypothesis by applying our tool to a vast set of real-world schemas.

\subsection{Experimental results}


\newcommand{\twolines}[2]{\begin{tabular}{c}#1\\(#2)\end{tabular}}

\newcommand{\fres}[2]{#1\%}

\begin{table*}[htb]
\centering
\small
\caption{\label{tab:results} Schema collections, correctness and completeness results, median/95th percentile/average runtime (in seconds).}
\vspace{-.25cm}
\libertineTabular 
\renewcommand{\arraystretch}{0.6} 
\begin{tabular}{lrccl*{8}{r}} \toprule
 \bf Collection 	& 	 {\bf \#Total }& {\begin{tabular}{l} \bf \#Sat/\\ \bf \#Unsat \end{tabular}}& \begin{tabular}{r}\bf Size (KB)\\ \bf Avg/Max \end{tabular} 	
		& 	\bf Tool	 &\bf Success    &	\bf Failure  
		&\begin{tabular}{r}\bf\!\!Errors\!\!\\ sat. \end{tabular} &  \begin{tabular}{r}\bf\!\!Errors\!\!\\ unsat. \end{tabular} 	
		& \begin{tabular}{r}\bf Med. \\ \bf Time \end{tabular} &  \begin{tabular}{r}\bf 95\% \\ \bf -tile \end{tabular} 
		 &  \begin{tabular}{r}\bf Avg. \\ \bf Time \end{tabular}  \\ 
\toprule
{GitHub} \cite{schema_corpus}	&		6,427 & 6,387/40	&	8.7/1,145 
		& Ours	
							& 99.08\%
								 & 0.92\%& 
								  \fres{0}{0} & \fres{0}0 	
								 	& 0.013 s & 0.600 s
									& 2.711 s 
									\\ 
					 		&&& & DG	&93.45\%  
								&4.89\%& \fres{1.21}{78}& \fres{0.45}{29}  
									 & 0.054 s & 0.103 s
									& 0.089 s 
									\\ 
		\midrule
%
\crevmeta{Kubernetes \cite{kuber}}{} 	& \crevmeta{1,092}{} & \crevmeta{1,087/5}{}& \crevmeta{24.0/1,310.7}{}
		& \crevmeta{Ours}{}	 
									& \crevmeta{100\%}{}  
									&\crevmeta{0\%}{}& 
									\crevmeta{0\%}{} & \crevmeta{0\%}{} 
										& \crevmeta{0.014 s}{} & \crevmeta{0.606 s}{} 
									& \crevmeta{0.605 s}{} 
										\\
							&&& & \crevmeta{DG}{}	&	\crevmeta{99.54\%}{}
									& \crevmeta{0\%}{} & \crevmeta{0\%}{} &  \crevmeta{0.46\%}{}
											& \crevmeta{0.078 s}{} & \crevmeta{0.144 s}{} 
									&\crevmeta{0.088 s}{}
											\\ 
			\midrule
\crevmeta{Snowplow \cite{snow}}{}  & \crevmeta{420}{} & \crevmeta{420/0}{}& \crevmeta{3.8/54.8}{}
		& \crevmeta{Ours}{}		
									&\crevmeta{99.52\%}{}  
									&\crevmeta{0.48\%}{}&
									\crevmeta{0\%}{} & \crevmeta{no unsat}{} 
										& \crevmeta{0.036 s}{} & \crevmeta{1.483 s}{} 
									&\crevmeta{0.892 s}{} 
										\\
							&&& & \crevmeta{DG}{}	&	\crevmeta{94.76\%}{}
									& \crevmeta{0\%}{}& \crevmeta{5.24\%}{} & \crevmeta{no unsat}{} 
											& \crevmeta{0.053 s}{} & \crevmeta{0.112 s}{} 
									& \crevmeta{0.062 s}{} 
 											\\ 
			\midrule
\crevmeta{WashingtonPost \cite{wp}}{} 	& \crevmeta{125}{} & \crevmeta{125/0}{}& 	\crevmeta{21.1/141.7}{}
		& \crevmeta{Ours}	
									& \crevmeta{100\%}{}  
									&\crevmeta{0\%}{}& 
									\crevmeta{0\%}{} & \crevmeta{no unsat}{} 
										& \crevmeta{0.021 s}{}  & \crevmeta{20.773 s}{} 
									&\crevmeta{3.622 s}{}
										\\
							&&& & \crevmeta{DG}{}	&	\crevmeta{96.8\%}{} 
									& \crevmeta{0\%}{}& \crevmeta{3.2\%}{} & \crevmeta{no unsat}{} 
											& \crevmeta{0.090 s}{} &  \crevmeta{0.181 s}{} 
									&\crevmeta{0.107 s}{} 
 											\\ 
			\toprule
Handwritten \cite{repro_package}  	& \crevmeta{233}{} & \crevmeta{195/38}{} &	\crevmeta{0.7/2.3}{}
		& \crevmeta{Ours}{}	
									& \crevmeta{100\%}{}
									& \crevmeta{0\%}{}
									& \crevmeta{0\%}{} & \crevmeta{0\%}{} 
										& \crevmeta{0.043 s}{} & \crevmeta{5.960 s}{} 
									&\crevmeta{2.454  s}{} 
										\\
							&&& & \crevmeta{DG}{}	&	\crevmeta{7.57\%}{} 
									& \crevmeta{36.87\%}{} & \crevmeta{48.99\%}{} & \crevmeta{6.57\%}{} 
											 & \crevmeta{0.072 s}{} &   \crevmeta{0.280 s}{} 
									&\crevmeta{0.091 s}{} 
											\\ 
			\midrule
{Containment-draft4}~\cite{DBLP:conf/er/AttoucheBCDFGSS21} & 1,331 & 450/881& 0.5/2.9
					&  Ours	
							& 		100\%  
							& 0\%& \fres{0}{0} &  \fres{0}{0}
										  & 0.002 s & 0.018 s 
									&0.005 s 
 \\ 
			 				&&& & DG	&		29.83\% 
							&28.85\%&   \fres{0.30}{4}  &   \fres{41.02}{336}   
									 & 0.051 s & 0.119 s
									&0.060 s 
									\\   
							&&& & CC	&	35.91\%  
									&62.96\%&  \fres{0.15}{2}&  \fres{0.98}{13}
										 & 0.003 s & 0.096 s
									&0.036  s 
										\\ %
			\bottomrule
\end{tabular}

\end{table*}

\subsubsection{Correctness and completeness}

\ifshort{
When testing each tool, we distinguish four outcomes: \emph{success}, when a result is returned and it is correct; \emph{failure}: when the code raises a run-time error or a timeout, that we set to 3,600 secs; \emph{logical error on satisfiable schema}, when the input schema~$S$ is satisfiable but the code returns either ``unsatisfiable'' or a witness that does not actually satisfy $S$; \emph{logical error on unsatisfiable schema}, when the input schema is unsatisfiable but a presumed witness is nevertheless returned.
}

\iflong{
In each run of each tool, we distinguish four outcomes:
\begin{compactitem}
 \item
\emph{success}, when a result is returned and it is correct;
\item 
 \emph{failure}: when the code raises a run-time error or a timeout, that we set at 3,600 secs (1 hour);
\item 
\emph{logical error on satisfiable schema}, when the input schema $S$ is satisfiable but the code returns either ``unsatisfiable'' or a witness that does not actually satisfy $S$;
\item 
 \emph{logical error on unsatisfiable schema}, when the input schema is unsatisfiable but a witness is nevertheless returned.
\end{compactitem}
}

\ifshort{We summarize the results of the experiments in Table~\ref{tab:results}.}
\iflong{We consider two kinds of experiments.
The first uses both the GitHub schemas and the hand-written schemas, comparing against the test data generator~DG.
The second uses the containment test suite and compares our tool with both the data generator (DG) and  the containment checker~(CC). 
We summarize the results in Table~\ref{tab:results}, together with the average and median runtimes.
}
%
%
%
\iflong{\paragraph{Our tool.}}
Our tool produces no logical error in any of our schema collections.
With the GitHub schemas, it fails with ``timeout'' for  0.56\% of sche\-mas (35 schemas),
and with ``out of memory'', when calling the automata library, for  0.36\% of schemas (23 schemas). 
\iflong{(We refer to Section~\ref{sec:problematic_schemas} for a breakdown of problematic schemas.)}
No failures arise in the other two schema collections, supporting hypothesis~H1.

\iflong{\paragraph{The data generator.}}
The DG tool successfully handles 93.45\% of the GitHub schemas, 
and has similar correctness ratio for the other real-world schemas
but it performs poorly regarding correctness on handwritten sche\-mas, and cannot be really used for inclusion checking,
since it does not detect unsatisfiability.
It is difficult to compare run-times between tools. Essentially, on most schemas the two tools have comparable times, evident when looking at the median times, but there is a small percentage of files where our tool takes a very long time, and this is reflected
on our disproportionately high average time.

\iflong{\paragraph{The containment checker.}}
The synthesized schemas show that our tool supports a much wider range of language features (hypothesis~H2),
which is natural since the CC tool targets a 
language subset, while completeness is core to our work.


\hide{\paragraph{Conclusion.}}
We can conclude that our tool advances the state-of-the-art for containment checking and witness generation, especially for sche\-mas that
present aspects of complexity (hypothesis H3).

\subsubsection{Runtime on real-world schemas}

\mrevmeta{
We next test hypothesis~H4, assessing runtime on real-world schemas.
In the three biggest collections, 95\% of the files are elaborated in less than 2.1 secs, with median 
$\leq$40 msecs, and average  $\leq$2.5 secs. 
The smaller Washington Post collection presents higher times, which 
will be discussed in Section~\ref{seq:qual}.
These results are coherent with hypothesis~H4
}{M.2, 1.3}

\hide{
\iflong{
\subsubsection{Runtime on real-world schemas}

We next test hypothesis~H4 in more detail, assessing runtime on real-world schemas.

\paragraph{Input size vs.\ runtime}
The scatterplots in Figure~\ref{fig:runtime} depict the elaboration time for each of the real-world schemas. We use a log-log
plot since both the execution times and sizes of the files span six orders of magnitudes, both with a very skewed distribution,
more readable in a log scale; this fact is evident by looking at the histograms on the top of the graph, whose
bell shape is a consequence of the log-scale used.

We first observe that great majority of dots are below the $10^3$ms lines --- i.e., the 95\% of the files can be elaborated in less
than 1~second,
which is coherent with hypothesis~H4. We also observe a clustering around a straight line with a slope not far from 1, which 
seems to indicate a size-runtime relation that is not exponential, but polynomial, so that even files in the Megabyte range may have reasonable time, below the 10 secs, or even tiny times,
in the 10 ms zone,
and the histogram of time distribution shows, in general, a median value at $10^1$ ms.
We observe that there are also many files whose runtime is in the $10^4-10^7$ ms range, and these files 
can have any size, although this is more common with big files.
More generally, we observe an extreme dispersion along the time axis for any zone of the size axis.
This shows that time does not really depend on the total size of the schema, but rather on the presence of certain 
operator arrangements that cause our algorithm to show its exponential nature \mrevmeta{and also to the argument taken by some operators, like $\xmaxL$ whose impact on the size of the Brics automaton has been statistically shown by an analysis that we performed.}{M.2}
Such arrangements can appear in any file, with bigger files having a higher probability of containing one.

\paragraph{Per-phase analysis}
To better understand the contribution of the different algorithm phases, we visualize the runtimes using boxplots in Figure~\ref{fig:boxplot};
as an optimization, the preparation and DNF-phases (P\&DNF) are implemented as a compound phase.
 
We see that the exponential P\&DNF phase has a wider span than the polynomial translation phase, with lower average but higher 
upper values,
but the costs of the two phases stays in practice comparable, which is again in line, on this set of schemas, with our hypothesis H4.

}
}

\hide{
\iflong{
\mrevmeta{
\subsubsection{Some indications about the impact of specific operators on the total execution time}
To better understand the impact of specific operators on the execution time, we opted for a statistical inference, using decision trees, and found that the most influential feature is the argument passed to $\xmaxL$.
We already knew that automata manipulation was expensive and has a cost that increases with the complexity of the regular expression but we found that, when the value of $\xmaxL$ exceeds 6,596, the execution time is above $10^3$ ms, no matter the schema at hand.}{M.2}
\mrevmeta{We confirmed this observation by analyzing the result of the Snowplow schemas where the two outliers emerge from the rest: these two schemas contains a $\xmaxL$ assertion whose argument is $10^6$ and exceed the 1 hour timeout limit to be analyzed.
}{}
\mrevmeta{Another potential reason incurring a high execution time is the use of $\xone$ with a long list of non-trivial arguments although this configuration may not be sufficient to create a blowup as we observed on a schema of Kubernetes whose root is a $\xone$ with a list of 600 arguments, and which takes 5 mn to process.
}{}
}
}

\bcolormeta
\subsection{Qualitative Insights}\label{seq:qual}
{
\mrevmeta{Several interesting insights can be extracted from an analysis of the space-time relationship 
for the GitHub collection, represented by
the scatterplot in Figure~\shortlong{\ref{fig:time}}{\ref{fig:time-github}}. The histograms at the top and at the right hand side indicate that 
schema size and run-time are distributed along 6 orders of magnitude, with a strong concentration on the low part of
both axes,
which forced us to use a log-log scale.
In the log-log plot, we observe a cloud with a slope of about 1, suggesting a linear correlation, but we also observe
that every file-size exhibits many outliers, and that long-running schemas can be found everywhere along the file-size 
axis. This clearly indicates that the runtime is affected more by the presence of specific combinations of operators,
which may take little space but cause exponential runtime, than by schema size.}{M.1, 1.6, 2.5, 3.2}

Indeed, our complexity analysis  shows that exponential complexity is triggered by some
specific operations, among which (1) object preparation, when different patterns overlap, requiring the generation of an exponential
number of \emph{choices} and of new variables;  (2) reduction to DNF; and  (3) pattern manipulation.

We tried to complement this theoretical knowledge with observations on the data.
We applied data-mining techniques to
correlate features of the schemas with the run-time. The feature that correlates more
clearly with very long run-time is the presence of a \qmaxL : $n$ statement with $n>65000$,
which induces the creation of a large automaton.
Other features with a strong correlation with high run-time are the presence of \qenum\ with extremely long
lists of arguments, that may then cause the generation of very big terms during DNF reduction,
and of \qone\ with long lists of argument, which again can generate big terms during DNF\iflong{, 
since \qone\ generates a conjunction during its translation}.

\iflong{We also resorted to visual inspection of problematic schemas, which indicated that nested objects with overlapping
patterns may also require a lot of time, as indicated by the theoretical analysis.}

The Washington Post collection required a specific analysis to explain its high 95\% percentile time and average time.
It is a smallish collection (125 schemas), where approximately 20\% of the files require around 20 secs 
for their elaboration.
All these files are very similar, with more than 2K nodes in their syntax trees and complex combinations of operators.
By selectively deleting specific subtrees, we could conclude that the high time is typically due to pattern overlapping between
an instance of \qpattProps\ and a corresponding instance of \qprops, confirming our theoretical knowledge of the 
strong influence of pattern overlapping over the complexity of
object preparation.
The small number of files in this collection and their high homogeneity explains the anomaly of the result.

Hence, the overall indication is that our algorithm fulfills its aim of proving that this exponential problem can be
successfully tackled on sizable real-world schema with a reasonable execution time, and that a careful analysis of the results of 
experiments over our vast and diverse dataset may guide further optimization efforts.
}{}
\ecolormeta


\iflong{
\mrevmeta{
Runtime for the other collections is comparable to that of GitHub with fewer timeouts for two Snowplow schemas, which contain a $\xmaxL$ assertion whose argument is $10^6$.
Another interesting observation is a schema from Kubernetes whose root consists in a $\xone$ with a list of 600 arguments, most of which are non-trivial, and which is elaborated in 5 mn. This  confirms that the use of $\xone$ may increase the running time but is not sufficient to create a blowup
}{}
}

\iflong{
\begin{figure*}[h]
 \begin{subfigure}[b]{0.25\textwidth}
         \centering
         \includegraphics[scale=0.3]{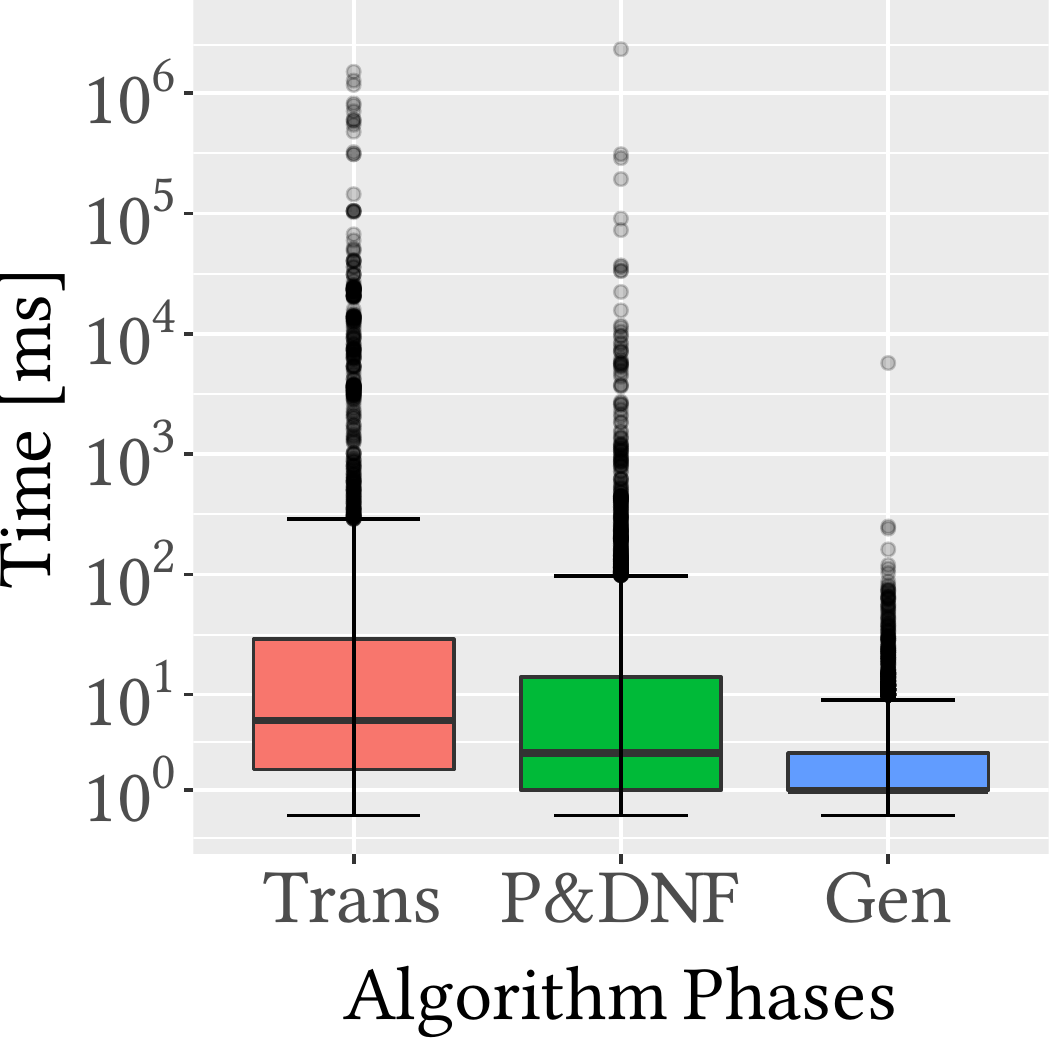}
         \caption{GitHub collection.}
         \label{fig:boxplot}
 \end{subfigure}
 \hfill
 \begin{subfigure}[b]{0.3\textwidth}
         \centering
         \includegraphics[scale=0.3]{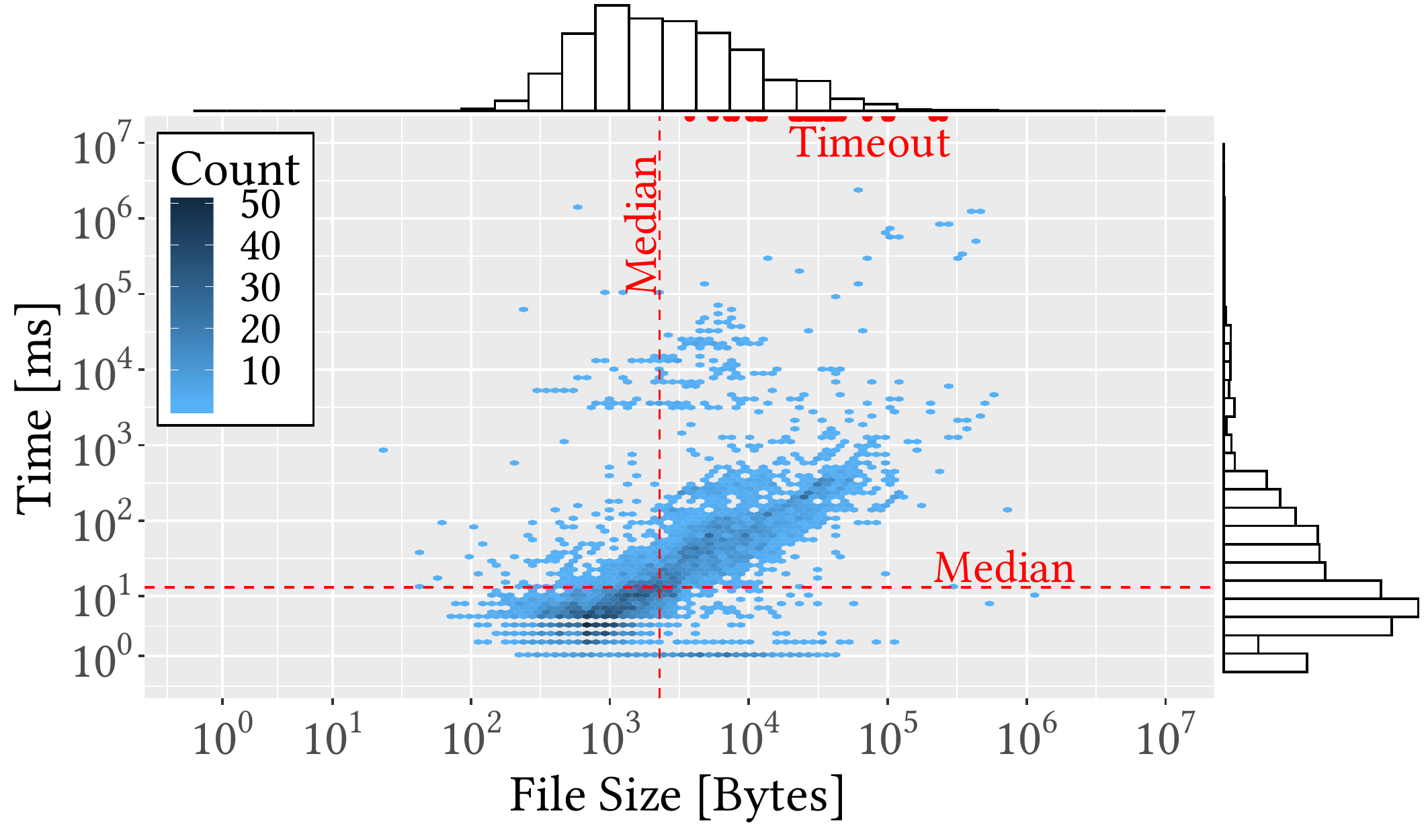}
         \caption{GitHub collection.}
         \label{fig:time-github}
 \end{subfigure}
 \hfill
 \begin{subfigure}[b]{0.3\textwidth}
         \centering
         \includegraphics[scale=0.3]{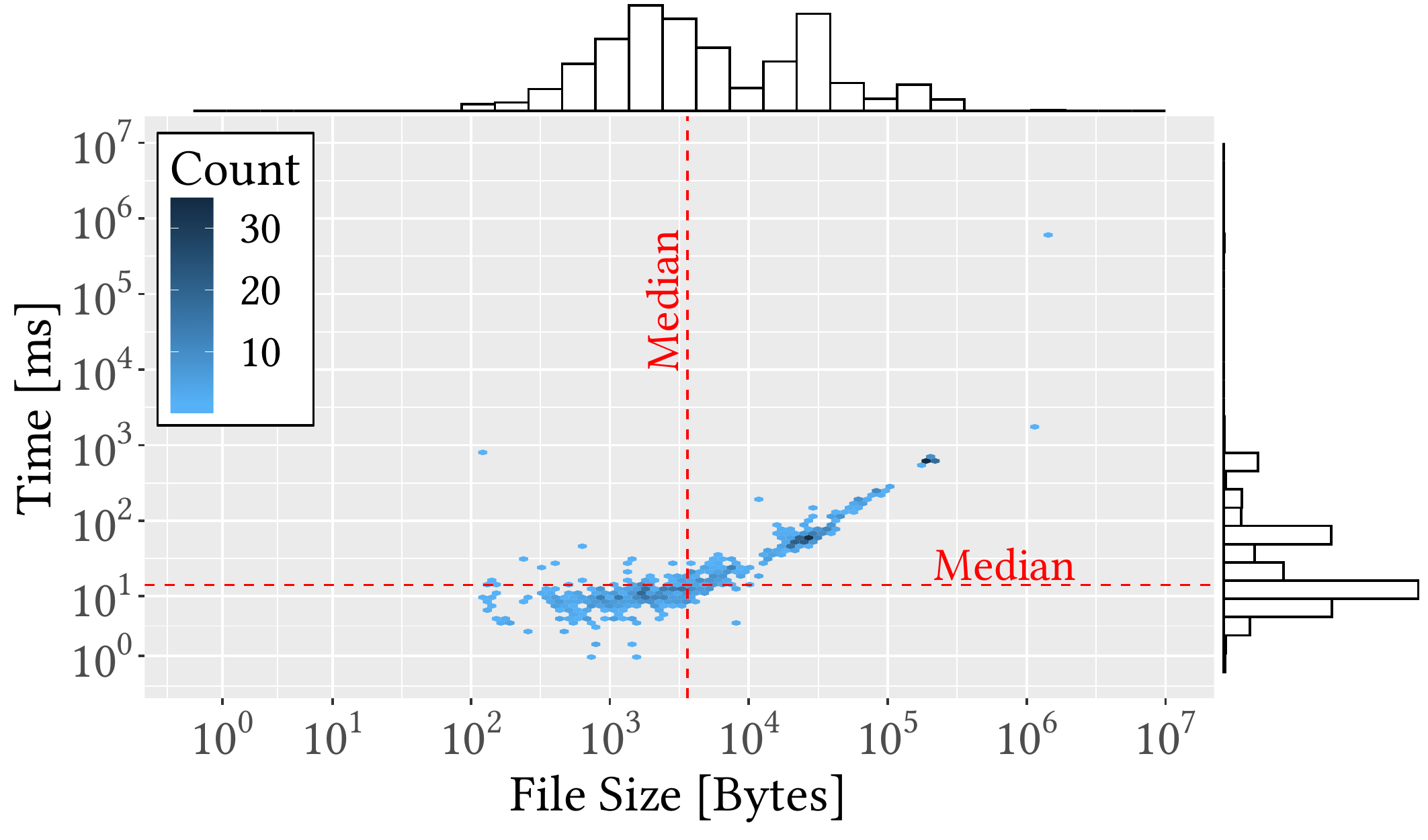}
         \caption{Kubernetes collection.}
         \label{fig:time-kuber}
 \end{subfigure}
 \hfill
 \begin{subfigure}[b]{0.37\textwidth}
         \centering
         \includegraphics[scale=0.3]{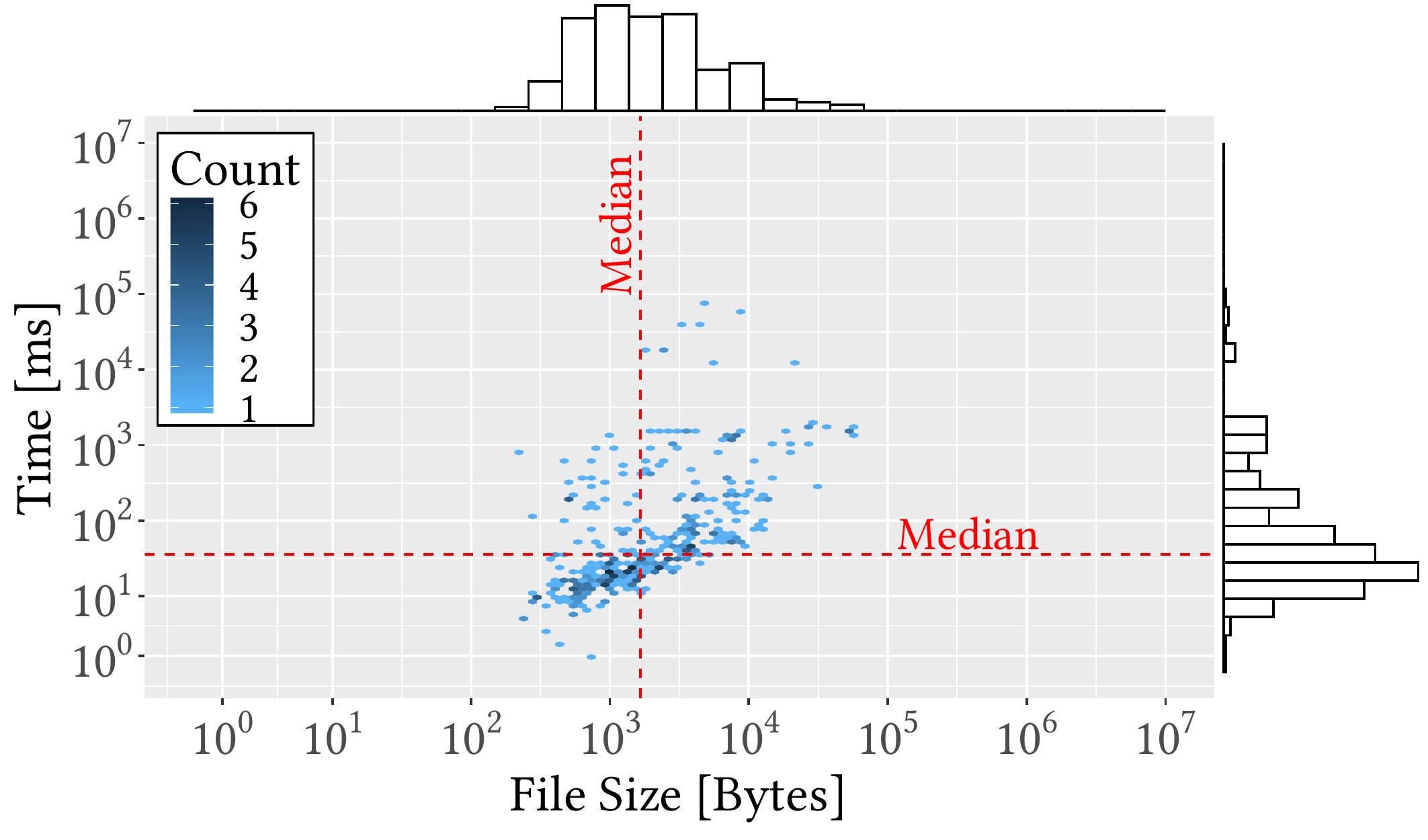}
         \caption{Snowplow collection.}
         \label{fig:time-snow}
 \end{subfigure}
 %
  \begin{subfigure}[b]{0.37\textwidth}
         \centering
         \includegraphics[scale=0.3]{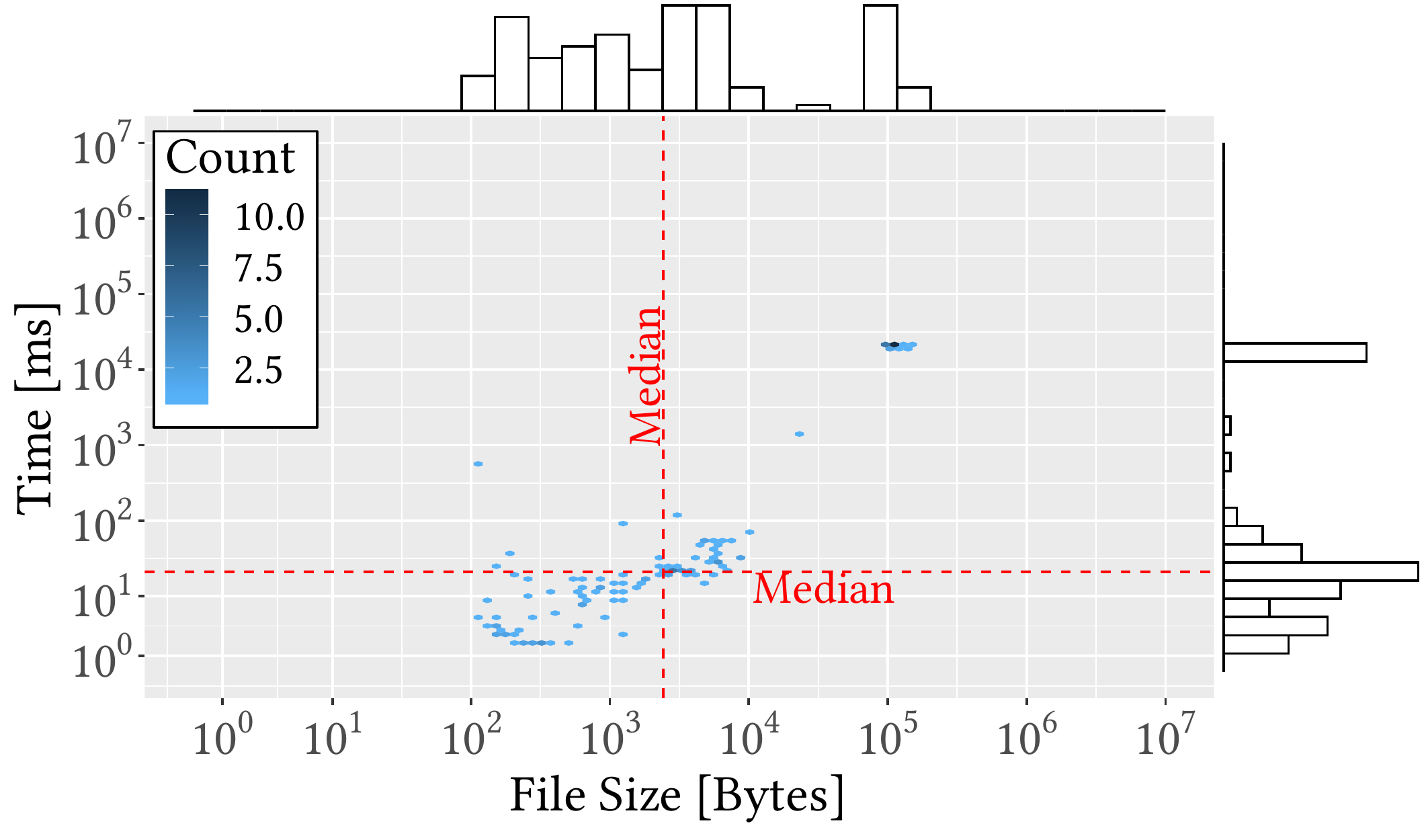}
         \caption{Washington Post collection.}
         \label{fig:time-wp}
 \end{subfigure}
  \caption{(a) Boxplots of processing times (in milliseconds, log scale) for the 3-phase witness generation algorithm, applied to GitHub schemas. 
 Boxes range from the lower to the upper quartile, horizontal line indicating the median. 
 Whiskers end at the 5th/95th percentile.
 Outliers above the whiskers are shown as individual dots, darker dots indicate overlapping values.
 (b-e)~Scatterplot showing size of the schema vs.\ time for generating a witness for the different schema collections. Along top and right edge, a stylized histogram shows the distribution.  Top right, the sizes of the files causing timeouts are shown in (b).}
\label{fig:runtime}
\end{figure*}
}


\ifshort{
\begin{figure}[h]
         \centering
         \includegraphics[width=0.48\textwidth]{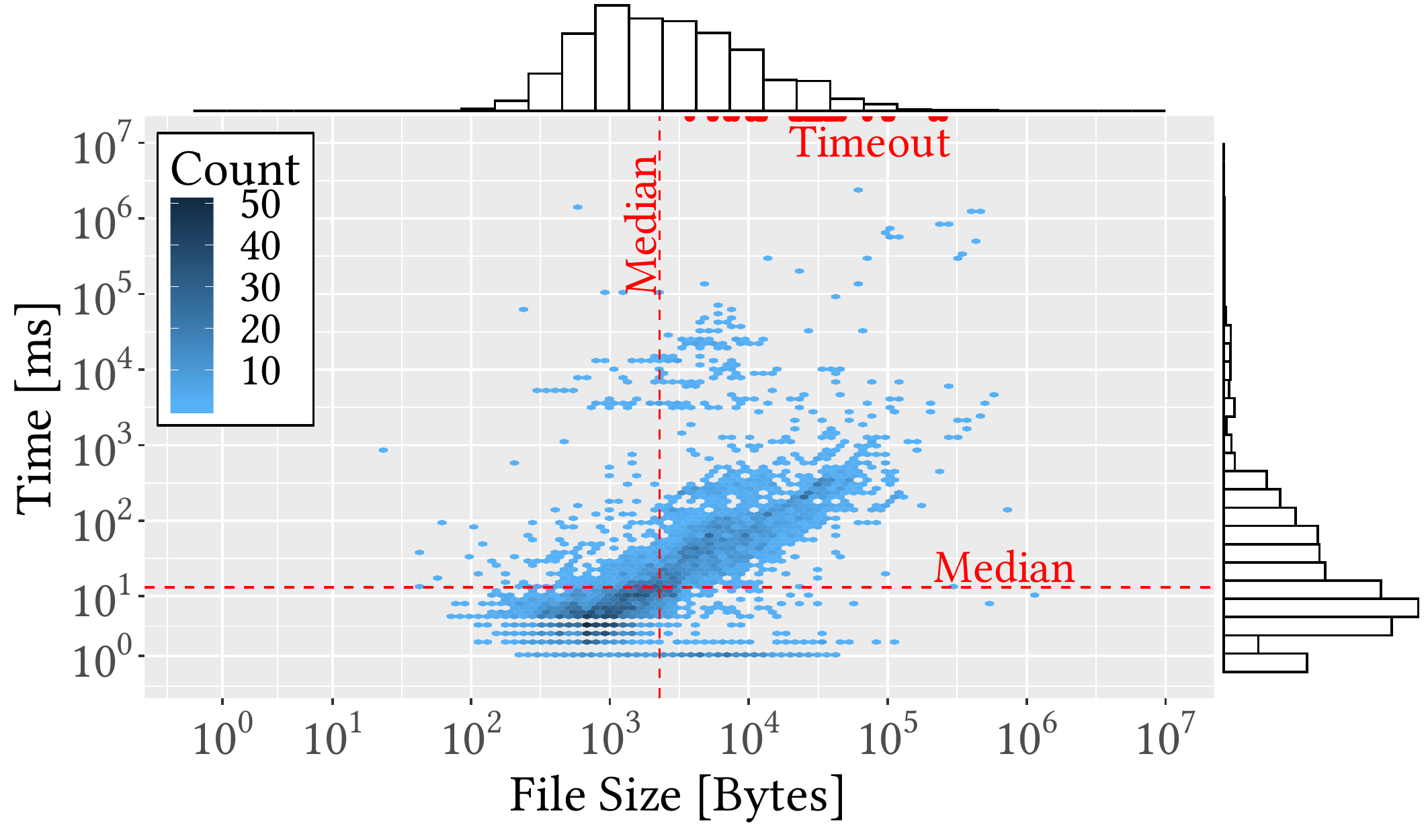}
          \caption{File size vs.\ runtime for GitHub schemas; log-log scatterplot with histograms,
		  highlighting the medians. Top right, the sizes of the files causing timeouts are shown.}

		  \label{fig:time}
\end{figure}
}

\hide{
\subsubsection{Additional results}
The full paper \cite{attouche2022witness} presents additional results:
First of all, we describe how different phases of the algorithm contribute to the total run-time. A surprising result is that the
polynomial phases --- described in Section~\ref{sec:firstphases} (minus GDNF normalization)  ---\GG{The P phases are not-elimination
and and-merge, but and-merge is not described any more...}
have on average a higher run-time  than the exponential phases (GDNF, preparation, and generation) although these
exponential phases have higher run-time in the most problematic schemas.
We analyze these most problematic schemas and we describe how they are characterized by certain operators ---
such as very long lists of properties or complex pattern expressions.

Finally, in the full paper we describe further lessons learned: 
patterns are important, and hence need special care; 
easy schemas are very common, and should be dealt with accordingly; 
the relevance of polynomial phases must not be underestimated;
in real-world schemas, \xone\ is almost invariably equivalent to \xany, allowing for an interesting optimization.
}

\iflong{
\subsubsection{Problematic schemas}
\label{sec:problematic_schemas}
Our data suggests that a very long runtime does not really depend of the size of the schema but on the presence of specific
arrangements of operators.

Our tool fails, with a timeout, only on 40  files, \crevmeta{during the phase which interleaves between preparation and DNF}{}.
In order to better understand which operator usages create problems to our algorithm, with a focus on those cases where the runtime is definitely too high, we inspected these schemas,
and verified that they all feature at least one of the following characteristics:
\begin{compactitem}
\item object specification with a very long list of properties (reaching 142 for some schema), leading to the object preparation examining a very high number of combinations ;
\item \mrevmeta{string assertions with an argument of $\xmaxL$ exceeding $10^6$ (Snowplow) or complex pattern expression combined with a relatively high argument for \xmaxL\/  (reaching 5,000): both situations lead to manipulating very large automata increasing the total cost of the entire analysis}{};
\item \mrevmeta{the use if recursive definitions  involving the root and a negation of a complex object definition, this entails a problem during object preparation and DNF construction}{}
\end{compactitem}
While these schemas present a tiny portion of the GitHub-crawled corpora, they turn out to be very useful for stress-testing our tool and for indicating optimization opportunities.

}

\iflong{
\subsection{Lessons learned}

The experiment was not only useful to verify our hypotheses, but lead us also to other relevant insights, which we summarize here.

\subsubsection{Patterns are important} Patterns appear in the {\xpatt} and {\xpattProps} operators,  and can be used to encode 
operators such as  {\xminL} , {\xmaxL} , and {\xaddProps}. Since these operators are not extremely common in real-world sche\-mas (see the empirical study in~\cite{DBLP:conf/er/BaaziziCGSS21}), it is easy
to overlook the practical relevance of patterns in {\jsonsch}, but we discovered that the high complexity of regular expression operations has noticeable impact on the performance of the algorithm. We now believe that, while it is a good idea to rely on a high-quality external library to deal with the general case, a robust tool for witness generation must also dedicate extra effort to the special cases that
arise in this specific application.

\subsubsection{Easy schemas are very common} 
Manual inspection reveals that most GitHub schemas are very simple, using a subset of the operators in a repetitive way, and 
especially the
 largest schemas tend to be simplistic, often having been automatically generated (as also observed in~\cite{DBLP:conf/er/MaiwaldRS19}).
This suggests that the average speed of any tool would greatly benefit from optimization targeted at this specific class of schemas.

\subsubsection{Polynomial phases can be relevant}
The boxplot shows that the polynomial phases of the algorithm take, on average, more time than the exponential phases.
Although we did hope that the exponential phase were manageable, this inversion was for us a surprise, and also a lesson:
do not underestimate the phases that appear inexpensive.

\subsubsection{\xone\ usually means \xany}
By a manual inspection of the schemas, we discovered that many schema designers define the different branches of
a \xone\ to be disjoint, as in 
$$\qone: [ \{ \qtype : \qnull \}, \{ \qtype : \qstr \}].$$
Hence, the designer is using \xone\ to tell the reader of the schema that the branches are disjoint, but if we substitute
that \xone\ with \xany, the semantics of the schema remains exactly the same.
This is extremely relevant, since \xone\ is a very common operator, and \xone\ is much more complex than \xany,
since it requires to compute the conjunction of each branch with the negation of all other branches.
We acted upon this observation, and implemented a very simple optimization, where we first rewrite any \xone\ to \xany, generate 
a witness for this simplified schema, check the witness against the original schema, and fall back on the complete algorithm only in the extremely rare case when the generated witness was not valid. This simple optimization proved extremely effective.

}

\section{Conclusions}\label{sec:concl}

{\jsonsch} is widely used in data-centric applications. 
\mrevmeta{The decidability and complexity of satisfiability and containment
were known, but no explicit algorithm had been defined, and it was not obvious whether the high asymptotic complexity of the problem was compatible
with a practical algorithm.
In this paper we have addressed this open problem. We have described an algorithm for witness generation, satisfiability, and containment, that 
is based on a specific combination of known and original techniques, to take into account the specific features of 
{\jsonsch} object and array operators, and the need to run in a reasonable time.
}{M.4, 2.6, 3.3}

Our extensive experiments prove the practical viability of the approach,
and provide insight into the actual behavior of the algorithm on real-world schemas.
These experiments 
are a necessary step for any redesign or re-factoring of the algorithm.

We have left the implementation of the $\xuniqIt$ operator out of the scope of the current paper in order to keep the size and complexity of this work under control, but the fundamental techniques that we have designed,
for object and array {\crcombination} and generation, still apply, with some important generalizations
that we believe deserve a dedicated analysis.

\iflong{
\newpage
\begin{small}
\paragraph{Acknowledgments}
The research has been partially supported by the MIUR project PRIN 2017FTXR7S ``IT-MaTTerS'' (Methods and Tools for Trustworthy Smart Systems)
and by the {\em Deutsche Forschungsgemeinschaft}\/ (DFG, German Research Foundation) -- 385808805.

We thank Dominik Freydenberger for proposing an algorithm to translate between ECMAScript and Brics REs.
We thank Avraham Shinnar for feedback on an earlier version of this paper.
We thank Stefan Klessinger for creating the charts and the reproduction package.
We thank the students who contributed to our implementation effort: Francesco Falleni, Cristiano Landi, Luca Escher, Lukas Ellinger, Christoph K\"ohnen, and Thomas Pilz. 
\end{small}
}
%

\bibliographystyle{ACM-Reference-Format}
\bibliography{references}

\end{document}
\endinput